\documentclass[aps,prl,twocolumn,superscriptaddress, nofootinbib]{revtex4-2}
\usepackage{graphicx,amssymb,amsmath,amsfonts,empheq}
\usepackage[dvipsnames]{xcolor}
\usepackage{color}
\usepackage{braket}
\usepackage{verbatim}
\usepackage{latexsym}
\usepackage{upgreek}
\usepackage{lipsum}
\usepackage{dsfont}
\usepackage{bm}
\usepackage{multirow}
\usepackage{enumitem}
\usepackage[normalem]{ulem}
\usepackage{url}
\usepackage{soul}
\usepackage{amsmath}
\usepackage{float}
\usepackage[export]{adjustbox}
\definecolor{orange(ryb)}{HTML}{FFA500}
\definecolor{lightorange(ryb)}{HTML}{FFB300}
\definecolor{dodgerblue}{HTML}{1E90FF}
\definecolor{lightdodgerblue}{HTML}{17b3ff}
\definecolor{crimson}{HTML}{FF4C4C}
\definecolor{pinkerton}{HTML}{EC368D}
\definecolor{forest}{HTML}{6DD189}

\usepackage[colorlinks=true,citecolor=dodgerblue,linkcolor=dodgerblue,urlcolor=dodgerblue,pdftitle={Emergent Forces}]{hyperref}

\usepackage{bbm}

\usepackage{tikz}
\usepackage{tikz-cd}
\usetikzlibrary{arrows}
\usetikzlibrary{intersections}
\usetikzlibrary{shapes.geometric}
\usetikzlibrary{decorations.pathmorphing, patterns,shapes}
\usetikzlibrary{decorations.markings}

\tikzset{
	partial ellipse/.style args={#1:#2:#3}{
		insert path={+ (#1:#3) arc (#1:#2:#3)}
	}
}

\tikzset{
	mid arrow/.style={postaction={decorate,decoration={
				markings,
				mark=at position .575 with {\arrow[#1]{stealth}}
	}}},
	near arrow/.style={postaction={decorate,decoration={
				markings,
				mark=at position .275 with {\arrow[#1]{stealth}}
	}}},
	far arrow/.style={postaction={decorate,decoration={
				markings,
				mark=at position .800 with {\arrow[#1]{stealth}}
	}}},
}

\pgfdeclarepatternformonly{south west lines}{\pgfqpoint{-0pt}{-0pt}}{\pgfqpoint{3pt}{3pt}}{\pgfqpoint{3pt}{3pt}}{
	\pgfsetlinewidth{0.4pt}
	\pgfpathmoveto{\pgfqpoint{0pt}{0pt}}
	\pgfpathlineto{\pgfqpoint{3pt}{3pt}}
	\pgfpathmoveto{\pgfqpoint{2.8pt}{-.2pt}}
	\pgfpathlineto{\pgfqpoint{3.2pt}{.2pt}}
	\pgfpathmoveto{\pgfqpoint{-.2pt}{2.8pt}}
	\pgfpathlineto{\pgfqpoint{.2pt}{3.2pt}}
	\pgfusepath{stroke}}

\makeatletter 
    
\renewcommand\onecolumngrid{
\do@columngrid{one}{\@ne}%
\def\set@footnotewidth{\onecolumngrid}
\def\footnoterule{\kern-6pt\hrule width 1.5in\kern6pt}%
}

\renewcommand\twocolumngrid{
        \def\footnoterule{
        \dimen@\skip\footins\divide\dimen@\thr@@
        \kern-\dimen@\hrule width.5in\kern\dimen@}
        \do@columngrid{mlt}{\tw@}
}%

\makeatother    

\makeatother  

\def\beq{\begin{equation}}
\def\eeq{\end{equation}}

\begin{document}
\title{Emergent Holographic Forces from Tensor Networks and Criticality}
\author{Rahul Sahay}
\affiliation{Department of Physics, Harvard University, Cambridge, MA 02138, USA}
\author{Mikhail D. Lukin}
\affiliation{Department of Physics, Harvard University, Cambridge, MA 02138, USA}
\author{Jordan Cotler}
\affiliation{Department of Physics, Harvard University, Cambridge, MA 02138, USA}
\affiliation{\it Society of Fellows, Harvard University, Cambridge, MA 02138, USA}

\begin{abstract}
The AdS/CFT correspondence stipulates a duality between conformal field theories and certain theories of quantum gravity in one higher spatial dimension.
However, probing this conjecture on contemporary classical or quantum computers is challenging.
We formulate an efficiently implementable
multi-scale entanglement renormalization ansatz
(MERA) model of AdS/CFT providing a mapping between a $(1+1)$-dimensional critical spin system and a $(2+1)$-dimensional bulk theory. 
Using a combination of numerics and analytics, we show that the bulk theory arising from this optimized tensor network furnishes excitations with attractive interactions. 
Remarkably, these excitations have one- and two-particle energies matching the predictions for matter coupled to AdS gravity at long distances, thus displaying key features of AdS physics.
We show that these potentials arise as a direct consequence of entanglement renormalization and discuss how this approach can be used to efficiently simulate bulk dynamics using realistic quantum devices.
\end{abstract}

\maketitle

\textbf{Introduction.}
One of the central problems in quantum many-body physics involves finding lattice models that realize novel physics emergent at long distances.
While simulating quantum gravity manifesting the holographic principle~\cite{hooft1993dimensional, susskind1995world} could provide new insights into the quantum nature of spacetime, such theories cannot directly emerge as the long-distance description of local lattice models since the latter violate holographic entropy bounds.
The AdS/CFT correspondence~\cite{maldacena1999large, witten1998anti} provides a surprising workaround: there are conformal field theories (CFTs) devoid of gravity that are dual to quantum gravity in Anti-de Sitter (AdS) space in one higher spatial dimension.
As such, a quantum simulation of a CFT using a local lattice Hamiltonian can, in principle, be related to a higher-dimensional gravitational system by an appropriate non-local mapping, providing a manifestation of the holographic principle~\cite{hooft1993dimensional, susskind1995world, susskind1998holographic}.
In this way, the AdS/CFT correspondence provides a pathway to probe quantum gravity through the study of quantum critical systems that emerge in synthetic quantum materials such as those realized in contemporary quantum simulators.

In practice, classical and quantum simulations of quantum gravity via AdS/CFT are challenging since the CFTs dual to ordinary gravity are strongly coupled and require large numbers of local degrees of freedom.
For this reason, it is desirable to have simplified models of AdS/CFT which capture salient features of quantum gravity and can be implemented on realistic experimental platforms.
Some candidates include the SYK model~\cite{sachdev1993gapless, kitaev2015, maldacena2016remarks, xu2020sparse, jafferis2022traversable} whose dual contains Jackiw-Teitelboim gravity, and the BFSS matrix model~\cite{banks1999m, itzhaki1998supergravity, hanada2014holographic, berkowitz2016precision, rinaldi2022matrix, pateloudis2023precision, maldacena2023simple} whose dual contains supergravity, although dynamical simulations of these models at suitable system sizes is still a challenging task~\cite{maldacena2023simple, kobrin2021_SYK, garciagracia2023sparsity}.

\begin{figure}[!t]
    \centering
    \includegraphics[width = 247 pt]{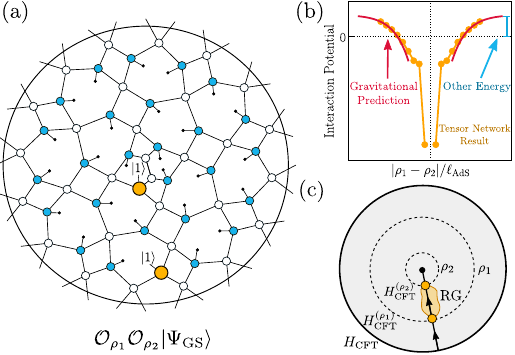}
    \caption{\textbf{Dualities from Holographic Tensor Networks.} (a) We depict a MERA representation of a CFT ground state, consisting of unitary disentanglers (shown in white) and isometric coarse-grainers (shown in blue).
    By modifying the isometric coarse-grainers (shown in yellow), we can insert excitations into the bulk.
    (b) The bulk excitations have an attractive interaction potential which reproduces features of bulk AdS physics.
    The dip in the potential is due to gravitational attraction, and the constant offset from zero is due to another force which we characterize.
    (c) The attractive potential can be understood using tensor network renormalization group methods.
    In particular, pushing the boundary Hamiltonian and the outer-most excitation towards the center of the network corresponds to transforming their operator representations via a renormalization group (RG) flow, which leads to the observed form of the potential.
    }
    \label{fig:dualities-from-MERA}
\end{figure}

Here we propose a toy model of holography amenable to experimental quantum simulations.
Our model---inspired by Chua et al.~\cite{chua_2017_holographic} and Qi~\cite{qi2013exact}---consists of a unitary circuit map between a $(1+1)$-dimensional critical spin chain and a $(2+1)$-dimensional ``bulk'' theory, built from a tensor network representation of the ground state of the chain [Fig.~\ref{fig:dualities-from-MERA}(a)].
We use this circuit to probe the bulk theory via extensive numerical simulations and analytic calculations, finding evidence that the bulk matches long-distance features of matter coupled to AdS gravity.

First, we numerically characterize the energetics of individual bulk excitations, finding that their potential energy matches the form of matter particles in an AdS background.
We show that the interactions between these excitations are attractive and fall off with distance.
Using entanglement renormalization for CFTs, we provide a quantum-mechanical understanding of the $1$ and $2$-particle potentials.
We find that the stress-energy tensor contributions to the potentials match independent AdS gravity calculations [schematically Fig.~\ref{fig:dualities-from-MERA}(b)~and~(c)].
Finally, we highlight that our holographic circuit can be realized in current quantum simulators and outline a specific approach using neutral-atom Rydberg tweezer arrays.
We note that some features of our gravity findings were anticipated by Refs.~\cite{Fitzpatrick:2012yx, fitzpatrick_2014_universality, chen_2018_ads3locality, anand_2018_exactop}, although these analyses do not apply directly to our specific CFT and moreover pertain to bulk states in a different orbital angular momentum regime than the ones we study.

\textbf{Duality Framework and Model.} To realize the duality between the CFT ground state of a qubit chain and a higher-dimensional theory, we first represent the state as a multi-scale entanglement renormalization ansatz (MERA) tensor network \cite{vidal_2008_class, vidal_2007_entanglementRG}.
Specifically, consider a $(1+1)$-dimensional periodic chain of $N = 2^{D}$ qubits, tuned to the critical point of a Hamiltonian $H$.
A MERA tensor network is designed to be a good variational ansatz for the system's ground state.
Below we review MERAs in a manner suggestive of holographic duality~\cite{swingle2012constructing}.

A MERA is defined by starting with a core state $|\psi_{\text{core}}\rangle$ of typically 4 qubits.
One grows the core state to eight qubits by bringing in four ancillas initialized as $|0\rangle^{\otimes 4}$ and coupling them to the existing qubits via unitaries
\begin{equation} \label{eq-vandw}
    \begin{tikzpicture}[scale = 0.5, baseline = {([yshift=-.5ex]current bounding box.center)}]
       \draw[color = black, ] (0, 0) -- (0.35355*2.2 , 0.35355*2.2);
       \draw[color = black, ] (0, 0) -- (0.35355*2.2 , -0.35355*2.2);
       \draw[color = black, ] (0, 0) -- (-0.35355*2.2 , 0.35355*2.2);
       \draw[color = black, ] (0, 0) -- (-0.35355*2.2 , -0.35355*2.2);
       \draw[color = black,  -stealth] (0, 0) -- (0.35355*1.7 , 0.35355*1.7);
       \draw[color = black, -stealth ] (0.35355*2.2 , -0.35355*2.2) -- (0.35355*1.1 , -0.35355*1.1);
       \draw[color = black, -stealth ] (0, 0) -- (-0.35355*1.7 , 0.35355*1.7);
       \draw[color = black, -stealth ] (-0.35355*2.2 , -0.35355*2.2) -- (-0.35355*1.1 , -0.35355*1.1);
       \filldraw[draw = black, fill = lightorange(ryb)] (0,0) circle (0.45);
        \node at (0.35355*3.2 , -0.35355*2.4) {\small $\ $};
       \node at (0, 0) {\small $w$};
   \end{tikzpicture}\hspace{-4mm}:\mathbb{C}^2 \otimes \mathbb{C}^2 \to \mathbb{C}^2 \otimes \mathbb{C}^2\,, \quad \begin{tikzpicture}[scale = 0.5, baseline = {([yshift=-.5ex]current bounding box.center)}]
       \draw[color = black, ] (0, 0) -- (0.35355*2.2 , 0.35355*2.2);
       \draw[color = black, ] (0, 0) -- (0.35355*2.2 , -0.35355*2.2);
       \draw[color = black, ] (0, 0) -- (-0.35355*2.2 , 0.35355*2.2);
       \draw[color = black, ] (0, 0) -- (-0.35355*2.2 , -0.35355*2.2);
       \draw[color = black,  -stealth] (0, 0) -- (0.35355*1.7 , 0.35355*1.7);
       \draw[color = black, -stealth ] (0.35355*2.2 , -0.35355*2.2) -- (0.35355*1.1 , -0.35355*1.1);
       \draw[color = black, -stealth ] (0, 0) -- (-0.35355*1.7 , 0.35355*1.7);
       \draw[color = black, -stealth ] (-0.35355*2.2 , -0.35355*2.2) -- (-0.35355*1.1 , -0.35355*1.1);
       \filldraw[draw = black, fill = lightorange(ryb)] (0,0) circle (0.45);
        \node at (0.35355*3.2 + 0.14 , -0.35355*2.4) {\small $\!\!\ket{0}$};
        \node at (0.35355*3.2 , 0.35355*2.4) {\small $\ $};
       \node at (0, 0) {\small $w$};
   \end{tikzpicture} = \begin{tikzpicture}[scale = 0.5, baseline = {([yshift=-.5ex]current bounding box.center)}]
   \draw[color = black, ] (0, 0) -- (0.35355*2.2 , 0.35355*2.2);
   \draw[color = black, ] (0, 0) -- (0 , -0.5*2.2);
   \draw[color = black, ] (0, 0) -- (-0.35355*2.2 , 0.35355*2.2);
   \draw[color = black,  -stealth] (0, 0) -- (0.35355*1.7 , 0.35355*1.7);
   \draw[color = black, -stealth ] (0, 0) -- (-0.35355*1.7 , 0.35355*1.7);
   \draw[color = black, -stealth ] (0, -0.5*2.2) -- (0 , -0.5*1.1);
    \draw[color = black, ] (0, 0) -- (0.35355*1.4 , -0.35355*1.4);
    \filldraw[fill = black] (0.35355*1.4 , -0.35355*1.4) circle (0.07);
    \filldraw[draw = black, fill = lightdodgerblue] (0,0) circle (0.35);
    \node at (0 , -0.35355*3.6) {\small $\ $};
    \node at (0, 0) {\small $v$};
\end{tikzpicture}\
\end{equation}
Here the tensor $v$ is an isometric map, and its leg ending in a dot is a reminder that $v$ arises from the unitary $w$ with a $\ket{0}$ contracted with one of its legs.
The state following this first stage is $w^{\otimes 4} (|0\rangle^{\otimes 4}\otimes |\psi_{\text{core}}\rangle) =  v^{\otimes 4}\ket{\psi_{\text{core}}}$.
Subsequently, local entanglement is generated by applying unitary entangling gates
\begin{equation}
    \begin{tikzpicture}[scale = 0.5, baseline = {([yshift=-.5ex]current bounding box.center)}]
       \draw[color = black, ] (0, 0) -- (0.35355*2.2 , 0.35355*2.2);
       \draw[color = black, ] (0, 0) -- (0.35355*2.2 , -0.35355*2.2);
       \draw[color = black, ] (0, 0) -- (-0.35355*2.2 , 0.35355*2.2);
       \draw[color = black, ] (0, 0) -- (-0.35355*2.2 , -0.35355*2.2);
       \draw[color = black,  -stealth] (0, 0) -- (0.35355*1.7 , 0.35355*1.7);
       \draw[color = black, -stealth ] (0.35355*2.2 , -0.35355*2.2) -- (0.35355*1.1 , -0.35355*1.1);
       \draw[color = black, -stealth ] (0, 0) -- (-0.35355*1.7 , 0.35355*1.7);
       \draw[color = black, -stealth ] (-0.35355*2.2 , -0.35355*2.2) -- (-0.35355*1.1 , -0.35355*1.1);
       \filldraw[draw = black, fill = white] (0,0) circle (0.35);
       \node at (0, 0) {\small $u$};
   \end{tikzpicture}: \mathbb{C}^{2} \otimes \mathbb{C}^{2} \to  \mathbb{C}^{2} \otimes \mathbb{C}^{2}\,
\end{equation}
resulting in $(u_{2,3} u_{4,5} u_{6,7} u_{8,1})v^{\otimes 4} |\psi_{\text{core}}\rangle$ where $u_{i,j}$ denotes $u$ applied to qubits $i$ and $j$. 
This procedure is iterated $D - 2$ times and can be notated by defining the unitaries $W_{2^{\ell-1}} \equiv  w^{\otimes 2^{\ell - 1}}$ and $U_{2^{\ell}} \equiv \prod_{i = 1}^{2^{\ell-1}} u_{2i, 2i + 1}$,  where the indices on the $u_{2i,2i+1}$ are understood to be modulo $2^\ell$.  
Repeated application of these unitaries to the core state and ancillas yields a state on $2^{D}$ qubits,
\begin{align}
 \label{E:MERAstate2}
 \ket{\Psi_{\text{MERA}}} = \prod_{\ell = 2}^{D - 1} U_{2^{{\ell + 1}}} W_{2^{\ell}}\Big(|0\rangle^{\otimes 2^{D-2}} \!\otimes |\psi_{\text{core}}\rangle\Big)\,,
\end{align}
where we have moved the ancillas $|0\rangle$ to the right.
We will subsequently denote the product of unitaries as $\mathcal{U}$. 
The resulting $2^D$-qubit state is depicted in Fig.~\ref{fig:dualities-from-MERA}(a).
In practice the tensors $w$ and $u$ are chosen to minimize the energy $\langle \Psi_\text{MERA}| H |\Psi_{\text{MERA}}\rangle$ so that $\ket{\Psi_{\text{MERA}}} \approx \ket{\Psi_{\text{GS}}}$.

It is often useful to consider the MERA circuit in reverse, wherein $v^\dagger = (\mathds{1} \otimes \langle 0|) w^\dagger$ is called a \textit{coarse-grainer} and $u^\dagger$ is called a \textit{disentangler}.
The reversed MERA circuit then implements real-space renormalization on the ``UV'' ground state $|\Psi_{\text{GS}}\rangle$ of $H$, repeatedly disentangling short-range entanglement and coarse-graining the resulting state, eventually yielding an ``IR'' core state.

To construct a holographic duality we observe that a MERA entails a unitary circuit $\mathcal{U}$ which maps the product state $|0\rangle^{\otimes 2^{D-2}} \!\!\otimes\!|\psi_{\text{core}}\rangle$ to $|\Psi_{\text{MERA}}\rangle \approx |\Psi_{\text{GS}}\rangle$.
We think of $|\Psi_{\text{GS}}\rangle$ as a state of the boundary CFT and $\ket{\Psi^{\text{bulk}}_{\text{GS}}} \equiv |0\rangle^{\otimes 2^{D-2}}\!\!\otimes\!\ket{\psi_{\text{core}}}$ as the dual state of a matter field in a bulk AdS-like theory on the disk.
Indeed, $\mathcal{U}^{\dagger}$ defines a generally non-local duality between the CFT described by Hamiltonian $H$ and a theory in one higher dimension with a product state ground state and Hamiltonian $H_{\text{bulk}} = \mathcal{U}^{\dagger} H\, \mathcal{U}$.
We will show that $\mathcal{U}$ instantiates a rich toy model of holography where the bulk manifests long-range behavior of matter coupled to AdS gravity.

\begin{figure}[!t]
    \centering
    \includegraphics[width = 247pt]{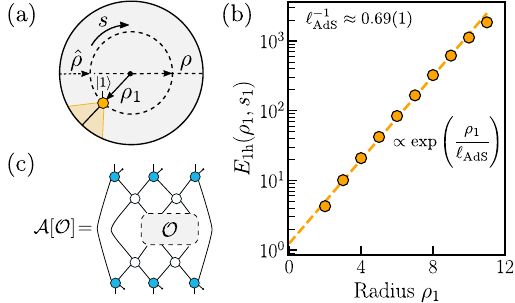}
    \caption{\textbf{Single Particle Phenomenology in the Bulk.} We use the MERA tensor network architecture to efficiently compute the energies of single excitations in the bulk theory for a system of $N = 2^{12}$ qubits.
    (a) We show the bulk coordinates $\rho$ and $s$, and depict the domain of influence of an excitation in the bulk (yellow) as we push it through the MERA circuit towards the boundary.
    (b) The numerically computed energy of a single particle is found to be mostly independent of angle and increases exponentially with radius $\rho$. 
    These findings recapitulate the exponential potential of a particle in AdS for $\rho \gtrsim  \ell_{\text{AdS}}$.
    By fitting to the form expected from gravity, we extract an effective curvature scale $\ell_{\text{AdS}}$ for the emergent spacetime.
    (c) The single particle energy can be understood microscopically by entanglement renormalization.
    Namely, the exponential behavior arises from the gapped spectrum of the MERA's ascending superoperator $\mathcal{A}$. 
 }
    \label{fig:single}
\end{figure}

\begin{figure}[!t]
    \centering
    \includegraphics[width = 247pt]{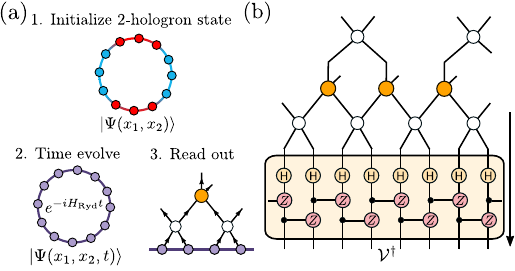}
    \caption{\textbf{Experimental Implementation.} The dual bulk physics presented in this work can be explored using contemporary quantum simulators.
    The procedure is outlined in (a) and consists of preparing a two-hologron state using the MERA circuit, time-evolving under a critical Hamiltonian using either analog or digital methods, and finally ``reading-out'' the bulk data by using the reversed MERA.
    (b) For a general critical Hamiltonian, the relevant MERA can be obtained through numerical optimization.
    However, for the transverse field Ising model, the MERA is related to the MERA used in the main text by a finite-depth local unitary $\mathcal{V}^{\dagger}$.
    }
    \label{fig:experimental}
\end{figure}

To be concrete, we consider a MERA for the ground state of the simple spin model
\begin{equation}\label{eq-H}
    H = \frac{N}{4\pi}\!\left( \sum_{s = 1}^{N} X_{s - 1} Z_s X_{s + 1} - \sum_{s = 1}^{N} X_{s} X_{s + 1}\right),
\end{equation} 
where $X,Y,Z$ are the Pauli matrices.
The normalization is chosen so that the continuum limit is that of a CFT on a circle of radius $2\pi$ (see Supplementary Materials~\cite{SM} for more details) and hence the low-lying spectrum of $H$ is that of the dilatation operator. 
The model has a $\mathbb{Z}_2 \times \mathbb{Z}_2^{T}$ symmetry generated by $\prod_s Z_s$ and an anti-unitary complex conjugation $\mathsf{K}$.
Moreover, $H$ is tuned to the critical point between a $\mathbb{Z}_2$-symmetry-breaking phase and a $\mathbb{Z}_2^{T}$-symmetry-protected topological phase, described by a $\mathbb{Z}_2 \times \mathbb{Z}_2^{T}$-enriched Ising CFT at long-distances~\cite{verresen_2021_SECFT}.
The Hamiltonian is related to the standard Ising model by a low-depth unitary circuit [shown later in Fig.~\ref{fig:experimental}(b)].
The model~\eqref{eq-H} has the convenient property that its ground state is well-approximated by a family of scale-invariant MERAs whose tensors can be determined analytically~\cite{SM, evenbly_2016_wavelets}, and we work with the lowest bond dimension representative of this family.
Here we study MERAs for finite (but large) spin chains by using the scale-invariant tensors of Ref.~\onlinecite{evenbly_2016_wavelets} for our $w$'s and $u$'s, and by numerically optimizing the core state $|\psi_{\text{core}}\rangle$.
In the Supplementary Materials~\cite{SM} we demonstrate that the MERA approximately reproduces many properties of the true ground state.
Using the MERA, we investigate features of the holographic mapping it provides.

\textbf{Single Particle Phenomenology.} To probe the bulk theory, we modify its ground state $\ket{\Psi_{\text{GS}}^{\text{bulk}}} \equiv \ket{0}^{\otimes 2^{D -2}} \!\otimes\! \ket{\psi_{\text{core}}}$ by flipping a bulk spin outside the core from $\ket{0}$ to $\ket{1}$ [c.f.~Fig.~\ref{fig:single}(a)].
The influence of the flipped bulk spin is spread by the bulk-to-boundary MERA circuit, and ultimately disrupts a portion of the CFT ground state characterized by the circuit lightcone of the bulk spin.
A similar approach was considered in Ref.~\onlinecite{chua_2017_holographic} for a gapped boundary theory where the authors coined bulk excitations as ``hologrons.'' 
In the Supplementary Materials~\cite{SM}, we explain there there is an ambiguity in the definition of the hologron but that our results are not sensitive to this ambiguity.
We investigate the energetics of single hologrons for a CFT boundary and provide both a gravitational interpretation and a microscopic quantum explanation of our numerical observations.

Let us consider the energy of a single hologron at position $x_1 = (\rho_1, s_1)$ in the bulk, where the first component is the ``radius'' indicating the circuit depth of the excitation away from the center of the tensor network and the second is the ``arclength'' coordinate labeling the ancillas at fixed radius.
In what follows, it will sometimes also be useful to use a reversed radial coordinate $\hat{\rho} \equiv D - \rho$, called the ``depth'' [c.f. Fig.~\ref{fig:single}(a)].
In our coordinates, the excitation energy of a single hologron is given by
\begin{equation} \label{eq-E1hexpression}
    E_{1\text{h}}(x_1) = \langle \mathcal{X}_{x_1} H_{\text{bulk}}\mathcal{X}_{x_1}\rangle_{\Psi^{\text{bulk}}_{\text{GS}}} - E_{\text{GS}} 
\end{equation}
where $\mathcal{X}_{x_1}, \mathcal{Y}_{x_1}, \mathcal{Z}_{x_1}$ are Pauli operators acting on the bulk site at position $x_1$, $\langle \cdots \rangle_{\psi}$ indicates an expectation value in the state $\ket{\psi}$, and $E_{\text{GS}} = \langle H \rangle_{\Psi_{\text{GS}}} = \langle H_{\text{bulk}} \rangle_{\Psi_{\text{GS}}^{\text{bulk}}}$ is the ground state energy.
Since the boundary states corresponding to a single hologron are manifestly MERA tensor network states, we can efficiently numerically evaluate Eq.~\eqref{eq-E1hexpression}.
We find that the excitation energy is largely independent of $s_1$ and increases exponentially with $\rho_1$ [see Fig.~\ref{fig:single}(b)].

\textit{Bulk AdS Interpretation.} An exponential potential is precisely the classical expectation for the energy of a matter particle in an AdS geometry.
To see this, recall that the spacetime metric for the global AdS geometry is
\begin{equation}
\label{E:metric1}
    -\cosh^2\!\left(\rho/\ell_{\text{AdS}}\right)c^2 dt^2 \!+ d\rho^2 + \ell_{\text{AdS}}^2 \sinh^2\!\left(\rho/\ell_{\text{AdS}}\right)d\theta^2
\end{equation}
where $\ell_{\text{AdS}}$ is the curvature scale.
The above is the unique maximally symmetric solution to Einstein's equations in $(2+1)$-dimensions with a negative cosmological constant $\Lambda = - 1/\ell_{\text{AdS}}^2$.
Our arclength parameter is given by $s = \ell_{\text{AdS}}\sinh(\rho/\ell_{\text{AdS}})\,\theta$.

Given the metric~\eqref{E:metric1}, one can compute the effective potential energy of a probe particle with small radial and orbital angular momentum, giving
\begin{align}
E_{\text{grav}}(x_1) = mc^2\cosh(\rho_1/\ell_{\text{AdS}})\,,
\end{align}
which agrees with the exponential obtained in the numerics.
This agreement motivates us to interpret the hologron as a matter excitation atop a background geometry defined by the tensor network.
The effective curvature scale of the geometry can be estimated by fitting $E_{1\text{h}}(\rho,s)$ to the predicted exponential form and we find $\ell_{\text{AdS}}^{-1}= 0.69(1) \approx \log(2)$.
Having $\ell_{\text{AdS}}$ be order $1$ is consistent with previous studies which argued that the MERA architecture can only access length scales greater than or equal to $\ell_{\text{AdS}}$ (see e.g.~\cite{bao2019beyond}).
As such, each bulk qubit represents an entire AdS-scale patch of space with the hologron being an AdS-scale excitation.
A comparison between the gravitational result and our numerics \cite{SM} suggests that the hologron mass is $mc^2 = 2.5(1)$.

\textit{Microscopic Origin.} A microscopic quantum understanding of the exponential potential can be obtained from entanglement renormalization, which arises from running the MERA circuit $\mathcal{U}$ in reverse.
Let us denote the microscopic energy density of the Hamiltonian $H$ by $h_s =X_{s - 1} Z_s X_{s + 1} - \frac{1}{2}(X_{s - 1} X_s + X_{s} X_{s + 1})$ so that $H = \frac{N}{4\pi} \sum_s h_s$.
Our strategy is to push $H$ up through the network to the depth of the hologron, and then to evaluate the energy at that depth.
To do so we leverage the ascension superoperator $\mathcal{A}$ ~\cite{vidal_2007_entanglementRG, Evenbly2013} [shown in Fig.~\ref{fig:single}(c)] which maps local operators at depth $\hat{\rho} \equiv D - \rho$ to depth $\hat{\rho} + 1$.
The ascension superoperator implements a discrete step of RG flow, and its eigenvalues are
$\lambda_{\alpha} = 2^{-\Delta_{\alpha}}$ where $\alpha$ labels local conformal primaries $\varphi_{\alpha}$ (for $\alpha = \mathds{1}, \varepsilon, \sigma$) and their descendants with scaling dimension $\Delta_{\alpha}$~\cite{SM, vidal_2007_entanglementRG, Evenbly2013}.

Iterating the ascension superoperator $\hat{\rho}$ times on $h_s$ gives 
\begin{equation}
\label{E:ascension1}
    \mathcal{A}^{\hat{\rho}}[h_s] = \varepsilon_s^{\text{GS}}\, \mathds{1} + c_{T}\, 2^{-\Delta_T \hat{\rho}} \varphi_{T} + c_{\bar{T}}\, 2^{-\Delta_{\bar{T}} \hat{\rho}} \varphi_{\bar{T}}   + \cdots
\end{equation}
where $T,\bar{T}$ label the holomorphic and anti-holomorphic components of the stress energy tensor, $\Delta_{T} = \Delta_{\bar{T}} = 2$, $\varepsilon_s^{\text{GS}}$ is the thermodynamic ground state energy density, $c_{T}, c_{\bar{T}}$ measures the overlap between $h_s$ and the stress energy tensor, and $\cdots$ refers to higher-order terms. 

With our ascension superoperator notations in hand, the energy of a single hologron at $x_1 = (\rho_1,s_1)$ is
\begin{equation}
\label{E:E1h1}
    E_{1\text{h}}(x_1) + E_{\text{GS}} = \frac{N}{4\pi}\sum_{s \,\in \,\text{bdy}} \langle \mathcal{A}^{\hat{\rho}_1-1}[h_s] \rangle_{\Psi^{\rho_1}_{1\text{h}}(x_1)} 
\end{equation}
where $\ket{\Psi^{\rho_1}_{1\text{h}}(x_1)}$ is the state of the MERA at radius $\rho_1$ with a single hologron at $x_1$.
Evaluating~\eqref{E:E1h1} we find
\begin{equation}
    E_{1\text{h}}(x_1) \approx \frac{N}{4\pi}\!\!\sum_{s \,\in\,\text{bdy} \,\cap\, \text{LC}(x_1)} \!\! \!\! (c_{T} + c_{\bar{T}})\,2^{-\Delta_T (\hat{\rho}_1 - 1)} f(s)
\end{equation}
where $\text{LC}(x_1)$ is the forward circuit lightcone of the hologron [the shaded region in Fig.~\ref{fig:single}(a)] and $f(s)$ comes from an expectation value involving $\varphi_T$.
Since the intersection of the light cone with the boundary has size $\sim 2^{\hat{\rho}_1}$ and $c_{T}$, $c_{\bar{T}}$, and $f(s)$ are order $1$, we find $E_{1\text{h}} \sim N\,2^{-(\Delta_T - 1) \hat{\rho}_1} \sim mc^2\, e^{\rho_1/\ell_{\text{AdS}}}$ where $\ell_{\text{AdS}}^{-1} = \log(2)$ and $mc^2$ is order $1$.
This analysis provides a quantum-mechanical explanation of the AdS potential (see the Supplementary Materials~\cite{SM} for more details).

\begin{figure}[!t]
    \centering
    \includegraphics[width =247 pt]{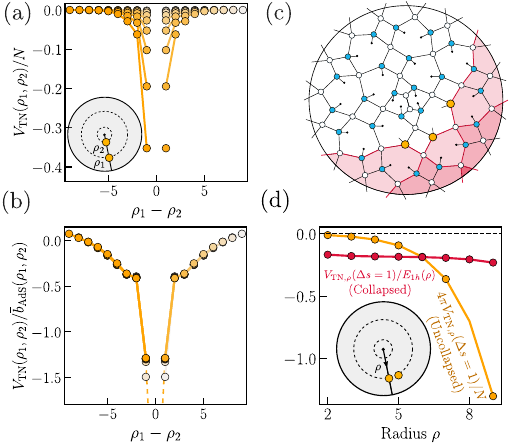}
    \caption{\textbf{Emergent Interaction Potentials in the Bulk.}
    (a) We show the interaction potential~\eqref{eq-potential} of two hologrons radially separated (see inset), numerically computed using a MERA for a system of $N = 2^{12}$ qubits as a function of $\rho_1 - \rho_2$.
    The different curves correspond to $\rho_2 \in [2, 11]$ (light to dark).
    The potential is attractive, quasi-local, and depends on the location of both particles (not strictly their distance).
    (b) A bulk AdS analysis predicts that the curves should collapse to a function of $|\rho_1 - \rho_2|$ if we divide by a simple function $\overline{b}_{\text{AdS}}(\rho_1, \rho_2) = \min\{E_{1\text{h}}(\rho_1),\,E_{1\text{h}}(\rho_2)\}$.
    We do so and remarkably the curves collapse to the predicted functional form.
    In (c, d), we investigate angularly separated hologrons at fixed radius [inset of (d)].
    (c) Owing to the lightcone structure of the MERA circuit, the domains of influence of angularly separated hologrons (shown in red) overlap only if they are within three sites of one another.
    Thus the angular potential is only non-trivial at arclength separation $\Delta s \leq 2$.
    (d) The angular potential at arclength separation $\Delta s = 1$ (shown in orange) is numerically found to be attractive and $\rho$-dependent.
    An AdS calculation predicts that the $\rho$-dependence can be largely eliminated by dividing by $E_{1\text{h}}(\rho)$, which we implement for the red curve.
    The curve flattens, consistent with the prediction.
    }
    \label{fig:interactions}
\end{figure}

\textbf{Emergent Forces.} We now consider interactions between hologrons.
Before presenting our numerical results, we consider the interaction potential for matter in AdS.

\textit{Bulk AdS Expectations.} The interaction potential between two particles located at $x_1 = (\rho_1, s_1)$ and $x_2 = (\rho_2, s_2)$ can be expressed as
\begin{equation}\label{eq-potential}
    V_{\text{int}}(x_1,x_2) = E_{2\text{h}}(x_1,x_2) - E_{1\text{h}}(x_1) - E_{1\text{h}}(x_2)\,.
\end{equation}
In the Supplementary Materials~\cite{SM} we consider two relativistic particles coupled to AdS.
We find that when the two particles are strictly radially separated ($s_1 = s_2$), have super-AdS scale separation $|\rho_1 - \rho_2| \gtrsim \ell_{\text{AdS}}$, and have small orbital angular momenta about $\rho = 0$, their interaction takes the form
\begin{equation} \label{eq-formofpotential}
    V_{\text{int}}(\rho_1, \rho_2) = b_{\text{AdS}}(\rho_1, \rho_2)\, \widetilde{V}_{\text{int}}(|\rho_1 - \rho_2|)\,,
\end{equation}
where $b_{\text{AdS}}(\rho_1,\rho_2) \approx \min\{e^{\rho_1/\ell_{\text{AdS}}}, e^{\rho_2/\ell_{\text{AdS}}}\}$ on super-AdS scales.
The prefactor $b_{\text{AdS}}(\rho_1,\rho_2)$ comes from a boost factor incurred by computing the potential in the rest frame of one of the two particles and then boosting back to the original frame. 
The gravitational contribution to $\widetilde{V}_{\text{int}}(|\rho_1-\rho_2|)$ is the attractive potential
\begin{align}
\label{E:Vtildegrav0}
\widetilde{V}_{\text{grav}}(|\rho_1 - \rho_2|) = - 8 G m^2 e^{-|\rho_1-\rho_2|/\ell_{\text{AdS}}}\,.
\end{align}

If the two particles are instead angularly separated and are at the same fixed $\rho$, it is useful to write the potential in terms of the arclength separation $|s_1 - s_2|$.
In the regime that $|s_1 - s_2|$ is fixed as $\rho/\ell_{\text{AdS}}$ is taken large, the angular potential takes the form
\begin{equation} \label{eq-angularpotential}
    V_{\text{int},\,\rho}(|s_1 - s_2|) = \cosh(\rho/\ell_{\text{AdS}}) \,\widetilde{V}_{\text{int},\,\rho}(|s_1-s_2|)\,.
\end{equation}
The prefactor $\cosh(\rho/\ell_{\text{AdS}}) \sim e^{\rho/\ell_{\text{AdS}}}/2$ arises from the radially-dependent energies of the particles.
We compute the gravitational contribution to $\widetilde{V}_{\text{int},\,\rho}(|s_1 - s_2|)$ in the Supplementary Materials~\cite{SM}.

\textit{Numerical Simulations.} 
With our expectations set, we numerically evaluate the 2-hologron potential using
\begin{align}
E_{2\text{h}}(x_1, x_2) = \langle \mathcal{X}_{x_1} \mathcal{X}_{x_2} H_{\text{bulk}}\mathcal{X}_{x_1} \mathcal{X}_{x_2}\rangle_{\Psi^{\text{bulk}}_{\text{GS}}} - E_{\text{GS}}\,.
\end{align}
We first consider two radially separated particles [as in the inset of Fig.~\ref{fig:interactions}(a)], and depict the tensor network interaction potential $V_{\text{TN}}(\rho_1, 0, \rho_2, 0)$ as a function of $\rho_1 - \rho_2$ for varying $\rho_2$ in Fig.~\ref{fig:interactions}(a).
Strikingly, the potential is both attractive and quasi-local.
Moreover if we normalize the curves in the figure by $\overline{b}_\text{AdS}(\rho_1,\rho_2) = \min\{E_{1\text{h}}(\rho_1),\,E_{1\text{h}}(\rho_2)\}$ which at super-AdS scales is proportional to $b_{\text{AdS}}(\rho_1,\rho_2)$, the curves collapse to a single function of distance $|\rho_1 - \rho_2|$ [Fig.~\ref{fig:interactions}(c)] in accordance with our bulk AdS prediction~\eqref{eq-formofpotential}.
Letting $\widetilde{V}_{\text{TN}} \equiv V_{\text{TN}}(\rho_1,\rho_2)/\overline{b}_{\text{AdS}}(\rho_1, \rho_2)$, we numerically find~\cite{SM}
\begin{align}
\label{E:numericalfit1}
\widetilde{V}_{\text{TN}}(|\rho_1 - \rho_2|) \approx C_1 - C_2\,e^{-|\rho_1-\rho_2|/\ell_{\text{AdS}}}
\end{align}
for $|\rho_1 - \rho_2| \gg  \ell_{\text{AdS}}$, where $C_1 = 0.080(8)$ and $C_2 = 13.6(8)$.
Notice that~\eqref{E:numericalfit1} is an attractive potential since it is minimized as $|\rho_1 - \rho_2|$ is taken small.

We additionally consider two angularly separated particles at fixed radius $\rho$ [as in the inset of Fig.~\ref{fig:interactions}(d)].
The angular interaction potential is guaranteed to be local and short-ranged due to the lightcone structure of the MERA.
In particular, viewing a binary MERA as a circuit from the core state to the boundary, any contiguous set of three sites at fixed radius has at most three contiguous bulk sites at each radius in its ``past lightcone,'' enabling the tensor network to be efficiently contractible.
Consequently, two angularly separated hologrons only interact if they are within two sites of one another [as shown in Fig.~\ref{fig:interactions}(c)].
Accordingly, we plot the nearest-neighbor interaction strength of two angularly separated hologrons as a function of $\rho$ in Fig.~\ref{fig:interactions}(d), finding it to be attractive (since it is negative).
Dividing the curve by $E_{1\text{h}}(\rho)$ which goes as $e^{\rho/\ell_{\text{AdS}}}$ at super-AdS scales, we see that the curve nearly flattens, consistent with our AdS prediction~\eqref{eq-angularpotential}.

\textit{Microscopic Origin.} Our numerical findings have several features that beg for a microscopic quantum understanding.
Below, we demonstrate that the full functional form of the potential can be understood through the lens of entanglement renormalization.
Here we will explain our main results, providing details in the Supplementary Materials~\cite{SM}.

Consider the radial potential~\eqref{eq-formofpotential} for $\rho_1 > \rho_2$.
To compute the interaction energy using entanglement renormalization, we simultaneously ascend the radius $\rho_1$ hologron and the boundary Hamiltonian to radius $\rho_2$.
An entanglement renormalization analysis gives
\begin{align} \label{eq-maintextVTN}
V_{\text{TN}}(\rho_1, \rho_2) \approx b_{\text{AdS}}(\rho_1, \rho_2)\sum_{\alpha \neq \mathds{1}}  C_{\alpha} \,e^{-(\Delta_{\alpha} - 1)\frac{|\rho_1 - \rho_2|}{\ell_{\text{AdS}}}}
\end{align}
where as before, $\alpha$ labels a conformal primary or descendant $\varphi_\alpha$.
The leading contributions to the sum are
\begin{align}
C_\varepsilon + (C_{T} + C_{\bar{T}} + C_{\partial \varepsilon} + C_{\bar{\partial} \varepsilon})\,e^{- |\rho_1-\rho_2|/\ell_{\text{AdS}}}\,.
\end{align}
As such, the long-range constant part of the potential $C_\varepsilon > 0$ is due to energy from $\varepsilon$-primary of the Ising CFT.
We additionally find in the Supplemental Materials~\cite{SM} that that $C_{T}, C_{\bar{T}}, C_{\partial \varepsilon}, C_{\bar{\partial} \varepsilon} < 0$ and $|C_{T}|, |C_{\bar{T}}| \gg |C_{\partial \varepsilon}|, |C_{\bar{\partial} \varepsilon}|$, meaning that there is an attractive force due to the stress tensor which dominates that from the $\varepsilon$ descendants. 
Since the boundary stress tensor is dual to the bulk graviton in AdS/CFT, the attractive part of the potential is evidently dominated by gravitational attraction.
Indeed the $b_{\text{AdS}}(\rho_1,\rho_2) \,(C_{T}+C_{\bar{T}})\,e^{-|\rho_1-\rho_2|/\ell_{\text{AdS}}}$ contribution to $V_{\text{TN}}$ matches our gravitational prediction~\eqref{E:numericalfit1}. 
Thus our analysis provides a quantitative, quantum-mechanical explanation of our numerical result~\eqref{E:numericalfit1}, and corroborates our bulk AdS predictions~\eqref{eq-formofpotential},~\eqref{E:Vtildegrav0}.

The angular potential~\eqref{eq-angularpotential} can be understood similarly.
Ascending the boundary Hamiltonian to radius $\rho$ affects the interaction energy multiplicatively as $e^{\rho/\ell_{\text{AdS}}}$, and stress tensor contributions lead to an attractive force. 

\textbf{Experimental Considerations.}
We now discuss how the bulk physics of our model can be probed experimentally with programmable quantum simulators.
There are several motivations for such experiments.
First, while the hologron energetics are informative, a comprehensive understanding of bulk dynamics necessitates exploring real-time evolution.
While many-body quantum dynamics is generally challenging to explore numerically, it is directly accessible on quantum devices.
Second, our numerics rely on the efficient simulability of MERAs at low bond dimension.
Using a quantum device it should be possible to access much larger bond dimensions, which would allow access to physics of the bulk AdS potentials at shorter distance scales as well as ameliorate non-CFT artifacts present at low bond-dimension~\cite{SM}.
Moreover, larger bond dimensions are also important for exploring CFTs beyond the Ising model. 
Finally, it would also be appealing to experimentally implement more general tensor network architectures without a strict light cone which more fully manifests bulk AdS symmetries~\cite{Evenbly:2017hyg, Milsted:2018san}, ameliorating non-AdS artifacts in the 2-hologron interaction potential emphasized in Fig.~\ref{fig:interactions}(c) (see also~\cite{Milsted:2018san}).

Our protocol for probing the dynamics of the bulk experimentally requires two ingredients: (1) the ability to perform arbitrarily non-local two-qubit gates and (2) the ability to simulate time-evolution under a Hamiltonian with a CFT ground state associated with the MERA $\mathcal{U}$.
More specifically,  our protocol consists of three steps [summarized in Fig.~\ref{fig:experimental}(a)].
We first experimentally prepare the boundary state associated with a bulk $2$-hologron state $\ket{\Psi(x_1, x_2)}$ using the digital gates of the MERA and an initial product state [(1)].
Subsequently, we use (2) to time-evolve the boundary state.
Finally, we use (1) to digitally ``read out'' the bulk by applying $\mathcal{U}^{\dagger}$.
The resulting state $\ket{\Psi_{\text{bulk}}(x_1, x_2, t)} \equiv \mathcal{U}^{\dagger}\ket{\Psi(x_1, x_2, t)}$ can be measured in the $z$-basis, or more generally characterized using shadow tomography (e.g.~\cite{van2022hardware}).
The measurement outputs would probe the bulk hologron dynamics.
The protocol can be generalized to more sophisticated entanglement renormalization schemes through a different choice of $\mathcal{U}$.

This approach can naturally be realized on several leading experimental platforms ranging from trapped ions to superconducting qubits~\cite{QuantinuumH2, QuantinuumH1_MERA, Wright_2019, Periwal_2021, Arute_Sycamore}.
Here we focus on a specific realization using
programmable Rydberg atom arrays~\cite{bluvstein2022quantum} which have emerged as promising platforms for both analog and digital quantum control~\cite{Semeghini21, ebadi2020_256, Samajdar_2020_theory, verresen2021prediction, keesling_2018_KZM, Bernien2017_51atom, sahay2023lake, giudici2022dynamical, ho_2019_scars, choi_2019_scars}.
Specifically, we note that neutral atoms can be coherently controlled with quantum logic gates and spatially reconfigured without significantly affecting stored entanglement~\cite{bluvstein2022quantum}, thereby enabling them to perform the necessary MERA-based state preparation and read-out stages of the protocol.
Moreover, the requisite time evolution can be instantiated by either digital Floquet engineering the transverse field Ising model Hamiltonian $H_{\text{TFIM}} = -\sum_{s} Z_{s} Z_{s + 1} + \sum_{s} X_s$ using periodic application of two-qubit gates or via analog evolution.
In the digital case, the MERA describing the simulated Hamiltonian's ground state is related to the MERA for the Hamiltonian of Eq.~\eqref{eq-H} by a finite-depth local unitary circuit $\mathcal{V}^{\dagger}$, shown at the bottom of Fig.~\ref{fig:experimental}(b).
In a realistic experiment, imperfect digital gate fidelities limit accessible circuit depths, motivating the exploration of a hybrid digital-analog approach.
In particular, a natural approach to analog time evolution could use the ``PXP'' Hamiltonian of the Rydberg chain, which is known to furnish an Ising CFT~\cite{keesling_2018_KZM, slagle2021CFT, Fendley_2004}.
In this PXP setting, a modified MERA accounting for local constraints~\cite{tagliacozzo2011gauge, cong2023enhancing} should be used due to the Rydberg blockade.

\textbf{Outlook.} In this work we have provided a concrete lattice instantiation of the holographic dictionary and studied its properties.
Our analysis indicates that the bulk theory provided by our tensor network model furnishes emergent features of matter coupled to AdS gravity.
We further demonstrated that these features can be quantitatively understood by using CFT data combined with real-space entanglement renormalization.
These findings motivate several interesting future directions.

Our techniques provide a new route for exploring holographic duality in concrete settings and reveal universal properties of RG flows of quantum many-body systems.
Along these lines, it would be interesting to generalize our model to other boundary theories, especially interacting ones, which may also exhibit emergent features of bulk AdS gravity.
Gapped theories will have a kind of Randall-Sundrum brane~\cite{Randall:1999vf} (an interior boundary of the bulk theory at which the RG flow terminates) in their bulk dual, which would be interesting to explore. 
Studying boundary gauge theories requires generalizing our framework to entanglement renormalization schemes for non-tensor product Hilbert spaces~\cite{tagliacozzo2011gauge}.
For exploring AdS/CFT in higher dimensions such as AdS$_4$/CFT$_3$, one would require higher-dimensional MERAs~\cite{vidal_2008_class} for which it would be prudent to leverage a quantum device.
It would also be valuable to connect our work to multi-region CFT entanglement entropies explored in quench dynamics~\cite{kusuki_2020_EE, Caputa_2019} and quantum simulators~\cite{joshi_2023_exploring}.

The most tantalizing future direction involves quantum simulation of bulk dynamics.
It is an interesting open question whether free CFTs, such as the Ising CFT studied here, can furnish bulk dynamics that bear features of gravity.
While the bulk dynamics for free CFTs arising from UV integrable Hamiltonians could potentially be simulated classically in certain regimes, efficiently simulating dynamics of UV non-integrable CFTs will require a quantum device.
MERAs for interacting CFTs would be good to explore since their bulk dynamics should be rich and perhaps more similar to gravitational dynamics.
Could we observe two hologrons move toward each other due to their mutual attraction?
Could we see a hologron `fall in' to a toy black hole, akin to~\cite{susskind2018things}?
One challenge is that hologrons may not be stable quasiparticles, but it may be possible to tune parameters of the MERA to increase stability.

Another interesting challenge is that any realistic quantum simulation will be susceptible to errors.
Note that our MERA circuit has a depth that is logarithmic in the number of qubits.
The corresponding tensor network also instantiates a partial form of quantum error correction which renders experimental protocols immune to certain kinds of errors~\cite{kim2017entanglement, kim2017robust}. 
In particular, in the Supplementary Materials~\cite{SM}, we show that while our approach is stable to individual errors, a finite error density preserves key qualitative features but affects the gravitational scaling collapse.
It would be interesting to explore if one can design MERA circuits with more robust error correction specifically suitable for holography~\cite{cong2023enhancing}.

It is plausible that a quantum simulation of our model or related ones could lead to new insights about holographic duality that cannot be gleaned by any feasible classical computation.
We emphasize that the success of such an endeavor is far from certain since the experiment could access features of our model that are neither accessible to current theory nor numerics.
Pessimistically, these features might not bear resemblance to bulk AdS physics.
More optimistically, there may be exciting opportunities to experimentally learn previously unknown features of holography that are ripe for discovery.

\textbf{Acknowledgments.}
We thank D.~Bluvstein, M.~Kalinowski, J.~Maldacena, U.~Metha, S.~Prembabu, X.~L.~Qi, T.~Schuster, M.~Strassler, A.~Strominger, and A.~Vishwanath for useful discussions.
We especially thank L.~Fitzpatrick and E.~Katz for discussions about the bulk AdS calculations, as well as S.~Divic, K.~Jensen, N. Maskara, and R.~Verresen for detailed comments on the manuscript.
R.S. acknowledges support from the U.S. Department of Energy, Office of Science, Office of Advanced Scientific Computing Research, Department of Energy Computational Science Graduate Fellowship under Award Number
DESC0022158.
J.C. is supported by a Junior Fellowship from the Harvard Society of Fellows.
This work was supported by the US Department of Energy [DE-SC0021013 and DOE Quantum Systems Accelerator Center (contract no. 7568717)],  the National Science Foundation, the Department of Defense Multidisciplinary University Research Initiative (ARO MURI, grant no. W911NF2010082), and the Harvard-MIT Center for Ultracold Atoms (NSF Physics Frontiers Center).

\bibliographystyle{apsrev4-1}
\bibliography{refs}

\pagebreak
\onecolumngrid
\appendix


\begin{center}
{\large \textbf{Supplementary Materials}}    
\end{center}

\section{Supplementary Materials A: Additional Details for Numerical Implementation of MERA Tensor Network}

Throughout the main text, we utilized a MERA tensor network optimized to approximate the ground state wavefunction of the Hamiltonian
\begin{equation} \label{eq-AppH}
    H = \frac{N}{4\pi}\! \left(\sum_s X_{s - 1}Z_{s} X_{s + 1} - \sum_{s} X_s X_{s + 1}\right) \,.
\end{equation}
In particular, our construction of the MERA consisted of using tensors obtained analytically from wavelet theory~\cite{evenbly_2016_wavelets} and numerically optimizing a core wavefunction $\ket{\psi_{\text{core}}}$.
For convenience, we report the analytically obtained tensors below:
\begin{align} \label{eq-explicittensors}
    w^{\dagger}_{x, x + 1} &= \frac{\sqrt{3} + \sqrt{2}}{4}  + \frac{\sqrt{3} - \sqrt{2}}{4} Z_x Z_{x + 1} + i \frac{1 + \sqrt{2}}{4} X_{x} Y_{x + 1} + i \frac{1 - \sqrt{2}}{4} Y_x X_{x + 1} =  \begin{tikzpicture}[scale = 0.5, baseline = {([yshift=-.5ex]current bounding box.center)}]
       \draw[color = red, ] (0, 0) -- (0.35355*2.2 , 0.35355*2.2);
       \draw[color = black, ] (0, 0) -- (0.35355*2.2 , -0.35355*2.2);
       \draw[color = black, ] (0, 0) -- (-0.35355*2.2 , 0.35355*2.2);
       \draw[color = black, ] (0, 0) -- (-0.35355*2.2 , -0.35355*2.2);
       \draw[color = red,  -stealth] (0, 0) -- (0.35355*1.7 , 0.35355*1.7);
       \draw[color = black, -stealth ] (0.35355*2.2 , -0.35355*2.2) -- (0.35355*1.1 , -0.35355*1.1);
       \draw[color = black, -stealth ] (0, 0) -- (-0.35355*1.7 , 0.35355*1.7);
       \draw[color = black, -stealth ] (-0.35355*2.2 , -0.35355*2.2) -- (-0.35355*1.1 , -0.35355*1.1);
       \filldraw[draw = black, fill = lightorange(ryb)] (0,0) circle (0.45);
        \node at (0.35355*3.2 , -0.35355*2.4) {\small $\ $};
       \node at (0, 0) {\small $w^{\dagger}$};
   \end{tikzpicture} \\
    u^{\dagger}_{x, x + 1} &= \frac{\sqrt{3} + 2}{4} + \frac{ \sqrt{3} - 2}{4} Z_x Z_{x + 1} + \frac{i}{4} X_{x} Y_{x + 1} + \frac{i}{4} Y_{x} X_{x + 1} = \begin{tikzpicture}[scale = 0.5, baseline = {([yshift=-.5ex]current bounding box.center)}]
       \draw[color = black, ] (0, 0) -- (0.35355*2.2 , 0.35355*2.2);
       \draw[color = black, ] (0, 0) -- (0.35355*2.2 , -0.35355*2.2);
       \draw[color = black, ] (0, 0) -- (-0.35355*2.2 , 0.35355*2.2);
       \draw[color = black, ] (0, 0) -- (-0.35355*2.2 , -0.35355*2.2);
       \draw[color = black,  -stealth] (0, 0) -- (0.35355*1.7 , 0.35355*1.7);
       \draw[color = black, -stealth ] (0.35355*2.2 , -0.35355*2.2) -- (0.35355*1.1 , -0.35355*1.1);
       \draw[color = black, -stealth ] (0, 0) -- (-0.35355*1.7 , 0.35355*1.7);
       \draw[color = black, -stealth ] (-0.35355*2.2 , -0.35355*2.2) -- (-0.35355*1.1 , -0.35355*1.1);
       \filldraw[draw = black, fill = white] (0,0) circle (0.45);
        \node at (0.35355*3.2 , -0.35355*2.4) {\small $\ $};
       \node at (0, 0) {\small $u^{\dagger}$};
   \end{tikzpicture}
\end{align}
Above, the ``hologron leg'' of the $w$ tensor is shown in red.\footnote{While these tensors can be found in Eq.~(1) of Ref.~\onlinecite{evenbly_2016_wavelets}, the convention for their directionality (i.e.~from UV to IR) can be found in Fig.~C1(e) of that work. While Ref.~\onlinecite{evenbly_2016_wavelets} primarily discusses these tensors in the context of the transverse field Ising model, their relevance to the $XZX - XX$ model is discussed in Section C of that paper's supplementary materials.}
Below, we provide additional numerical details that qualify the degree to which our tensor network faithfully approximates the ground state of the model~\eqref{eq-AppH} and demonstrate the robustness of our numerical results to varying the gauge of the analytically obtained tensors.

\subsection{Choice of Hamiltonian Normalization}

In the main text, there is a prefactor of $N/4\pi$ in front of the Hamiltonian.
Here we provide some comments on this normalization and its motivation.
Such a normalization is unusual when studying lattice Hamiltonians since it makes the ground state energy scale quadratically with the number of qubits squared as opposed to linearly with the number of qubits, as is familiar from basic thermodynamic considerations.
However, the $N/4\pi$ normalization is consistent with a continuum limit of a $(1 + 1)$-dimensional CFT on a circle of radius $2\pi$. 
To see this, recall that the low-energy Hamiltonian of the Ising CFT on the circle of radius $L$ is given by 
\begin{equation}
    H_{\text{CFT}} = \int_0^{L} dx\, v\,  [T(x) + \bar{T}(x)] = \int_0^{L} dx\, v\, \left[ :-i\gamma_R(x) \partial_x \gamma_R(x): + :i\gamma_L(x) \partial_x \gamma_L(x):\right]
\end{equation}
which from normal ordering has a vacuum energy of zero.
If we write the Hamiltonian in terms of momentum modes, we find
\begin{equation}
    H_{\text{CFT}} = -i \sum_{n = 0 }^{\infty} \left(\frac{2\pi n v}{L}\right) [:\gamma_{R, n} \gamma_{R, -n}: - :\gamma_{L, n} \gamma_{L, -n}:] \,.
\end{equation}
Now if we want to put the above Hamiltonian on a lattice with lattice constant $a = L/N \ll L$, then the only permissible positive momenta are: $k = 2\pi n/L \in [0, \pi/a]$.
Thus the infinite sum is truncated down to a sum from at most $0$ to $N/2$.
Furthermore, the fermions are required to live in a well-defined band that is periodic in the Brillouin zone.
To accommodate this band requirement we take
\begin{equation}
    \left( \frac{2 \pi n}{L}\right) \to \frac{1}{a} \sin\left( \frac{2 \pi n a}{L} \right) \approx  \pm \frac{2\pi n}{L}\ \quad \text{if}\  \ na \ll L \text{  or } \frac{(N - 2n)}{2}a \ll L  \,.
\end{equation}
The former condition corresponds to the right movers and the latter condition corresponds to the left movers.
Thus the lattice model corresponds to
\begin{equation}
    H_{\text{lattice}} = - \frac{i}{a} \sum_{n = 0}^{N/2} v\sin\left(\frac{2\pi n a}{L} \right) :\gamma_{n} \gamma_{-n}:\, \approx H_{\text{CFT}}
\end{equation}
provided that we only examine excitations with wavevector $2\pi n a/L \ll 1$ (or equivalently $n \ll N$).
Recalling that $a = L/N$ and taking $L = 2\pi$ and $v = 2$, we recover the normalization $N/4 \pi$ which we used in the main text.

\subsection{Relation Between Bulk and Boundary Velocities}

Throughout the main text, we make comparisons between bulk and boundary quantities.
As such, it is important to understand the relationship between the appropriate units on both sides of the duality.
To do so, recall our metric for global AdS$_3$ which we recapitulate here:
\begin{align}
\label{E:globalmetric2}
ds^2 = - \cosh^2(\rho/\ell_{\text{AdS}})\,c^2 dt^2 + d\rho^2 + \ell_{\text{AdS}}^2\sinh^2(\rho/\ell_{\text{AdS}})\,d\theta^2\,.
\end{align}
To understand the relation between the bulk metric and the boundary metric associated with the CFT, we pass to conformal coordinates by taking $\sinh(\rho/\ell_{\text{AdS}}) = \tan(\phi)$ and arrive at the metric
\begin{align}
\label{E:conformalmetric2}
ds^2 =  \frac{1}{\cos^2(\phi)}\left(-c^2 dt^2 + \ell_{\text{AdS}}^2\,d\phi^2 + \ell_{\text{AdS}}^2 \sin^2(\phi)\,d\theta^2\right).
\end{align}
To obtain the metric of the CFT boundary, we strip off the $1/\cos^2(\phi)$ conformal factor consider fixed $\phi = \pi/2$.  As such, the boundary metric is evidently
\begin{align}
ds_{\text{bdy}}^2 = - c^2 dt^2 +  \ell_{\text{AdS}}^2\, d\theta^2
\end{align}
which is the standard metric on a circle of circumference $2 \pi \ell_{\text{AdS}}$.
Given this, we can relate the bulk speed of light $c$ to the boundary speed of light (denoted by $v$) by
\begin{equation}
    \boxed{c = \frac{2\pi \ell_{\text{AdS}}}{L}\,v}
\end{equation}
where $L$ is the circumference of the circle used for the boundary CFT.

\subsection{Quality of Approximation of the Ground State}

Here we study the degree to which our MERA tensor network faithly reproduces the true ground state of~\eqref{eq-AppH}, but without the multiplicative factor of $N/4\pi$ to match with previous literature.
To do so, we first examine the overlap of our MERA wavefunction with the true wavefunction at small system sizes.
Subsequently, we compare the MERA-obtained energy density to the energy density obtained in exact diagonalization as a function of system size, finding the MERA's energy density to be close to that of the true ground state.
Finally, we show that correlation functions of the MERA decay in a manner consistent with the Ising CFT.

We start by investigating the overlap between our MERA state and the ground state found through exact diagonalization.
Since the $4$-qubit core state $\ket{\psi_{\text{core}}}$ of the MERA is variationally optimized, our MERA perfectly overlaps with the $N = 4$ ground state.
At larger system sizes of $N = 8, 16$, we find that the overlap densities remain large, being $\approx 0.998$ and $\approx 0.952$ respectively.
These large overlaps at small system sizes motivate our procedure for determining a finite-size MERA wavefunction from the scale-invariant tensors of Evenbly and White~\cite{evenbly_2016_wavelets}.

\begin{figure}[H]
    \centering
    \includegraphics[width = 0.9\textwidth]{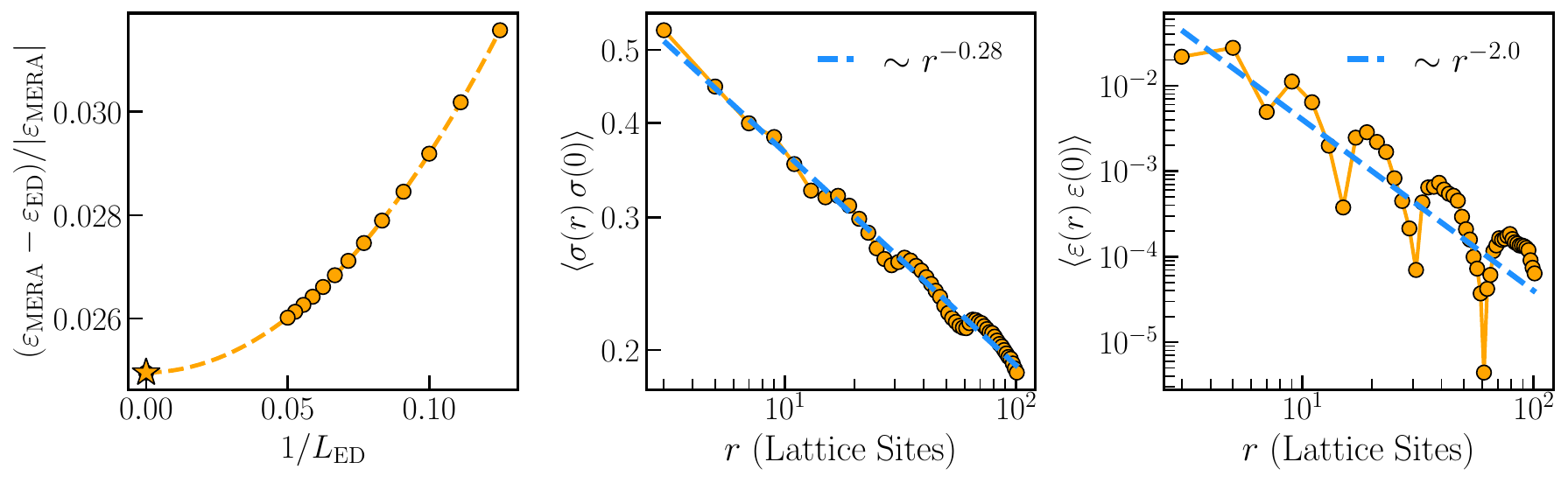}
    \caption{\textbf{Quality of Approximation of the Ground State.} We characterize the energetics and correlations of the MERA tensor network used to approximate the ground state of Eq.~\eqref{eq-H}.
    In the left panel, we depict the fractional error of the energy density of the MERA at $L_{\text{MERA}} = 2^{10}$ as compared with the energy density found in exact diagonalization at $L_{\text{ED}} = 8$ to $20$ and the exact thermodynamic energy density at $L_{\text{ED}} \to \infty$ (shown with a star). We find that the fractional error of the energy density of the MERA is around $3.2\%$ at this system size.
    In the center and right panels we depict correlation functions computed with a MERA of size $L_{\text{MERA}} = 2^{10}$ for $\sigma(r) = X_r$, a lattice proxy for the $\sigma$ primary of the Ising CFT, and $\varepsilon(r) = X_r Z_{r + 1} X_{r + 2} + \frac{1}{2}(X_rX_{r + 1} + X_{r + 1} X_{r + 2})$, a lattice proxy for the $\varepsilon$ primary.
    Both are found to be algebraically decaying as expected with exponents that are close to the theoretical expectation.}
    \label{fig:quality}
\end{figure}

In addition to small system size overlaps, we note that the energy density of the MERA can be computed in the thermodynamic limit (see Supplementary Materials D for more details) and is given by $\varepsilon_{\text{MERA}} \approx -1.24222$, deviating from the known thermodynamic energy density of $\varepsilon_{\text{ED}}(\infty) \approx -1.27323$ of $\approx 2.4\%$.
Furthermore, we theoretically show in Supplementary Materials D that the energy density for a finite size MERA deviates from the thermodynamic MERA by at most $1/N$ corrections.
Finally, in the leftmost panel of Fig.~\ref{fig:quality}, we compare the the energy density of a MERA prepared for a system of size $L_{\text{MERA}} = 2^{10} = 1024$ with the energy density found via exact diagonalization at system size $L_{\text{ED}}$ (see left panel of Fig.~\ref{fig:quality}).
We find that the fractional error $(\varepsilon_{\text{MERA}} - \varepsilon_{\text{ED}}(L_{\text{ED}}))/|\varepsilon_{\text{MERA}}|$ decreases with system size as expected and approaches the thermodynamic deviation of $\approx 2.4\%$.

Finally, we examine correlation functions computed with the MERA state.
Our objective is to verify that our MERA reproduces the correlation function of the two local primary operators of the Ising CFT: $\sigma, \varepsilon$.
This can be done by examining correlation functions of the lattice operators $\sigma(r) \sim X_r$ and $\varepsilon(r) \sim X_r Z_{r + 1} X_{r + 2} + \frac{1}{2}(X_rX_{r + 1} + X_{r + 1} X_{r + 2})$.
To understand the salience of the operators, we observe that the first operator is charged under $\mathbb{Z}_2$ symmetry, analogous to the $\sigma$ primary.
As such the leading order decay of its correlation function with distance $r$ is predicted to be algebraic $\sim r^{-2 \Delta_{\sigma}} = r^{-0.25}$.
The second operator is neutral under the $\mathbb{Z}_2$-symmetry.
However, it is odd under the Kramers-Wannier duality $X_r Z_r X_{r + 1} \leftrightarrow -X_r X_{r + 1}$, which exchanges the SPT and spontaneous symmetry-breaking phase in the model of Eq.~\eqref{eq-H} and is a non-invertible symmetry of the critical point.
The properties of the second operator are characteristic of the $\varepsilon$-primary of the Ising CFT.
As a consequence, the leading order decay of its correlation function is predicted to be algebraic with the exponent of the $\varepsilon$-primary $\sim r^{-2 \Delta_{\varepsilon}} = r^{-2}$.
We show the numerically obtained correlation functions for both in the center and right panel of Fig.~\ref{fig:quality}, finding $2\Delta_{\sigma} \approx 0.28$ and $2\Delta_{\varepsilon} \approx 2.0$, which are in the vicinity of the CFT expectations ($2\Delta_{\sigma} = 0.25$ and $2\Delta_{\varepsilon} = 2.0$ respectively) and are in agreement with the numerical findings of Ref.~\cite{evenbly_2016_wavelets} (see Supplementary Materials D for more details).
We remark that the oscillatory features of the correlation function arise as a consequence of the fact that the MERA explicitly breaks lattice translation invariance down to $\log(N)$.

\section{Supplementary Materials B: Additional Numerical Data for Hologron Energetics}

In the main text, we presented an analysis of the functional form of 1-particle and 2-particle hologron energetics for a MERA tensor network describing the ground state of Eq.~\eqref{eq-H} for a system of $2^{12}$ qubits.
Here we present additional numerical data showing that this analysis is unchanged when considering smaller system sizes, arbitrary ``gauges'' for the definition of the hologron, and imperfectly numerically optimized MERAs.

\subsection{Hologrons Energetics at Smaller System Sizes}

\begin{figure}
    \centering
    \includegraphics[width = 0.95\textwidth]{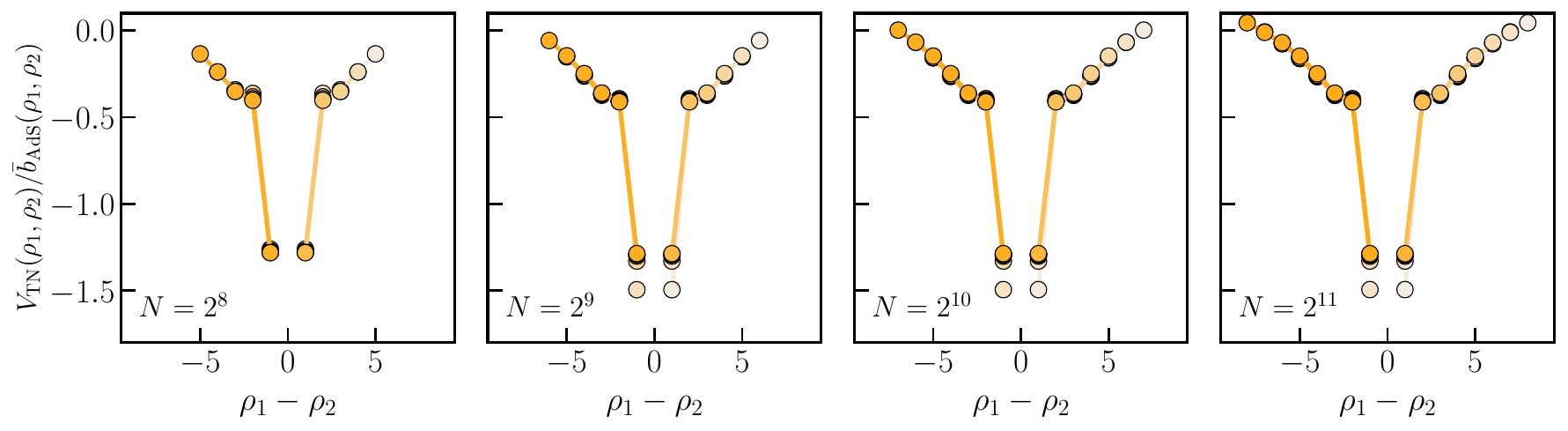}
    \caption{\textbf{Collapsed Hologron Potentials Across System Sizes.} We depict the hologron potentials at smaller system sizes than presented in the main text ($\log_2(N) = 8$ to $11$ from left to right). 
    We find qualitatively identical behavior across system sizes, even at a system size known to be experimentally accessible in programmable Rydberg atom arrays $(N = 2^8 = 256)$.
    The above suggests that the gravitational physics discussed in the main text can be probed in contemporary quantum simulation experiments.
    }
    \label{fig:sys-size}
\end{figure}

We start by showing that our hologron energetics are robust to working at smaller system sizes, which are easier to access in contemporary quantum simulators than the relatively large $2^{12}$ qubits probed in the main text.
In particular, in Fig.~\ref{fig:sys-size}, we depict the ``collapsed'' radial interaction potential for log system size $\log_2(L) = 8$ to $11$ (from left to right).
We see that for all system sizes, the interaction potential collapses by dividing by the single hologron energy, consistent with the AdS prediction.
As an added remark, we note that the plots below nicely suggest that as one probes larger boundary system sizes, more of the emergent AdS bulk potential is accessible.
Indeed, while a non-trivial collapse of the gravitational potential is already accessible at $N = 2^8 = 256$ qubits (which has been achieved previously in programmable Rydberg atom arrays \cite{ebadi2020_256, bluvstein2023logical}), Fig.~\ref{fig:sys-size} motivates experimentally probing larger systems which would likely enable probing cleaner signatures of emergent forces.

\subsection{Hologron Interaction Potentials with Arbitrary Hologron Gauge}

Recall that when optimizing a MERA to capture the critical ground state of a Hamiltonian, only the isometric $v$ tensor (See Eq.~\eqref{eq-vandw}) is specified\footnote{Up to the standard gauge transformation of tensor networks where one inserts a matrix and its inverse at each contracted leg.}  rather than the unitary $w$ tensor.
This leaves the definition of the hologron, which corresponds to inserting a $\ket{1}$ into one of the legs of the $w$ tensor, ambiguous.
Indeed, if we call the isometric tensor corresponding to the hologron $\widetilde{v}$ defined as:
\begin{equation} \label{eq-vtilde}
    \widetilde{v} = \begin{tikzpicture}[scale = 0.5, baseline = {([yshift=-.5ex]current bounding box.center)}]
       \draw[color = black, ] (0, 0) -- (0.35355*2.2 , 0.35355*2.2);
       \draw[color = black, ] (0, 0) -- (0.35355*2.2 , -0.35355*2.2);
       \draw[color = black, ] (0, 0) -- (-0.35355*2.2 , 0.35355*2.2);
       \draw[color = black, ] (0, 0) -- (-0.35355*2.2 , -0.35355*2.2);
       \draw[color = black,  -stealth] (0, 0) -- (0.35355*1.7 , 0.35355*1.7);
       \draw[color = black, -stealth ] (0.35355*2.2 , -0.35355*2.2) -- (0.35355*1.1 , -0.35355*1.1);
       \draw[color = black, -stealth ] (0, 0) -- (-0.35355*1.7 , 0.35355*1.7);
       \draw[color = black, -stealth ] (-0.35355*2.2 , -0.35355*2.2) -- (-0.35355*1.1 , -0.35355*1.1);
       \filldraw[draw = black, fill = lightorange(ryb)] (0,0) circle (0.45);
        \node at (0.35355*3.2 + 0.14 , -0.35355*2.4) {\small $\!\!\ket{1}$};
        \node at (0.35355*3.2 , 0.35355*2.4) {\small $\ $};
       \node at (0, 0) {\small $w$};
   \end{tikzpicture} = \begin{tikzpicture}[scale = 0.5, baseline = {([yshift=-.5ex]current bounding box.center)}]
   \draw[color = black, ] (0, 0) -- (0.35355*2.2 , 0.35355*2.2);
   \draw[color = black, ] (0, 0) -- (0 , -0.5*2.2);
   \draw[color = black, ] (0, 0) -- (-0.35355*2.2 , 0.35355*2.2);
   \draw[color = black,  -stealth] (0, 0) -- (0.35355*1.7 , 0.35355*1.7);
   \draw[color = black] (0, 0) -- (0.35355*2. , -0.35355*2.);
   \filldraw[fill = black] (0.35355*2. , -0.35355*2.) circle (0.07) ;
   \draw[color = black, -stealth ] (0, 0) -- (-0.35355*1.7 , 0.35355*1.7);
   \draw[color = black, -stealth ] (0, -0.5*2.2) -- (0 , -0.5*1.1);
   \filldraw[draw = black, fill = lightorange(ryb)] (0,0) circle (0.5);
    \node at (0 , -0.35355*3.6) {\small $\ $};
    \node at (0, 0) {\small $\widetilde{v}$};
\end{tikzpicture}: \mathbb{C}^2 \to \mathbb{C}^2 \otimes \mathbb{C}^2
\end{equation}
Then, the only constraint that $\widetilde{v}$ must satisfy is orthogonality to $v$: $v^{\dagger} \widetilde{v} = \widetilde{v}^{\dagger} v = 0$.
This means that if we express $\widetilde{v} = |\psi_0\rangle \langle 0| + |\psi_1\rangle \langle 1|$, where $\ket{\psi_0}$ and $\ket{\psi_1}$ are orthogonal two-qubit states in the image of $\widetilde{v}$, then any unitary rotation $G$ between $\ket{\psi_0}$ and $\ket{\psi_1}$ taking $\widetilde{v}' = G\, \widetilde{v}$ is a ``hologron gauge transformation'': a transformation of $w$ that changes the definition of the hologron while leaving the ground state invariant.

\begin{figure}
    \centering
    \includegraphics[width = 0.3\textwidth]{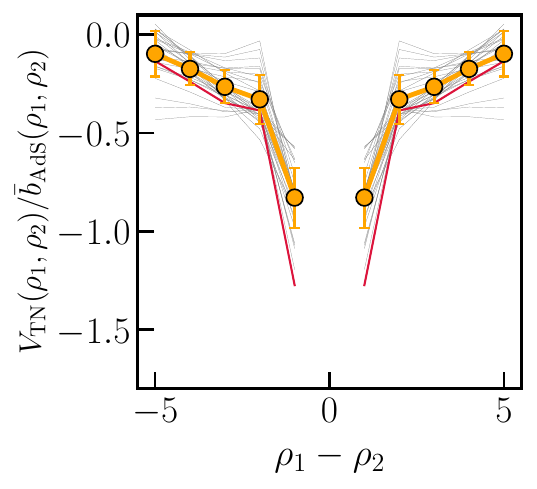}
    \caption{\textbf{Hologron Potentials in a Random Hologron Gauge.} We compute the collapsed hologron potentials for a MERA describing a chain of $2^8$ qubits with a random hologron gauge [see discussion around Eq.'s~\eqref{eq-vtilde}~and~\eqref{eq-Gthetaphi}].
    For each random gauge, we observe a good collapse for each random gauge (not shown).
    As such, only the aggregate collapsed potential is reported (found from averaging all $\rho_2$ data points at a given value of $\rho_1 - \rho_2$).
    In the plot above, we show in light gray $30$ different realizations of the hologron potentials for different random gauges.
    The average of these is shown in orange and is found to be attractive and a standard deviation for this average is also reported for each data point.
    In red we show the potential shown in the main text, which is computed by using the ``symmetric gauge'' of the tensors of Eq.~\eqref{eq-explicittensors}.}
    \label{fig:RandomGauge}
\end{figure}

Below, we demonstrate that the conclusions of the main text are largely independent of gauge. 
To do so, we compute the two-particle interaction collapsed the AdS boost factor for an ensemble of unitary MERAs with random but scale and translation invariant hologron gauges.
The ensemble is defined by taking the $w$ tensor in Eq.~\eqref{eq-explicittensors} and varying the tensors by a random hologron gauge transformation: 
\begin{equation} \label{eq-Gthetaphi}
    G(\boldsymbol{\theta}, \varphi) = v v^{\dagger} + e^{i \varphi} \,\widetilde{v} \left[e^{i \theta_x X + i \theta_y Y + i \theta_z Z} \right] \widetilde{v}^{\dagger},
\end{equation}
where $\theta_{x, y, z}$ are uniformly distributed between $[0, 2\pi)$.
Since the value of $\varphi$  cannot affect the computation of the energy, it is henceforth set to zero.
Note that $G$ crucially has the property that $Gv = v$ and $v^{\dagger} G^{(\dagger)} \widetilde{v} = \widetilde{v}^{\dagger} G^{(\dagger)} v  =  0$.
For all elements of this ensemble, we find good numerical collapse and as such, only the result of this collapse is shown in Fig.~\ref{fig:RandomGauge}.
On average, and for a majority of cases, the potential is attractive.
However, we point out that there is a small fraction of gauge choices where the potential is neither monotonically increasing nor decreasing (and so are neither attractive nor repulsive).

\subsubsection{Natural Symmetric Gauges for Hologrons}

We remark that, in the presence of symmetries, there are certain natural gauge choices for the hologron that are most compatible with the symmetries.
In particular, for the Hamiltonian of Eq.~\eqref{eq-AppH}, the global $\mathbb{Z}_2$ symmetry of the Hamiltonian generated by $\prod_s Z_s$, turns into a ``virtual'' symmetry of the tensors $u$ and $v$ where the tensors are conjugating by a product of $Z$'s on each leg of the tensor.
A natural gauge choice for the hologron is one where the $w$ tensor also enjoys this virtual symmetry, best expressed as the following diagrammatic equality:
\begin{equation}
    \begin{tikzpicture}[scale = 0.5, baseline = {([yshift=-.5ex]current bounding box.center)}]
       \draw[color = black, ] (0, 0) -- (0.35355*2.2 , 0.35355*2.2);
       \draw[color = black, ] (0, 0) -- (0.35355*2.2 , -0.35355*2.2);
       \draw[color = black, ] (0, 0) -- (-0.35355*2.2 , 0.35355*2.2);
       \draw[color = black, ] (0, 0) -- (-0.35355*2.2 , -0.35355*2.2);
       \draw[color = black,  -stealth] (0, 0) -- (0.35355*1.7 , 0.35355*1.7);
       \draw[color = black, -stealth ] (0.35355*2.2 , -0.35355*2.2) -- (0.35355*1.1 , -0.35355*1.1);
       \draw[color = black, -stealth ] (0, 0) -- (-0.35355*1.7 , 0.35355*1.7);
       \draw[color = black, -stealth ] (-0.35355*2.2 , -0.35355*2.2) -- (-0.35355*1.1 , -0.35355*1.1);
       \filldraw[draw = black, fill = lightorange(ryb)] (0,0) circle (0.45);
        \node at (0.35355*3.1 , -0.35355*2.4) {\small $Z$};
        \node at (0.35355*3.1 , 0.35355*2.4) {\small $Z $};
        \node at (-0.35355*3.1 , -0.35355*2.4) {\small $Z$};
        \node at (-0.35355*3.1 , 0.35355*2.4) {\small $Z $};
       \node at (0, 0) {\small $w$};
   \end{tikzpicture} =\     \begin{tikzpicture}[scale = 0.5, baseline = {([yshift=-.5ex]current bounding box.center)}]
       \draw[color = black, ] (0, 0) -- (0.35355*2.2 , 0.35355*2.2);
       \draw[color = black, ] (0, 0) -- (0.35355*2.2 , -0.35355*2.2);
       \draw[color = black, ] (0, 0) -- (-0.35355*2.2 , 0.35355*2.2);
       \draw[color = black, ] (0, 0) -- (-0.35355*2.2 , -0.35355*2.2);
       \draw[color = black,  -stealth] (0, 0) -- (0.35355*1.7 , 0.35355*1.7);
       \draw[color = black, -stealth ] (0.35355*2.2 , -0.35355*2.2) -- (0.35355*1.1 , -0.35355*1.1);
       \draw[color = black, -stealth ] (0, 0) -- (-0.35355*1.7 , 0.35355*1.7);
       \draw[color = black, -stealth ] (-0.35355*2.2 , -0.35355*2.2) -- (-0.35355*1.1 , -0.35355*1.1);
       \filldraw[draw = black, fill = lightorange(ryb)] (0,0) circle (0.45);
       \node at (0, 0) {\small $w$};
   \end{tikzpicture} \qquad \qquad \begin{tikzpicture}[scale = 0.5, baseline = {([yshift=-.5ex]current bounding box.center)}]
       \draw[color = black, ] (0, 0) -- (0.35355*2.2 , 0.35355*2.2);
       \draw[color = black, ] (0, 0) -- (0.35355*2.2 , -0.35355*2.2);
       \draw[color = black, ] (0, 0) -- (-0.35355*2.2 , 0.35355*2.2);
       \draw[color = black, ] (0, 0) -- (-0.35355*2.2 , -0.35355*2.2);
       \draw[color = black,  -stealth] (0, 0) -- (0.35355*1.7 , 0.35355*1.7);
       \draw[color = black, -stealth ] (0.35355*2.2 , -0.35355*2.2) -- (0.35355*1.1 , -0.35355*1.1);
       \draw[color = black, -stealth ] (0, 0) -- (-0.35355*1.7 , 0.35355*1.7);
       \draw[color = black, -stealth ] (-0.35355*2.2 , -0.35355*2.2) -- (-0.35355*1.1 , -0.35355*1.1);
       \filldraw[draw = black, fill = white] (0,0) circle (0.35);
        \node at (0.35355*3.1 , -0.35355*2.4) {\small $Z$};
        \node at (0.35355*3.1 , 0.35355*2.4) {\small $Z $};
        \node at (-0.35355*3.1 , -0.35355*2.4) {\small $Z$};
        \node at (-0.35355*3.1 , 0.35355*2.4) {\small $Z $};
       \node at (0, 0) {\small $u$};
   \end{tikzpicture} =\     \begin{tikzpicture}[scale = 0.5, baseline = {([yshift=-.5ex]current bounding box.center)}]
       \draw[color = black, ] (0, 0) -- (0.35355*2.2 , 0.35355*2.2);
       \draw[color = black, ] (0, 0) -- (0.35355*2.2 , -0.35355*2.2);
       \draw[color = black, ] (0, 0) -- (-0.35355*2.2 , 0.35355*2.2);
       \draw[color = black, ] (0, 0) -- (-0.35355*2.2 , -0.35355*2.2);
       \draw[color = black,  -stealth] (0, 0) -- (0.35355*1.7 , 0.35355*1.7);
       \draw[color = black, -stealth ] (0.35355*2.2 , -0.35355*2.2) -- (0.35355*1.1 , -0.35355*1.1);
       \draw[color = black, -stealth ] (0, 0) -- (-0.35355*1.7 , 0.35355*1.7);
       \draw[color = black, -stealth ] (-0.35355*2.2 , -0.35355*2.2) -- (-0.35355*1.1 , -0.35355*1.1);
       \filldraw[draw = black, fill = white] (0,0) circle (0.35);
       \node at (0, 0) {\small $u$};
   \end{tikzpicture}
\end{equation}
The above implies that the isometries $v, \widetilde{v}$ are forced to satisfy: $ (Z \otimes Z) v Z = v$ and $ (Z \otimes Z) \widetilde{v} Z = -\widetilde{v}$.
Under this constraint, the possible gauge transformations described in Eq.~\eqref{eq-Gthetaphi} are restricted to satisfy:
\begin{equation}
    - (Z \otimes Z) \cdot [G(\boldsymbol{\theta}, \varphi)\widetilde{v}] 
 \cdot Z = [G(\boldsymbol{\theta}, \varphi)\widetilde{v}] \implies (Z\otimes Z) G(\boldsymbol{\theta}, \varphi) (Z\otimes Z)  \widetilde{v} =  G(\boldsymbol{\theta}, \varphi)\widetilde{v}
\end{equation}
Hence, $(Z \otimes Z) G(\boldsymbol{\theta}, \varphi)(Z \otimes Z) = G(\boldsymbol{\theta}, \varphi)$, constraining $\theta_x = \theta_y = 0$.
Imposing a natural action of the time-reversal symmetry of Eq.~\eqref{eq-AppH} on the tensors motivates choosing their coefficients to be real, implying that $v = v^*$ and $\widetilde{v} = \widetilde{v}^*$.
The remaining allowed gauge transformations are those for which $G^*(\theta_z, \varphi) = G(\theta_z, \varphi)$.
The allowed gauge transformations then have $\theta_z = n \pi/2$ for $n \in \mathbb{Z}_4$ and $\varphi$ chosen to make $e^{i n \pi Z/2}$ real.

To fully fix the gauge of the tensors of Eq.~\eqref{eq-explicittensors}, an additional off-diagonal constraint is imposed that does not correspond to any boundary symmetry.
This off-diagonal constraint is expressed as: 
\begin{equation}
\begin{tikzpicture}[scale = 0.5, baseline = {([yshift=-.5ex]current bounding box.center)}]
   \draw[color = black, ] (0, 0) -- (0.35355*2.2 , 0.35355*2.2);
   \draw[color = black, ] (0, 0) -- (0.35355*2.2 , -0.35355*2.2);
   \draw[color = black, ] (0, 0) -- (-0.35355*2.2 , 0.35355*2.2);
   \draw[color = black, ] (0, 0) -- (-0.35355*2.2 , -0.35355*2.2);
   \draw[color = black,  -stealth] (0, 0) -- (0.35355*1.7 , 0.35355*1.7);
   \draw[color = black, -stealth ] (0.35355*2.2 , -0.35355*2.2) -- (0.35355*1.1 , -0.35355*1.1);
   \draw[color = black, -stealth ] (0, 0) -- (-0.35355*1.7 , 0.35355*1.7);
   \draw[color = black, -stealth ] (-0.35355*2.2 , -0.35355*2.2) -- (-0.35355*1.1 , -0.35355*1.1);
   \filldraw[draw = black, fill = lightorange(ryb)] (0,0) circle (0.45);
    \node at (0.35355*3.1 , -0.35355*2.4) {\small $X$};
    \node at (0.35355*3.1 , 0.35355*2.4) {\small $X$};
    \node at (-0.35355*3.1 , -0.35355*2.4) {\small $Y$};
    \node at (-0.35355*3.1 , 0.35355*2.4) {\small $Y$};
   \node at (0, 0) {\small $w$};
\end{tikzpicture} =\     \begin{tikzpicture}[scale = 0.5, baseline = {([yshift=-.5ex]current bounding box.center)}]
   \draw[color = black, ] (0, 0) -- (0.35355*2.2 , 0.35355*2.2);
   \draw[color = black, ] (0, 0) -- (0.35355*2.2 , -0.35355*2.2);
   \draw[color = black, ] (0, 0) -- (-0.35355*2.2 , 0.35355*2.2);
   \draw[color = black, ] (0, 0) -- (-0.35355*2.2 , -0.35355*2.2);
   \draw[color = black,  -stealth] (0, 0) -- (0.35355*1.7 , 0.35355*1.7);
   \draw[color = black, -stealth ] (0.35355*2.2 , -0.35355*2.2) -- (0.35355*1.1 , -0.35355*1.1);
   \draw[color = black, -stealth ] (0, 0) -- (-0.35355*1.7 , 0.35355*1.7);
   \draw[color = black, -stealth ] (-0.35355*2.2 , -0.35355*2.2) -- (-0.35355*1.1 , -0.35355*1.1);
   \filldraw[draw = black, fill = lightorange(ryb)] (0,0) circle (0.45);
   \node at (0, 0) {\small $w$};
\end{tikzpicture} \qquad \qquad \begin{tikzpicture}[scale = 0.5, baseline = {([yshift=-.5ex]current bounding box.center)}]
   \draw[color = black, ] (0, 0) -- (0.35355*2.2 , 0.35355*2.2);
   \draw[color = black, ] (0, 0) -- (0.35355*2.2 , -0.35355*2.2);
   \draw[color = black, ] (0, 0) -- (-0.35355*2.2 , 0.35355*2.2);
   \draw[color = black, ] (0, 0) -- (-0.35355*2.2 , -0.35355*2.2);
   \draw[color = black,  -stealth] (0, 0) -- (0.35355*1.7 , 0.35355*1.7);
   \draw[color = black, -stealth ] (0.35355*2.2 , -0.35355*2.2) -- (0.35355*1.1 , -0.35355*1.1);
   \draw[color = black, -stealth ] (0, 0) -- (-0.35355*1.7 , 0.35355*1.7);
   \draw[color = black, -stealth ] (-0.35355*2.2 , -0.35355*2.2) -- (-0.35355*1.1 , -0.35355*1.1);
   \filldraw[draw = black, fill = white] (0,0) circle (0.35);
    \node at (0.35355*3.1 , -0.35355*2.4) {\small $X$};
    \node at (0.35355*3.1 , 0.35355*2.4) {\small $X$};
    \node at (-0.35355*3.1 , -0.35355*2.4) {\small $Y$};
    \node at (-0.35355*3.1 , 0.35355*2.4) {\small $Y$};
   \node at (0, 0) {\small $u$};
\end{tikzpicture} =\     \begin{tikzpicture}[scale = 0.5, baseline = {([yshift=-.5ex]current bounding box.center)}]
   \draw[color = black, ] (0, 0) -- (0.35355*2.2 , 0.35355*2.2);
   \draw[color = black, ] (0, 0) -- (0.35355*2.2 , -0.35355*2.2);
   \draw[color = black, ] (0, 0) -- (-0.35355*2.2 , 0.35355*2.2);
   \draw[color = black, ] (0, 0) -- (-0.35355*2.2 , -0.35355*2.2);
   \draw[color = black,  -stealth] (0, 0) -- (0.35355*1.7 , 0.35355*1.7);
   \draw[color = black, -stealth ] (0.35355*2.2 , -0.35355*2.2) -- (0.35355*1.1 , -0.35355*1.1);
   \draw[color = black, -stealth ] (0, 0) -- (-0.35355*1.7 , 0.35355*1.7);
   \draw[color = black, -stealth ] (-0.35355*2.2 , -0.35355*2.2) -- (-0.35355*1.1 , -0.35355*1.1);
   \filldraw[draw = black, fill = white] (0,0) circle (0.35);
   \node at (0, 0) {\small $u$};
\end{tikzpicture}
\end{equation}
Under this constraint, we have that $(Y \otimes X) v = \widetilde{v}\, Y$ and $(Y \otimes X) \widetilde{v} = v\, Y$.
Thus we find
\begin{equation}
    (Y \otimes X)[G(\theta_z)\widetilde{v}] Y = v\,, \qquad (Y \otimes X) [G(\theta_z)v] Y = (Y \otimes X) v Y = G(\theta_z) \widetilde{v}\,.
\end{equation}
Note that the second equation fixes $\theta_z =0$ because $(Y \otimes X) \widetilde{v}\, Y = v$.

\subsection{Hologrons Potentials in Fully Numerically Optimized MERAs}

\begin{figure}
    \centering
    \includegraphics[width = 480 pt]{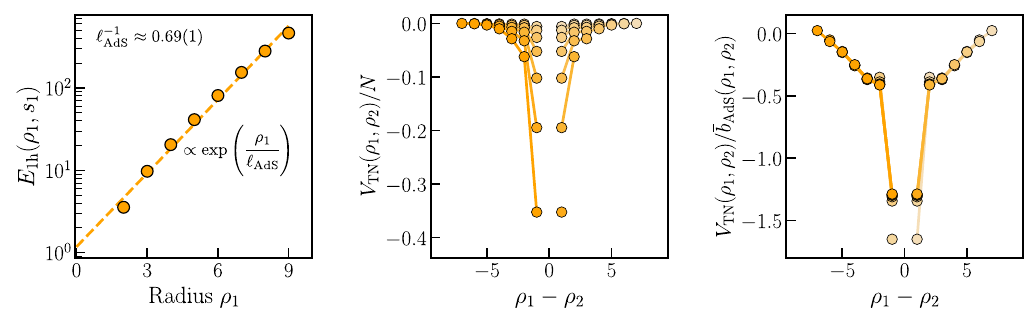}
    \caption{\textbf{Numerically Optimized MERA}. Here we compute the $1$-hologron potential, as well as collapsed and uncollapsed $2$-hologron potential for a numerically optimized MERA with $N = 2^9 = 512$.  We find essentially the same results as the exact MERA in the main text, establishing that we can find a suitable MERA for our duality by optimization alone.}
    \label{fig:XZX_opt}
\end{figure}

In the majority of our numerical simulations, we used the $w$ and $u$ MERA tensors obtained analytically from wavelet theory as discussed above.
Here we show that similar results can be obtained if the tensors are obtained from numerical optimization.
In particular, in Fig.~\ref{fig:XZX_opt}, we show the 1-hologron, 2-hologron, and collapsed potential obtained from a MERA whose tensors are fully numerically optimized.
We find qualitatively identical and quantitatively similar results to those obtained from the analytic tensors suggesting both a robustness of our results to slight deformations of the gates that occur from numerical optimization.

\section{Supplementary Materials C: AdS Gravity calculations}
\label{App:gravity}

Here we derive the potential energies of one and two particles in AdS$_3$.  We begin by introducing some of the necessary tools before presenting our main results.

\subsection{Embedding coordinates and geodesics}

It is useful to construct AdS$_3$ as a hyperboloid in the embedding space $\mathbb{R}^{2,2}$.
The reason we do so is that our subsequent analysis will exploit nonlinear symmetries of AdS$_3$ such as boost symmetries, which can be linearly realized on $\mathbb{R}^{2,2}$.
We can then perform computations in embedding space where AdS$_3$ is viewed as an embedded submanifold.
We will subsequently pass back to an intrinsic geometric description solely in terms of AdS$_3$.

The metric on AdS$_3$ is induced by pulling back the embedding space measure $\eta = \text{diag}(-1,-1,1,1)$ to the hyperboloid.  We parameterize the hyperboloid by $X^A$, satisfying
\begin{align}
X^A \eta_{AB} X^A= - \ell_{\text{AdS}}^2\,.
\end{align}
A branch of the hyperboloid can be covered by the coordinates
\begin{align}
X^A(t,\rho, \theta) &= \ell_{\text{AdS}}\Big( \cos(ct/\ell_{\text{AdS}})\,\cosh(\rho/\ell_{\text{AdS}}),\, \sin(ct/\ell_{\text{AdS}})\,\cosh(\rho/\ell_{\text{AdS}}),\\
& \qquad \qquad \qquad \qquad \qquad \qquad \qquad \qquad \qquad \qquad \qquad  \cos(\theta)\,\sinh(\rho/\ell_{\text{AdS}}),\,\sin(\theta)\,\sinh(\rho/\ell_{\text{AdS}})\Big)\,. \nonumber
\end{align}
The induced metric $dX^A \eta_{AB} dX^B = - dX_1^2 - dX_2^2 + dX_3^3 + dX_4^2$ provides the global AdS$_3$ metric
\begin{align}
\label{E:globalmetric1}
ds^2 = - \cosh^2(\rho/\ell_{\text{AdS}})\,c^2 dt^2 + d\rho^2 + \ell_{\text{AdS}}^2\sinh^2(\rho/\ell_{\text{AdS}})\,d\theta^2
\end{align}
where (upon taking the universal cover) $-\infty < t <\infty$, $0 \leq \rho < \infty$, and $0 \leq \theta < 2 \pi$.  We can recover the $(t,\rho, \theta)$ coordinates from $X^A$ using e.g.
\begin{align}
(t, \rho, \theta) = \left(\frac{\ell_{\text{AdS}}}{c}\,\text{arctan}\!\left(\frac{X^2}{X^1}\right),\,\ell_{\text{AdS}}\,\text{arccosh}\!\left(\frac{1}{\ell_{\text{AdS}}}\sqrt{(X^1)^2 + (X^2)^2}\right),\,\text{arctan}\!\left(\frac{X^4}{X^3}\right)\right)\,,
\end{align}
and being careful about domains and trigonometric inverses.

Let us compute a radially oscillating geodesic in the coordinates~\eqref{E:globalmetric1}.  Note that the timelike geodesic equation is
\begin{align}
\ddot{X}^A = - X^A \quad \text{subject to} \quad X^A \eta_{AB} X^A= - \ell_{\text{AdS}}^2\,.
\end{align}
So the solutions are of the form (see e.g.~\cite{kaplan2016lectures})
\begin{align}
X^A(t) = v_c^A \, \cos(ct/\ell_{\text{AdS}}) + v_s^A\,\sin(ct/\ell_{\text{AdS}})\,.
\end{align}
The solution corresponding to the static particle at $\rho = 0$ is
\begin{align}
X_{\text{origin}}^A(t) = (\ell_{\text{AdS}}\, \cos(ct/\ell_{\text{AdS}}),\,\ell_{\text{AdS}}\, \sin(ct/\ell_{\text{AdS}}),\,0,\,0)
\end{align}
corresponding to 
\begin{align}
v_c^A = (\ell_{\text{AdS}},0,0,0)\,,\quad v_s^A = (0,\ell_{\text{AdS}},0,0)\,.
\end{align}
A more interesting solution is
\begin{align}
X_{\rho_*}^A(t) = (\ell_{\text{AdS}}\, \cos(ct/\ell_{\text{AdS}})\,\cosh(\rho_*/\ell_{\text{AdS}}),\,\ell_{\text{AdS}}\, \sin(ct/\ell_{\text{AdS}}),\,\ell_{\text{AdS}}\,\cos(ct/\ell_{\text{AdS}})\,\sinh(\rho_*/\ell_{\text{AdS}}),\,0)
\end{align}
corresponding to
\begin{align}
v_c^A = (\ell_{\text{AdS}}\, \cosh(\rho_*/\ell_{\text{AdS}}),\,0,\,\ell_{\text{AdS}}\,\sinh(\rho_*/\ell_{\text{AdS}}),\,0)\,,\quad v_s^A = (0,\ell_{\text{AdS}},0,0)\,.
\end{align}
This describes a particle with no angular momentum and located along the line given by $\theta = 0, \pi$.  It is oscillating between $(\rho_*, 0)$ and $(\rho_*, \pi)$.  (Indeed, notice that that $X^4 = 0$ along the trajectory.)

The geodesics can be alternatively derived from the energy of a single particle in the AdS$_3$ background, given by
\begin{align}
\label{E:effectiveH1}
&E_{\text{grav}} = \cosh(\rho_*/\ell_{\text{AdS}})    \sqrt{m^2 c^4 + P_\rho^2 c^2 + \frac{P_\theta^2 c^2}{\ell_{\text{AdS}}^2\sinh^2(\rho_*/\ell_{\text{AdS}})}}\,.
\end{align}
In the regime that $P_\rho, P_\theta$ are small,\footnote{This regime can be contrasted with e.g.~a circular orbit at a fixed radius $\rho_*$, wherein the angular momentum is $P_\theta = \pm mc\,\ell\sinh^2(\rho_*/\ell)$ and so the energy is $E = mc^2 \cosh^2(\rho_*/\ell)$.} the energy becomes
\begin{align}
\boxed{E_{\text{grav}} = mc^2 \cosh(\rho_*/\ell_{\text{AdS}})\,.}
\end{align}
For $\rho_* \gtrsim \ell_{\text{AdS}}$ the above is proportional to $e^{\rho_*/\ell_{\text{AdS}}}$, which agrees with the form of our single-particle energies in the main text.

Going forward we will want to construct a boost which maps our oscillating particle to a stationary trajectory at the origin. To this end, let us find an embedding space $SO(2,2)$ transformation $M$ which boosts us from $X_{\rho_*}^A(t)$ to $X_{\text{origin}}^A(t)$. Since we must map the $v_c^A$ and $v_s^A$ from one solution to those of the other solution, we have some partial information.  It is easy to guess that
\begin{align}
M_{\rho_* \to 0} = \begin{pmatrix}
\cosh(\rho_*/\ell_{\text{AdS}}) & 0 & - \sinh(\rho_*/\ell_{\text{AdS}}) & 0 \\
0 & 1& 0 & 0 \\
- \sinh(\rho_*/\ell_{\text{AdS}}) & 0 & \cosh(\rho_*/\ell_{\text{AdS}}) & 0 \\
0 & 0 & 0 & 1
\end{pmatrix}\,,
\end{align}
which does what we want.

\subsection{Two particle radial potential}

Now we consider two particles (labeled by $1$ and $2$) with rest mass $m$ along a $\theta = 0, \pi$ radial geodesic.  They are initialized at $(\rho_1, 0)$ and $(\rho_2,0)$, respectively, we we suppose that $1 \ll \rho_2/\ell_{\text{AdS}} \ll \rho_1/\ell_{\text{AdS}}$.  If we boost by $M_{\rho_2 \to 0}$ to coordinates $(\tilde{t},\tilde{\rho},\tilde{\theta})$, then particle $2$ becomes stationary at $\tilde{\rho} = 0$, and particle $1$ travels along a radial geodesic with maximal radial excursion $\rho_1 - \rho_2$.  At time $\tilde{t} = 0$ in these boosted coordinates, we can obtain the interaction energy between the two particles as follows.

First we consider the energy $E_1^{\text{source 2}}$ of particle 1 due to the fields sourced by particle 2.  Here we do not assume that the fields sourced by particle 2 are gravitational.  The quantity $E_1^{\text{source 2}}$ will only depend on the distance from particle 1 to $\rho = 0$ in our boosted coordinates, i.e.~the location of particle 2.  This distance is $|\rho_1 - \rho_2|$, and so we write $E_1^{\text{source 2}} = E_1^{\text{source 2}}(|\rho_1 - \rho_2|)$.  We can also consider the energy $E_1^{\text{no source}}$ of particle 1 in the absence of particle 2.  Similarly $E_1^{\text{no source}} = E_1^{\text{no source}}(|\rho_1 - \rho_2|)$.  Then in our boosted coordinates, the interaction energy between particles 1 and 2 is
\begin{align}
\widetilde{V}_{\text{int}}(|\rho_1 - \rho_2|) = E_1^{\text{source 2}}(|\rho_1 - \rho_2|) - E_1^{\text{no source}}(|\rho_1 - \rho_2|)\,.
\end{align}

Now we would like to go from our current coordinates $(\tilde{t},\tilde{\rho},\tilde{\theta})$ back to the unboosted coordinates $(t,\rho,\theta)$. Recall that energies transform like the $0$-component of a momentum $P_\mu$.  Since at time $\tilde{t} = 0$ the $1$- and $2$- components of the momentum of particle 1 is equal to zero,\footnote{Our arguments also work in the regime that $1$- and $2$-components of the momentum of particle 1 are merely small.} we have
\begin{align}
V_{\text{int}}(\rho_1, \rho_2) \equiv \left(\left.\frac{\partial t(\tilde{t},\tilde{\rho},\tilde{\theta})}{\partial \tilde{t}}\right|_{(\tilde{t},\tilde{\rho},\tilde{\theta}) = (0,\rho_1 - \rho_2, 0)} \right)^{-1}\,\widetilde{V}_{\text{int}}(|\rho_1 - \rho_2|)\,.
\end{align}
Above, the left-hand side is the interaction energy in the unboosted coordinates.

To compute $\frac{\partial t(\tilde{t},\tilde{\rho},\tilde{\theta})}{\partial \tilde{t}}$, we observe that the embedding space coordinates
\begin{align}
\label{E:XprimeA1}
X^A(t,\rho, \theta) = \ell_{\text{AdS}}&\!\left(  \cos\!\left(\frac{ct}{\ell_{\text{AdS}}}\right) \cosh\!\left(\frac{\rho}{\ell_{\text{AdS}}}\right),\, \sin\!\left(\frac{ct}{\ell_{\text{AdS}}}\right) \cosh\!\left(\frac{\rho}{\ell_{\text{AdS}}}\right), \cos(\theta) \sinh\!\left(\frac{\rho}{\ell_{\text{AdS}}}\right),\,\sin(\theta) \sinh\!\left(\frac{\rho}{\ell_{\text{AdS}}}\right)\right)
\end{align}
can be written in terms of $\tilde{t},\tilde{\rho},\tilde{\theta}$ as
\begin{align}
\label{E:XprimeA2}
&\Big(\ell_{\text{AdS}} \cosh(\tilde{\rho}/\ell_{\text{AdS}}) \cosh(\tilde{\rho}_1/\ell_{\text{AdS}})
   \cos(c\tilde{t}/\ell_{\text{AdS}}) + \ell_{\text{AdS}} \cos (\tilde{\theta}) \sinh(\tilde{\rho}/\ell_{\text{AdS}})
   \sinh(\tilde{\rho}_1/\ell_{\text{AdS}}), \,\ell_{\text{AdS}} \cosh(\tilde{\rho}/\ell_{\text{AdS}}) \sin(c\tilde{t}/\ell_{\text{AdS}})\,, \nonumber \\
   & \qquad \quad \ell_{\text{AdS}} \cosh(\tilde{\rho}/\ell_{\text{AdS}}) \sinh
   (\tilde{\rho}_1/\ell_{\text{AdS}}) \cos(c\tilde{t}/\ell_{\text{AdS}})+ \ell_{\text{AdS}} \cos (\tilde{\theta})
   \sinh(\tilde{\rho}/\ell_{\text{AdS}}) \cosh(\tilde{\rho}_1/\ell_{\text{AdS}})\,,\,\ell_{\text{AdS}} \sin(\tilde{\theta})\sinh(\tilde{\rho}/\ell_{\text{AdS}})\Big) 
\end{align}
That is, by equating~\eqref{E:XprimeA1} with~\eqref{E:XprimeA2} we can solve for $t,\rho,\theta$ in terms of $\tilde{t}, \tilde{\rho}, \tilde{\theta}$.  Then a tedious but straightforward calculation gives
\begin{align}
\label{E:boostfactor1}
V_{\text{int}}(\rho_1, \rho_2) = \frac{\cosh(\rho_1/\ell_{\text{AdS}})}{\cosh((\rho_1-\rho_2)/\ell_{\text{AdS}})}\,\widetilde{V}_{\text{int}}(|\rho_1 - \rho_2|)\,.
\end{align}
Expanding out the boost factor for $1 \ll \rho_2/\ell_{\text{AdS}} \ll \rho_1/\ell_{\text{AdS}}$, we obtain
\begin{align}
\label{E:Vintradial1}
\boxed{V_{\text{int}}(\rho_1, \rho_2) = \min\{e^{\rho_1/\ell_{\text{AdS}}}, e^{\rho_2/\ell_{\text{AdS}}}\}\, \widetilde{V}_{\text{int}}(|\rho_1-\rho_2|)}
\end{align}
which is the result reported in the main text.

\section{Two particle angular potential}

Let us repeat the same procedure with two particles at positions $(\rho,0)$ and $(\rho, \delta \theta)$ at time $t = 0$.  In the regime $\delta \theta \to 0$ and $\sinh(\rho/\ell_{\text{AdS}}) \to \infty$ with $\delta \theta\,\sinh(\rho/\ell_{\text{AdS}})$ fixed, we find
\begin{align}
\label{E:Vintangular1}
\boxed{V_{\text{int},\,\rho}(|s_1 - s_2|) = \cosh(\rho/\ell_{\text{AdS}}) \,\widetilde{V}_{\text{int},\,\rho}(|s_1-s_2|)}
\end{align}
where $|s_1 - s_2| = \ell_{\text{AdS}}\sinh(\rho/\ell_{\text{AdS}}) \, \delta\theta$ is the angular arclength at radius $\rho$.

\section{Gravity contribution to radial potential}

Here we compute the gravitational contribution to the radial potential~\eqref{E:Vintradial1}.  To this end, we again consider two particles with rest mass $m$ along a $\theta = 0, \pi$ radial geodesic, initialized at $(\rho_1, 0)$ and $(\rho_2,0)$.   As usual, we consider the super-AdS scale regime $1 \ll \rho_2/\ell_{\text{AdS}} \ll \rho_1/\ell_{\text{AdS}}$.  As before, boosting by $M_{\rho_2 \to 0}$ the second particle is rendered stationary at $\rho = 0$ and the first particle travels along a radial geodesic with maximal excursion $\rho_2 - \rho_1$.

In our boosted coordinates in which the second particle is at rest at $\rho = 0$, we can consider the energy of a probe particle (a proxy for the first partcle) in the background generated by the second particle.  This will be a BTZ background with metric
\begin{align}
\label{E:BTZ1}
ds^2 = - \left(\cosh^2(\rho/\ell_{\text{AdS}}) - \frac{8 G m}{c^2}\right) c^2 dt^2 + \frac{\cosh(\rho/\ell_{\text{AdS}})^2}{\cosh^2(\rho/\ell_{\text{AdS}}) - \frac{8 G m}{c^2}}\,d\rho^2 + \ell_{\text{AdS}}^2\sinh^2(\rho/\ell_{\text{AdS}})\,d\theta^2\,.
\end{align}
Then the energy of a probe particle in this background is
\begin{align}
\label{E:effectiveHBTZ1}
&E = \sqrt{\cosh^2(\rho/\ell_{\text{AdS}}) - \frac{8 G m}{c^2}}   \sqrt{m^2 c^4 + \frac{\cosh^2(\rho/\ell_{\text{AdS}}) - \frac{8 G m}{c^2}}{\cosh^2(\rho/\ell_{\text{AdS}})}\,P_\rho^2 c^2 + \frac{P_\theta^2 c^2}{\ell_{\text{AdS}}^2\sinh^2(\rho/\ell_{\text{AdS}})}}\,.
\end{align}
We can think of $E$ as the temporal component of a $3$-momentum $P_\mu = (E/c, P_\rho, P_\theta)$, which evidently satisfies $g^{\mu \nu} P_\mu P_\nu = - m^2 c^2$.  As such, if $\rho/\ell_{\text{AdS}} \gg 1$ and $P_\rho, P_\theta$ are small, the energy of the particle at position $\rho$ is\footnote{This is analogous to how in $d+1$ spacetime dimensions, the energy of a momentum-less particle at radius $r$ in the Schwarzschild background is $E = mc^2 \sqrt{1 - \frac{r_s}{r^{d-2}}}$.  Here $r_s$ is the Schwarzschild radius.  Then the Newtonian corresponds to expanding in $r_s/r^{d-2} \ll 1$, giving $E \approx mc^2 - \frac{1}{2}\,\frac{r_s}{r^{d-2}}$ which gives the Newtonian potential.  In our setting, we are instead considering the super-AdS regime where $\rho/\ell_{\text{AdS}} \gg 1$.}
\begin{align}
\label{E:Eapprox1}
E \approx mc^2\cosh(\rho/\ell_{\text{AdS}}) - \frac{4 G m^2}{\cosh(\rho/\ell_{\text{AdS}})}\,.
\end{align}
In our current coordinates, the first particle is initially at $\rho = \rho_1 - \rho_2$.  As such, the interacting part of~\eqref{E:Eapprox1} is\footnote{We note that in the sub-AdS regime where $|\rho_1 - \rho_2|/\ell_{\text{AdS}} \ll 1$, the potential becomes $- 4 G m^2 (1 - \frac{1}{2 \ell_{\text{AdS}}^2}\,(\rho_1 - \rho_2)^2 + \cdots)$ where the leading term is evidently a negative constant.  We are not interested in the sub-AdS regime, but rather in the super-AdS regime.}
\begin{align}
\label{E:Vtildegrav1}
\widetilde{V}_{\text{grav}}(|\rho_1 - \rho_2|) = - \frac{4 G m^2}{\cosh((\rho_1 - \rho_2)/\ell_{\text{AdS}})}\,.
\end{align}

We would like to boost back to our original coordinates wherein the first particle is at $\rho = \rho_1$.  Boosts in vacuum AdS$_3$ are not symmetries of the BTZ background (i.e.~they do not preserve the BTZ metric), but nonetheless we can use the same change of coordinates that would have corresponded to a (inverse) boost in vacuum AdS$_3$.

As such, comparing~\eqref{E:Vtildegrav1}
 with~\eqref{E:boostfactor1} we find
\begin{align}
V_{\text{grav}}(\rho_1, \rho_2) = -\frac{\cosh(\rho_1/\ell_{\text{AdS}})}{\cosh((\rho_1-\rho_2)/\ell_{\text{AdS}})}\,\frac{4 G m^2}{\cosh((\rho_1 - \rho_2)/\ell_{\text{AdS}})}
\end{align}
which for $1 \ll \rho_2/\ell_{\text{AdS}} \ll \rho_1/\ell_{\text{AdS}}$ becomes
\begin{align}
\boxed{V_{\text{grav}}(\rho_1, \rho_2) = - \min\{e^{\rho_1/\ell_{\text{AdS}}},e^{\rho_2/\ell_{\text{AdS}}}\}\,8 G m^2\,e^{- |\rho_1 - \rho_2|/\ell_{\text{AdS}}}}
\end{align}
This is our desired result.

\section{Gravity contribution to angular potential}

Here we compute the gravity contribution to the angular potential~\eqref{E:Vintangular1}.  For this we locate the particles at $(\rho,0)$ and $(\rho, \delta \theta)$ at time $t = 0$, and consider the regime $\delta \theta \to 0$ and $\sinh(\rho/\ell_{\text{AdS}}) \to \infty$ with $\delta \theta\,\sinh(\rho/\ell_{\text{AdS}})$ fixed.  Then to second order in $m$ we have
\begin{align}
mc^2 \cosh(\rho/\ell_{\text{AdS}}) + \cosh(\rho/\ell_{\text{AdS}})\,4 G m^2\,\!\left(\frac{1}{2}\,\delta \theta^2\sinh^2(\rho/\ell_{\text{AdS}})\left(1 + \frac{1}{\sqrt{1 + \frac{1}{4}\,\delta\phi^2 \cosh^2(\rho/\ell_{\text{AdS}})}}\right)  - 1 \right)\,.
\end{align}
At large $\rho/\ell_{\text{AdS}}$ the interaction term goes as
\begin{align}
\boxed{V_{\text{grav},\,\rho}(|s_1-s_2|) = \cosh(\rho/\ell_{\text{AdS}})\,4 G m^2\!\left(\frac{1}{2 \ell_{\text{AdS}}^2}\,|s_1-s_2|^2\left(1 + \frac{1}{\sqrt{1 + \frac{1}{4 \ell_{\text{AdS}}^2}\,|s_1-s_2|^2}}\right)  - 1 \right)}
\end{align}
where as before $|s_2 - s_1| = \ell_{\text{AdS}}\sinh(\rho/\ell_{\text{AdS}}) \, \delta\theta$ is the angular arclength at radius $\rho$.

\subsection{Additional Comments on Related Literature}

Here we provide some additional comments on the conjecture in~\cite{Castro:2011zq} and its relation with our work.  The conjecture stipulates that the Ising CFT is dual to pure Einstein gravity with $G = 3\,\frac{c^3}{\hbar}\,\ell_{\text{AdS}}$.  To support the conjecture, the authors consider the Euclidean gravity torus partition which they compute by a sum over saddle points with 1-loop fluctuations.  For $G = 3\,\frac{c^3}{\hbar}\,\ell_{\text{AdS}}$, they find
\begin{align}
\label{E:Castro1}
Z_{\text{grav}}^{\text{saddles}}(\tau) = 8\,Z_{\text{Ising}}(\tau)
\end{align}
where $\tau$ is the modular parameter of the torus.  Although one might expect that~\eqref{E:Castro1} should hold without a factor of $8$, the authors of~\cite{Castro:2011zq} say that the $8$ is unimportant since it may be due to an unknown normalization of the gravity path integral measure.  In the decade or so since~\cite{Castro:2011zq} was published, there are now calculations (e.g.~\cite{Saad:2019lba, Cotler:2020ugk, Cotler:2018zff}) which are sensitive to the prefactor of the gravity path integral measure, and so in hindsight the factor of $8$ in~\eqref{E:Castro1} is cause for concern.  We further remark that bootstrap arguments guarantee $Z_{\text{grav}}^{\text{saddles}}(\tau) \propto Z_{\text{Ising}}(\tau)$ if the gravity partition function (i) has $c = \frac{1}{2}$, (ii) is modular invariant, and (iii) corresponds to unitary quantum mechanics (i.e.~has a state-counting interpretation).  While conditions (i) and (ii) are guaranteed by $G = 3\,\frac{c^3}{\hbar}\,\ell_{\text{AdS}}$ and the symmetries of AdS$_3$ gravity, condition (iii) is far from clear.  Said differently, the remarkable finding of~\cite{Castro:2011zq} is that for pure Einstein gravity satisfying conditions (i) and (ii), condition (iii) also seems to hold.  As such, the gravity theory at least appears to emulate unitary quantum mechanics.  Similar relations to~\eqref{E:Castro1} have been established for higher-genus surfaces~\cite{Jian:2019ubz}, which are similarly remarkable.

In the torus partition function setting, a puzzle is why it should be valid to `approximate' the gravity torus partition function by a sum over saddles and their 1-loop fluctuations.  Indeed, $G = 3\,\frac{c^3}{\hbar}\,\ell_{\text{AdS}}$ constitutes strong coupling, and so off-shell configurations in the gravity path integral (insofar as the path integral is well-defined) should be very important.  As such, if the conjecture of~\cite{Castro:2011zq} holds, then perhaps we should expect
\begin{align}
Z_{\text{grav}}^{\text{saddles}}(\tau) + Z_{\text{grav}}^{\text{non-saddles}}(\tau) \stackrel{?}{=}
Z_{\text{Ising}}(\tau)\,.
\end{align}
If this equation is to be consistent with~\eqref{E:Castro1}, then we would need $Z_{\text{grav}}^{\text{non-saddles}}(\tau) = -7\,Z_{\text{Ising}}(\tau)$.  We also note that the conjecture of~\cite{Castro:2011zq} only pertains to partition functions on surfaces, and so does not make predictions about the bulk dual of Ising CFT operator insertions.

\section{Supplementary Materials D: Hologron Energetics From the Entanglement Renormalization}

Here we provide additional details for the derivation of hologron energetics from entanglement renormalization.
The central object of the derivation is the so-called ascension superoperator of the MERA, whose definition and properties will be discussed at length below.
We start by first discussing how the MERA finds the ground state energy of the Hamiltonian via real-space entanglement renormalization.
Subsequently, we apply insights from this case to derive the exponential decay of the hologron energy as a function of its distance from the boundary of the network.
After this, we turn our attention to deriving the properties of the 2-hologron radial potential.
We use entanglement renormalization to derive the functional form of the potential, quoted in the main text as Eq.~\eqref{eq-maintextVTN}.
In doing so, we demonstrate that the long-distance attractive interaction in the tensor network is dominated by contributions to the stress-tensor (dual to the graviton in AdS/CFT).
We conclude by deriving the form of the angular potential.

\subsection{Setup for MERA Energetics}

As a warm-up to understanding hologron energetics, we first present some intuition for how the MERA computes the ground state energy of the Hamiltonian.
Setting notation, we express the Hamiltonian as a sum of local terms
\begin{equation}
H = \frac{N}{4\pi} \sum_s \hat{h}_s^{(0)}   
\end{equation}
where we define $\hat{h}_s^{(0)}$ to be a local term centered at $s$, which sums to the Hamiltonian.
Notation in hand, the ground state energy in a MERA state $\ket{\psi}$ can be expressed as
\begin{equation} \label{eq-energyexpression}
    E_{\text{GS}} = \frac{N}{4\pi} \sum_s \varepsilon_s \equiv \frac{N}{2\pi v}\sum_s  \bra{\psi} \hat{h}_s^{(0)}\ket{\psi} = \frac{N}{4\pi}\sum_s \left\langle \mathcal{A}_{s, D - 2} \!\left[ \mathcal{A}_{s, D - 3} \!\left[ \cdots \mathcal{A}_{s, 1} [\hat{h}_s^{(0)}] \right] \right]\, \right\rangle_{\!\psi_{\text{core}}}.
\end{equation}
In the above expression, $\psi_{\text{core}}$ is the core state of the MERA and we introduced the $3$-site ascension superoperator $\mathcal{A}_{s, \hat{\rho}}$.
Generally, the ascension superoperators $\mathcal{A}_{s, \hat{\rho}}$ are defined as
\begin{equation}
    \mathcal{A}_{s, \hat{\rho}}: \mathcal{B}_{k\leq 3}\!\left(\mathbb{C}^{2^{D - \hat{\rho}}}\right) \longrightarrow \mathcal{B}_{k\leq 3}\!\left(\mathbb{C}^{2^{D - (\hat{\rho} + 1)}}\right)
\end{equation}
which map the space $\mathcal{B}_{k\leq 3}\!\left(\mathbb{C}^{2^{D - \hat{\rho}}}\right)$ of $k \leq 3$-local operators on the Hilbert space at depth $\hat{\rho} \equiv D - \rho$ to the space $\mathcal{B}_{k\leq 3}\!\left(\mathbb{C}^{2^{D - (\hat{\rho} + 1)}}\right)$ of $k \leq 3$-local operators on the Hilbert space at depth $\hat{\rho} - 1$.
Depending on the position of the boundary operator $s$ and $\hat{\rho}$, there are two possible ascension superoperators: 
\begin{equation} \label{eq-3siteascension}
\mathcal{A}_L[\mathcal{O}^{(\hat{\rho})} ] = \includegraphics[scale = 1, valign = c]{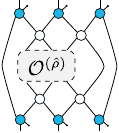} \qquad \mathcal{A}_R[\mathcal{O}^{(\hat{\rho})} ] = \includegraphics[scale = 1, valign = c]{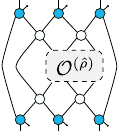}
\end{equation}
These ascension superoperators naturally implement a renormalization group transformation: they act on an operator $\mathcal{O}$ and produce an operator that acts at lower energies and longer distances.
In Eq.~\eqref{eq-energyexpression}, the ascension superoperator is iterated $D - 2$ times to ``ascend'' or ``renormalize'' the operator to the top of the network (i.e.~to the longest length scale present in the system), after which its expectation is taken from the core wavefunction.
Exploiting the translation symmetry of the Hamiltonian (with crucially no assumptions of the translation invariance of the state), we can simplify the expression of the energy to have that
\begin{equation} \label{eq-average3siteascension}
    E_{\text{GS}} = \frac{N}{4\pi}\sum_s \langle \mathcal{A}^{D -2}[ \hat{h}_s^{(0)}] \rangle_{\psi_{\text{core}}} \quad \text{where} \quad \mathcal{A} = \frac{1}{2}(\mathcal{A}_L + \mathcal{A}_R)
\end{equation}
and $\mathcal{A}$ is the average ascension superoperator.

To gain an understanding of the action of the average ascension superoperator, recall that it is linear on the space  $\mathcal{B}_{k\leq 3}\!\left(\mathbb{C}^{2^{D - \hat{\rho}}}\right)$ and can be diagonalized as \cite{vidal_2007_entanglementRG, Evenbly2013}
\begin{equation}
    \mathcal{A}[\, \mathcal{O} \, ] = \sum_{\alpha} \lambda_{\alpha} \varphi^{R}_{\alpha} \text{tr}\left(\varphi^{L}_{\alpha} \mathcal{O} \right)\,, \qquad \text{tr}\left( \varphi_{\alpha}^{L} \varphi_{\beta}^{R} \right) = \delta_{\alpha \beta}\,,
\end{equation}
where $\{\lambda_{\alpha}\}$ is the spectrum of the superoperator and $\varphi^{L/R}_{\alpha}$ are its left and right eigenoperators\footnote{Conventionally, each right eigenoperator is normalized to have norm $1$ when viewed as a vector in the doubled Hilbert space except for $\varphi_{\mathds{1}}^{R}$, which is chosen to simply be the identity for convenience.}.
It has been shown in previous work~\cite{vidal_2007_entanglementRG, Evenbly2013} that the eigenvalues of $\mathcal{A}$ encode the scaling dimensions of operators in the CFT.
Namely, for a binary MERA, $\Delta_{\alpha} = -\log_2(\lambda_{\alpha})$, where $\Delta_{\alpha}$ is the scaling dimension of the ``lattice scaling operator'': $\varphi^R_{\alpha}$.
These scaling dimensions are shown on the left panel of Fig.~\ref{fig:ksiteascension}.
The connection between the spectral decomposition of the ascension superoperator and the scaling operators and dimensions of the CFT naturally provides us with an understanding of how the ground state energy of the MERA is computed.
In particular, for the case of the Ising CFT, the energy density operators can be expressed as
\begin{equation}
    \hat{h}_s^{(0)} = X_{s-1} Z_{s} X_{s + 1} - \frac{1}{2} \left( X_{s - 1} X_{s} + X_{s} X_{s + 1} \right)\,.
\end{equation}
Such operators only have support on the conformal family associated with the $\mathds{1}$-primary\footnote{The restricted support is because the operators are (1) each charge neutral under the $\mathbb{Z}_2$ symmetry (preventing support on $\sigma$) and (2) neutral under the non-invertible $\mathbb{Z}_2$ Kramers-Wannier duality (preventing support on $\varepsilon$).} and thus we can rewrite the operators as\
\begin{equation}
    \hat{h}_s^{(0)} = \varepsilon^{\text{GS}}_s +  c_T\,  \varphi_{T}^R (s) + c_{\bar{T}}\,  \varphi_{\bar{T}}^R (s) + \cdots 
\end{equation}
where $c_{\alpha} \equiv \text{tr}\!\left( \varphi_{\alpha}^{L}\,  \hat{h}_s^{(0)} \right)$, the $\cdots$ indicate terms with higher scaling dimensions, and $\varepsilon_s^{\text{GS}} =  \text{tr}\!\left( \varphi_{\mathds{1}}^{L}\,  \hat{h}_s^{(0)} \right)$ is the ground state energy density.
We will later discuss a procedure to extract the $c_{\alpha}$ coefficients systematically where we will find that $c_{T} = c_{\bar{T}}$, consistent with the energy density being the continuum operator $\propto T(z, \bar{z}) + \bar{T}(z, \bar{z})$ up to an additive constant.
Having resolved $\hat{h}_s^{(0)}$ in terms of the right eigenvectors of $\mathcal{A}$, we have
\begin{equation} \label{eq-h-ascension}
    \mathcal{A}[\hat{h}^{(0)}_s] =  \varepsilon_s^{\text{GS}} + 2^{-\Delta_T} 
 c_T\, \varphi_{T}^R (s) + 2^{-\Delta_{T}} 
 c_{\bar{T}}\, \varphi_{\bar{T}}^R (s)\, +  \cdots  
\end{equation}
where $\Delta_{T} = 2$.
To make this expression more amenable to recursive application, we take $\mathcal{A}[\hat{h}_s^{(0)}] = \varepsilon_s^{\text{GS}} + 2^{-\Delta_T}\hat{h}_s^{(1)}$ and successive applications to be 
\begin{equation} \label{eq-recursiveascension}
    \mathcal{A}[\hat{h}_s^{(k)}] = 2^{-\Delta_T} \hat{h}_{s}^{(k + 1)} \quad \text{where} \quad \hat{h}_s^{(k + 1)} =  c_T\,  \varphi_{T}^R (s) + c_{\bar{T}}\,  \varphi_{\bar{T}}^R (s) + 2^{\Delta_T k} \mathcal{O}\!\left( 2^{-(\Delta_{T} + 1) k} \right)\,.
\end{equation}
As such, we have that the net ground state energy density is given by
\begin{equation} \label{eq-hierarchal-energydensity}
    \varepsilon_s = \varepsilon^{\text{GS}}_s + 2^{-\Delta_T (D - 2)}  \langle \hat{h}_s^{(D-2)} \rangle_{\psi_{\text{top}}} \longrightarrow \varepsilon^{\text{GS}}_s
\end{equation}
as we go to the thermodynamic limit of $D = \log_2(N) \to \infty$.

\subsection{Single Hologron Energetics}

Given the understanding of MERA energetics explained above, we are prepared to derive the excitation energy of a single hologron.
To do so, let us note that since a hologron insertion at $(\hat{\rho}_1, s_1)$ has as a strictly finite forward light cone $\text{LC}(\hat{\rho}_1, s_1)$ (e.g. the region on the boundary which is influenced by the hologron insertion), we have
\begin{equation}
   \frac{4\pi}{N}(E_{1\text{h}} + E_{\text{GS}}) = \sum_{ s \notin \text{bdy} \cap \text{LC}(\hat{\rho}_1, s_1)} \varepsilon_s + \sum_{s \in \text{bdy} \cap \text{LC}(\hat{\rho}_1, s_1)} \langle \hat{h}^{(0)}_s  \rangle_{\psi_{1\text{h}}(\rho_{1})}
\end{equation}
where the expectation value is taken relative to $\ket{\psi_{1\text{h}}(\rho_1)}$: the ground state MERA with a single hologron insertion at $(\hat{\rho}_1, s_1)$.
Up to some small boundary contributions (which are fractionally negligible), the expectation value in the lightcone can be evaluated by first ascending $\hat{h}_s^{(0)}$ via the average ascension superoperator up to the depth below the hologron $\hat{\rho}_1 - 1$.
The result is 
\begin{align}
   \frac{4 \pi }{N}(E_{1\text{h}} + E_{\text{GS}}) &= \sum_{ s \notin \text{bdy} \cap \text{LC}(\hat{\rho}_1, s_1)} \varepsilon_s + \sum_{s \in \text{bdy} \cap \text{LC}(\hat{\rho}_1, s_1)} \langle \mathcal{A}^{\hat{\rho}_1 - 1}[\hat{h}^{(0)}_s] \rangle_{\psi_{1\text{h}}^{\rho_1}(\rho_1)}\\
    &=\sum_{ s \notin \text{bdy} \cap \text{LC}(\hat{\rho}_1, s_1)} \varepsilon_s + \sum_{s \in \text{bdy} \cap \text{LC}(\hat{\rho}_1, s_1)} \left[ \varepsilon^{\text{GS}}_s + 2^{-\Delta_T (\hat{\rho}_1 - 1)} \langle \hat{h}_s^{(\hat{\rho}_1 - 1)} \rangle_{\psi_{1\text{h}}^{\rho_1}(\rho_1)} \right]\,,
\end{align}
where $\psi_{1\text{h}}^{\rho_1}(\rho_{1})$ is a state obtained by taking $\psi_{\text{core}}$ and acting $(D - 2 - \hat{\rho}_1)$ layers of the MERA circuit, with the layer corresponding to radius $\rho_1$ containing a hologron insertion.
Now note that
\begin{equation}
\ket{\psi_{1\text{h}}^{\rho_1}(\rho_1)} = \widetilde{\mathcal{X}}_{\rho_1, s_1}\ket{\psi^{\rho_1}} \quad \text{ where } \quad    
\widetilde{\mathcal{X}}_{\rho_1, s_1} = \begin{tikzpicture}[scale = 0.35, baseline = {([yshift=-.5ex]current bounding box.center)}]
       \draw[color = black, ] (0 - 2.5, 0) -- (0.35355*2.2 - 2.5 , 0.35355*2.2);
       \draw[color = black, ] (0 - 2.5, 0) -- (0.35355*2.2 - 2.5 , -0.35355*2.2);
       \draw[color = black, ] (0 - 2.5, 0) -- (-0.35355*2.2 - 2.5 , 0.35355*2.2);
       \draw[color = black, ] (0 - 2.5, 0) -- (-0.35355*2.2 - 2.5 , -0.35355*2.2);
       \draw[color = black,  -stealth] (0 - 2.5, 0) -- (0.35355*1.7 - 2.5 , 0.35355*1.7);
       \draw[color = black, -stealth ] (0.35355*2.2 - 2.5 , -0.35355*2.2) -- (0.35355*1.1 - 2.5 , -0.35355*1.1);
       \draw[color = black, -stealth ] (0 - 2.5, 0) -- (-0.35355*1.7 - 2.5 , 0.35355*1.7);
       \draw[color = black, -stealth ] (-0.35355*2.2 - 2.5 , -0.35355*2.2) -- (-0.35355*1.1 - 2.5 , -0.35355*1.1);
       \filldraw[draw = black, fill = white] (0 - 2.5,0) circle (0.35);
       \draw[color = black, ] (0, 0) -- (0.35355*2.2 , 0.35355*2.2); 
       \draw[color = black, ] (0, 0) -- (0.35355*2.2 , -0.35355*2.2); 
       \draw[color = black, ] (0, 0) -- (-0.35355*2.2 , 0.35355*2.2); 
       \draw[color = black, ] (0, 0) -- (-0.35355*2.2 , -0.35355*2.2); 
       \draw[color = black,  -stealth] (0, 0) -- (0.35355*1.7 , 0.35355*1.7);
       \draw[color = black, -stealth ] (0.35355*2.2 , -0.35355*2.2) -- (0.35355*1.1 , -0.35355*1.1);
       \draw[color = black, -stealth ] (0, 0) -- (-0.35355*1.7 , 0.35355*1.7);
       \draw[color = black, -stealth ] (-0.35355*2.2 , -0.35355*2.2) -- (-0.35355*1.1 , -0.35355*1.1);
       \filldraw[draw = black, fill = white] (0,0) circle (0.35);
       \draw[color = black, ] (0 - 2.5/2, 0 + 1.23) -- (0.35355*2.2 - 2.5/2, -0.35355*2.2 + 1.23);
       \draw[color = black, ] (0 - 2.5/2, 0 + 1.23) -- (-0.35355*2.2- 2.5/2 , 0.35355*2.2 + 1.23);
       \draw[color = black, ] (0 - 2.5/2, 0 + 1.23) -- (-0.35355*2.2 - 2.5/2 , -0.35355*2.2 + 1.23);
       \draw[color = black] (0 - 2.5/2, 0 + 1.23) -- (0.35355*1.7 - 2.5/2 , 0.35355*1.7 + 1.23);
       \draw[color = black,  -stealth] (0 - 2.5/2, 0 + 1.23) -- (-0.35355*1.7 - 2.5/2 , 0.35355*1.7 + 1.23);
       \filldraw[draw = black, fill = lightorange(ryb)] (0 - 2.5/2,0 + 1.23) circle (0.35);
       \draw[color = black, ] (0 - 2.5/2, 0 + 2.8) -- (0.35355*2.2 - 2.5/2 , 0.35355*2.2 + 2.8);
       \draw[color = black, ] (0 - 2.5/2, 0 + 2.8) -- (0.35355*1.8 - 2.5/2, -0.35355*1.8 + 2.8);
       \draw[color = black, ] (0 - 2.5/2, 0 + 2.8) -- (-0.35355*2.2- 2.5/2 , 0.35355*2.2 + 2.8);
       \draw[color = black, ] (0 - 2.5/2, 0 + 2.8) -- (-0.35355*2.2 - 2.5/2 , -0.35355*2.2 + 2.8);
       \draw[color = black,  -stealth] (0 - 2.5/2, 0 + 2.8) -- (0.35355*1.7 - 2.5/2 , 0.35355*1.7 + 2.8);
       \draw[color = black,  -stealth] (0 - 2.5/2, 0 + 2.8) -- (-0.35355*1.7 - 2.5/2 , 0.35355*1.7 + 2.8);
       \draw[color = black, -stealth ] (-0.35355*2.2 - 2.5/2 , -0.35355*2.2 + 2.8) -- (-0.35355*1.1 - 2.5/2 , -0.35355*1.1 + 2.8);
       \draw[color = black] (0.35355*1.8 - 2.5/2 , -0.35355*1.8 + 2.8) -- (0.35355*1.1 - 2.5/2 , -0.35355*1.1 + 2.8);
       \node at (0.35355*2.2 - 2.5/2  + 0.3, -0.35355*2.2 + 2.8) {\normalsize $X$};
       \filldraw[draw = black, fill = lightorange(ryb)] (0 - 2.5/2,0 + 2.8) circle (0.35);
       \draw[color = black, ] (0 - 2.5, 0 + 4.05 ) -- (0.35355*2.2 - 2.5 , 0.35355*2.2 + 4.05 );
       \draw[color = black, ] (0 - 2.5, 0 + 4.05 ) -- (0.35355*2.2 - 2.5 , -0.35355*2.2 + 4.05 );
       \draw[color = black, ] (0 - 2.5, 0 + 4.05 ) -- (-0.35355*2.2 - 2.5 , 0.35355*2.2  + 4.05);
       \draw[color = black, ] (0 - 2.5, 0 + 4.05) -- (-0.35355*2.2 - 2.5 , -0.35355*2.2 + 4.05);
       \draw[color = black,  -stealth] (0 - 2.5, 0 + 4.05) -- (0.35355*1.7 - 2.5 , 0.35355*1.7 + 4.05);
       \draw[color = black, -stealth ] (0 - 2.5, 0 + 4.05 ) -- (-0.35355*1.7 - 2.5 , 0.35355*1.7 + 4.05 );
       \draw[color = black, -stealth ] (-0.35355*2.2 - 2.5 , -0.35355*2.2 + 4.05 ) -- (-0.35355*1.1 - 2.5 , -0.35355*1.1 + 4.05 );
       \filldraw[draw = black, fill = white] (0 - 2.5,0 + 4.05 ) circle (0.35);
       \draw[color = black, ] (0, 0 + 4.05) -- (0.35355*2.2 , 0.35355*2.2 + 4.05); 
       \draw[color = black, ] (0, 0 + 4.05) -- (0.35355*2.2 , -0.35355*2.2 + 4.05); 
       \draw[color = black, ] (0, 0 + 4.05) -- (-0.35355*2.2 , 0.35355*2.2 + 4.05); 
       \draw[color = black, ] (0, 0 + 4.05) -- (-0.35355*2.2 , -0.35355*2.2 + 4.05); 
       \draw[color = black,  -stealth] (0, 0 + 4.05) -- (0.35355*1.7 , 0.35355*1.7 + 4.05);
       \draw[color = black, -stealth ] (0.35355*2.2 , -0.35355*2.2 + 4.05) -- (0.35355*1.1 , -0.35355*1.1 + 4.05);
       \draw[color = black, -stealth ] (0, 0 + 4.05) -- (-0.35355*1.7 , 0.35355*1.7 + 4.05);
       \filldraw[draw = black, fill = white] (0,0 + 4.05) circle (0.35);
       \draw[color = black] (-0.35355*2.2 - 2.5 , 0.35355*2.2) -- (-0.35355*2.2 - 2.5 , -0.35355*2.2 + 4.05);
       \draw[color = black] (0.35355*2.2, 0.35355*2.2) -- (0.35355*2.2 , -0.35355*2.2 + 4.05);
       \node at (3.02 + 0.5, 0) {$u^{\dagger} \otimes u^{\dagger}$};
       \node at (4.05 + 0.5, 2.025) {$ w (\mathds{1} \otimes X) w^{\dagger} $};
       \node at (2.6 + 0.5, 4.05) {$u \otimes u$};
   \end{tikzpicture}
\end{equation}
and $\ket{\psi^{\rho_1}}$ is obtained by taking $\psi_{\text{core}}$ and acting $(D - 2 - \hat{\rho}_1)$ layers of the MERA circuit, with the outermost layer \textit{not} containing a hologron insertion.
Thus we have
\begin{align}
\frac{4\pi}{N}(E_{1\text{h}} + E_{\text{GS}}) &=\sum_{ s \notin \text{bdy} \cap \text{LC}(\hat{\rho}_1, s_1)} \varepsilon_s + \sum_{s \in \text{bdy} \cap \text{LC}(\hat{\rho}_1, s_1)} \left[ \varepsilon^{\text{GS}}_s + 2^{-\Delta_T (\hat{\rho}_{1} - 1)} \langle \widetilde{\mathcal{X}}_{\rho_1, s_1} \hat{h}_s^{(\hat{\rho}_1 - 1)} \widetilde{\mathcal{X}}_{\rho_1, s_1} \rangle_{\psi^{\rho_1}} \right]\\
&= \frac{4\pi}{N}\,E_{\text{GS}} + \sum_{s \in \text{bdy} \cap \text{LC}(\hat{\rho}_1, s_1)} \left[2^{-\Delta_T (\hat{\rho}_1 - 1)} \langle \widetilde{\mathcal{X}}_{\rho_1, s_1} \hat{h}_s^{(\hat{\rho}_1 - 1)} \widetilde{\mathcal{X}}_{\rho_1, s_1} - \hat{h}_s^{(\hat{\rho}_1 - 1)} \rangle_{\psi^{\rho_1}} \right] \,.
\end{align}
From the above equation, we can read off that the excitation energy of the hologron is given by
\begin{align}
    E_{1\text{h}} &=  \frac{N}{4\pi} \sum_{s \in \text{bdy} \cap \text{LC}(\hat{\rho}_1, s_1)} \left[2^{-\Delta_T (\hat{\rho}_{1} - 1)}\langle \widetilde{\mathcal{X}}_{\rho_1, s_1} \hat{h}_s^{(\hat{\rho}_1 - 1)} \widetilde{\mathcal{X}}_{\rho_1, s_1} - \hat{h}_s^{(\hat{\rho}_1 - 1)} \rangle_{\psi^{\rho_1}} \right]\\
    & \approx \frac{N}{\pi} \times 2^{-\Delta_T \hat{\rho}_1} \times \langle \widetilde{\mathcal{X}}_{\rho_1, s_1} \hat{h}_s^{(\hat{\rho}_1 - 1)} \widetilde{\mathcal{X}}_{\rho_1, s_1} - \hat{h}_s^{(\hat{\rho}_1 - 1)} \rangle_{\psi^{\rho_1}}\\
    &\approx \frac{1}{\pi}\, e^{\rho_1/\ell_{\text{AdS}}} \times \text{tr}\!\left(\varphi_{\mathds{1}}^L \widetilde{\mathcal{X}}_{\rho_1, s_1} \hat{h}_s^{(\hat{\rho}_1 - 1)} \widetilde{\mathcal{X}}_{\rho_1, s_1} \right) = \frac{1}{2} mc^2\,  e^{\rho_1/\ell_{\text{AdS}}}, \qquad \rho_1 \gg \log(2)
\end{align}
where we used the fact that the size of lightcone scales as $2^{\hat{\rho}_1}$ and we remind the reader that $\hat{\rho} = D - \rho$.
Furthermore, in the last step, we used the fact that, in the MERA, expectation values with no hologron insertions reduce to:
\begin{equation}
    \langle \widetilde{\mathcal{X}}_{\rho_1, s_1} \hat{h}_s^{(\hat{\rho}_1 - 1)} \widetilde{\mathcal{X}}_{\rho_1, s_1} - \hat{h}_s^{(\hat{\rho}_1 - 1)} \rangle_{\psi^{\rho_1}} =  \text{tr}\!\left(\varphi_{\mathds{1}}^L \widetilde{\mathcal{X}}_{\rho_1, s_1} \hat{h}_s^{(\hat{\rho}_1 - 1)} \widetilde{\mathcal{X}}_{\rho_1, s_1} \right)
\end{equation}
in the large number of layers ($\rho$ in this case) limit.
This is because other primaries that arise when we resolve $\widetilde{\mathcal{X}}_{\rho_1, s_1} \hat{h}_s^{(\hat{\rho}_1 - 1)} \widetilde{\mathcal{X}}_{\rho_1, s_1} - \hat{h}_s^{(\hat{\rho}_1 - 1)}$ in the right eigenbasis of $\mathcal{A}$ will decay to zero as we ascend this operator from $\hat{\rho}_1 - 1$ to the core state.
From the above, we predict that $\ell_{\text{AdS}}$ is given by
\begin{equation}
    \ell_{\text{AdS}}^{-1} = (\Delta_T - 1) \log(2) = \log(2)\,.
\end{equation}
and the mass energy $mc^2$ is given by
\begin{align}
    \frac{1}{2}\,mc^2 &=  \frac{1}{\pi}\,\text{tr}\left(\varphi_{\mathds{1}}^L \widetilde{\mathcal{X}}_{\rho_1, s_1} \hat{h}_s^{(\hat{\rho}_1 - 1)} \widetilde{\mathcal{X}}_{\rho_1, s_1} \right) \approx \frac{1}{\pi} \,c_{T, \bar{T}}\,\text{tr}\left(\varphi_{\mathds{1}}^L \widetilde{\mathcal{X}}_{\rho_1, s_1} \varphi_{T + \bar{T}}^R \widetilde{\mathcal{X}}_{\rho_1, s_1} \right) \approx 0.92\,.
\end{align}
where we discuss our procedure for extracting $\varphi_{T + \bar{T}}^R$ from the ascension superoperator (which is non-trivial since such operators have the same scaling dimension as $\varphi_{\partial \varepsilon}$ say and are hard to distinguish from eigendecomposition alone) later in this section of the supplementary.
This result is close to the value of the mass-energy $mc^2/2 =  1.25$ that we numerically computed and quoted in the main text.
The slight discrepancy between the two can presumably be accounted for by refining some of the approximations we made.
%

\subsection{Radial Hologron Potential}

We now turn to explain the form of the radial hologron potential.
To do so, we will need to consider MERAs with hologrons inserted at positions $(\hat{\rho}_{1}, s_h)$ and $(\hat{\rho}_{2}, s_h)$, where (without loss of generality) $\hat{\rho}_{1} < \hat{\rho}_{2}$ (i.e.~the second hologron is closer to the center of the network). 
From our discussions of the single-hologron case, we already know that the expression for the two-hologron excitation energy is given by
\begin{align}
    \frac{4\pi}{N} (E_{2\text{h}}  + E_{\text{GS}}) = \sum_{s \notin \text{bdy} \cap \text{LC}_{\mathbf{2}}} \varepsilon_s  &+ \sum_{s \in \text{bdy} \cap \text{LC}_\mathbf{1}} \left[ \varepsilon^{\text{GS}}_s + 2^{-\Delta_T( \hat{\rho}_{1} - 1)} \langle \widetilde{\mathcal{X}}_\mathbf{1} \hat{h}_s^{(\hat{\rho}_{1} - 1)} \widetilde{\mathcal{X}}_\mathbf{1} \rangle_{\psi^{\rho_{1}}_{1\text{h}}(\rho_{2})} \right] \\
    & + \sum_{s \in \text{bdy} \cap \text{LC}_\mathbf{2} \cap \overline{\text{LC}_\mathbf{1}}} \left[ \varepsilon_s^{\text{GS}} + 2^{-\Delta_T( \hat{\rho}_{2} - 1)} \langle\hat{h}_s^{(\hat{\rho}_{2} - 1)} \rangle_{\psi^{\rho_{2}}_{1\text{h}}(\rho_{2})} \right]
\end{align}
where $\text{LC}_{\mathbf{1}} \equiv \text{LC}(\hat{\rho}_{1}, s_{h})$, $\text{LC}_{\mathbf{2}} \equiv \text{LC}(\hat{\rho}_{2}, s_{h})$, and $\mathcal{X}_\mathbf{1} \equiv \mathcal{X}_{\rho_{1}, s_h}$.
Henceforth, we use the subscript notation $\mathbf{1} \equiv (\hat{\rho}_{1}, s_{h})$ and $\mathbf{2} \equiv (\hat{\rho}_{2}, s_{h})$.
By adding and subtracting the Hamiltonian density $h_s^{(\rho_{1} - 1)}$ in the first line, we arrive at
\begin{align}    
    \frac{4\pi}{N} (E_{2\text{h}}(\rho_{1}, \rho_{2})  + E_{\text{GS}}) &= \frac{4 \pi}{N}  (E_{\text{GS}} + E_{1\text{h}}(\rho_{2}))  + \sum_{s \in \text{bdy} \cap \text{LC}_{\mathbf{1}}} \left[ 2^{-\Delta_T( \hat{\rho}_{1} - 1)} \langle \widetilde{\mathcal{X}}_{\mathbf{1}} \hat{h}_s^{(\hat{\rho}_{1} - 1)} \widetilde{\mathcal{X}}_\mathbf{1} - h_s^{(\hat{\rho}_{1} - 1)} \rangle_{\psi^{\rho_{1}}_{1\text{h}}(\rho_{2})} \right]
\end{align}
where we ascended the ``added'' Hamiltonian density up to the location of the second hologron, thereby yielding $E_{1\text{h}}(\rho_{2}) + E_{\text{GS}}$.
We further remind ourselves that the state $\ket{\psi^{\rho_{1}}(\rho_{2})}$ corresponds to the state obtained from acting $\rho_{1}$  MERA layers on the core state with a hologron insertion at layer $\rho_{2} < \rho_{1}$.
To progress, it is necessary to ascend
\begin{equation}
    \delta_{\mathbf{1}, s} \equiv  \widetilde{\mathcal{X}}_{\mathbf{1}} \hat{h}_s^{(\hat{\rho}_{1} - 1)} \widetilde{\mathcal{X}}_{\mathbf{1}} - h_s^{(\hat{\rho}_{1} - 1)}
\end{equation}
up to the location of the second hologron.
However, we recognize that the above operator is no longer $\leq 3$-local.
Indeed, $\mathcal{X}_1$ can be oriented relative to $\hat{h}_s$ in one of six ways: two of which are $6$-local, two of which are $5$-local, and two of which are $4$-local.
As such, to deal with the operator $\delta_{\mathbf{1}, s}$, we need to understand some general properties of higher-body ascension superoperators.

.

\subsubsection{Averaged $k$-site Ascension Superoperators}

We had previously discussed how the spectrum of the average $3$-site ascension superoperator was related to the scaling dimensions and scaling operators of the CFT. 
In this section, we show that these results generalize when we try to ascend $k > 3$-local operators.

\begin{figure}
    \centering
    \includegraphics[width = 494 pt]{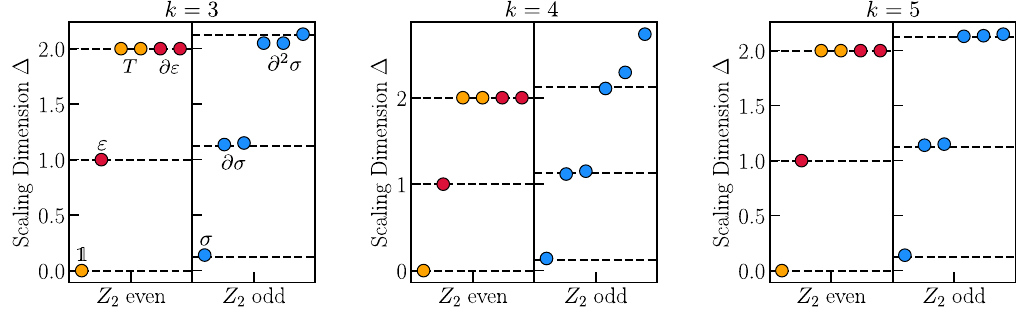}
    \caption{\textbf{Scaling Dimensions from Averaged $k$-site Ascension Superoperators.} In the above, the scaling dimensions computed from averaged $k$-site ascension superoperators are shown.
    In particular, the scaling dimensions are color-coded according to which conformal family they correspond to, with yellow being $\mathds{1}$, red being $\varepsilon$, and blue being $\sigma$.
    We remark that for all values of $k$, the scaling dimensions approximately match CFT predictions for the primaries and lowest descendants, though deviations are often seen at higher descendants.
    }
    \label{fig:ksiteascension}
\end{figure}

Let us consider the set of $k$-site ascension supoperators.
A simple leg counting exercise implies that there are $k - 1$ distinct operators, which we label $\mathcal{A}^{[k, j]}$ $j = 0,...,k - 2$.
For example, for $k = 3$, $ \mathcal{A}_L = \mathcal{A}^{[3, 0]}$ and $\mathcal{A}_R = \mathcal{A}^{[3, 1]}$.
As a consequence, we can define the average $k$-site ascension superoperator to be: 
\begin{equation}
    \mathcal{A}^{[k]} = \frac{1}{k - 1} \sum_{j  = 0}^{k - 2} \mathcal{A}^{[k, j]} \,.
\end{equation}
To verify that the relationship between the conformal tower and the spectrum of the ascension superoperator is maintained for $\mathcal{A}^{[k > 3]}$, we numerically compute its spectral decomposition $\{ \lambda_{\alpha}, \varphi_{\alpha}^L, \varphi_{\alpha}^{R}\}$ and plot $-\log_2(\lambda_{\alpha})$ resolved by the $\mathbb{Z}_2$-charge of $\varphi_{\alpha}^{L/R}$.
The results are shown in Fig.~\ref{fig:ksiteascension} for $\mathcal{A}^{[k = 3, 4, 5]}$.
We find that the spectra for each nicely approximate the lowest scaling dimensions of the conformal tower.

Given the average ascension superoperator at a given size, we can readily resolve $\delta_{\mathbf{1}, s}$ under the lowest dimension scaling operators that overlap with it for each configuration of $s$ and $s_1$.
Unsurprisingly, since $\delta_{\textbf{1}, s}$ is $\mathbb{Z}_2$-charge neutral, it lacks any support on the $\sigma$ conformal family.
As such, we find that for each $s$ and $s_1$
\begin{equation}
    \delta_{\mathbf{1}, s} =  \varepsilon^{h}_{\mathbf{1}, s} + b_{\varepsilon}\, \varphi_{\varepsilon}^R + \left(b_{T}\, \varphi_{T}^R + b_{\bar{T}}\, \varphi_{\bar{T}}^R \right) + \left(b_{\partial \varepsilon}\, \varphi_{\partial \varepsilon}^R + b_{\bar{\partial} \varepsilon}\, \varphi_{\bar{\partial} \varepsilon}^R \right) +  \cdots
\end{equation}
where $\cdots$ are higher order terms, $\varepsilon^{h} = c_{T, \bar{T}} \,\text{tr}(\varphi^L_{\mathds{1}} \delta_{\mathbf{1}s})$, and   $b_{\alpha} = c_{T, \bar{T}}\,\text{tr}(\varphi^L_{\alpha} \delta_{\mathbf{1}s})$.
Note that unlike $\hat{h}$, which only has support on objects in the $\mathds{1}$-conformal family (notably the stress tensor), here there is additional support on the $\varepsilon$-conformal family.
Consequently, under the appropriate average ascension superoperator $\mathcal{A}^{[k]}$: 
\begin{align}
    \mathcal{A}^{[k]}[\delta_{\mathbf{1}, s}] &= \varepsilon^{h}_{\mathbf{1}, s} + 2^{-\Delta_\varepsilon} b_\varepsilon\, \varphi_{\varepsilon}^R (\mathbf{1}, s) + 2^{-\Delta_{T}}\, \sum_{\alpha \in \{T, \bar{T}, \partial \varepsilon, \bar{\partial} \varepsilon\}} b_{\alpha} \varphi_{\alpha}^{R} + \cdots
\end{align}
where $\Delta_{\varepsilon} = 1$ and one should contrast the above against the average ascension of the hamiltonian density in Eq.~\eqref{eq-h-ascension}.
Once again, for ease of iteration, we will take
\begin{equation}
    \mathcal{A}^{[k]}[\delta_{\mathbf{1}, s}] =  \varepsilon_{\mathbf{1}, s}^h +  2^{-\Delta_{\varepsilon}} \delta^{(1)}_{\mathbf{1}, s}\,, \qquad \mathcal{A}^{[k]}[\delta^{(\hat{\rho})}_{\mathbf{1}, s}] = 2^{- \Delta_{\varepsilon}}\delta^{(\hat{\rho} + 1)}_{\mathbf{1}, s}
\end{equation}
where $\delta_{\mathbf{1}, s}^{(\hat{\rho})}$ is given by
\begin{align}
    \delta_{\mathbf{1}, s}^{(\hat{\rho})} &= b_{\varepsilon}\, \varphi_{\varepsilon}^R + 2^{-(\Delta_{T} - \Delta_{\varepsilon})\hat{\rho}} \left(b_{T}\, \varphi_{T}^R + b_{\bar{T}}\, \varphi_{\bar{T}}^R \right) + 2^{-(\Delta_{\partial \varepsilon} - \Delta_{\varepsilon})\hat{\rho}}\left(b_{\partial \varepsilon}\, \varphi_{\partial \varepsilon}^R + b_{\bar{\partial} \varepsilon}\, \varphi_{\bar{\partial} \varepsilon}^R \right) +  \cdots \\
    &= \sum_{\alpha \neq \mathds{1}} b_{\alpha}\, 2^{-(\Delta_{\alpha} - \Delta_{\varepsilon}) \hat{\rho}} \varphi_{\alpha}^R \,.
\end{align}
Note that, under repeated iteration, we have
\begin{align}
    \mathcal{A}^{\hat{\rho}_{2} - \hat{\rho}_{1}}[ \delta_{\mathbf{1}, s}] &= \varepsilon_{\mathbf{1}, s}^h + 2^{-\Delta_{\varepsilon} (\hat{\rho}_{2} - \hat{\rho}_{1})} \delta_{\mathbf{1}, s}^{(\hat{\rho}_{2} - \hat{\rho}_{1})} \,.
\end{align}

\subsubsection{Selectively-Averaged Ascension Superoperators}

\begin{figure}
    \centering
    \includegraphics[width = 494 pt]{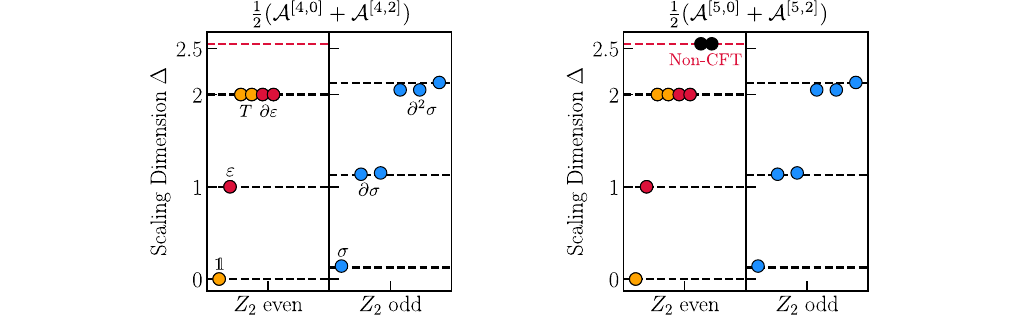}
    \caption{\textbf{Scaling Dimensions from Selectively Averaged $k$-site Ascension Superoperators.} In the above, we show the scaling dimensions computed from selectively averaged ascension superoperators, which naturally appear when ascending products of the Hamiltonian and the hologron operator.
    In particular, on the left, scaling dimensions extracted from the ascension superoperator $\frac{1}{2} \left(\mathcal{A}^{[4, 0]} + \mathcal{A}^{[4, 2]} \right)$  are shown to nicely matches the conformal tower, with no eigenvalues inconsistent with the CFT up to the third descendents.
    Here, the scaling dimensions are color-coded into their conformal families with yellow being $\mathds{1}$, red being $\varepsilon$, and blue being $\sigma$.
    However, on the right, we see that selectively averaging the $k = 5$ ascension superoperator only over even sites $\frac{1}{2} \left(\mathcal{A}^{[4, 0]} + \mathcal{A}^{[4, 2]} \right)$ yields an unphysical (non-CFT) scaling dimension of $\Delta \approx 2.55$.
    }
    \label{fig:selecavg}
\end{figure}

With the above properties of the average ascension superoperator established, it appears as though we can naively proceed with the calculation of the radial potential.
However, there is an important subtlety that we have thus far glossed over.
Unlike the case of the bare Hamiltonian $H$, which ascended via the average ascension superoperator due to the translation invariance of the Hamiltonian terms [see discussion between Eqs.~\eqref{eq-3siteascension}~and~\eqref{eq-average3siteascension}], the operator $\delta_{\mathbf{1}, s}$ cannot be made to ascend via the appropriate average ascension superoperator (even if we compute the potential averaged all angular locations $s_{1}$).
The reason is because, from the definition of $\mathcal{X}$, the leftmost support of $\delta_{\textbf{1}, s}$ always ends on the left leg of a unitary disentangler. 
Instead, depending on the location $(\textbf{1}, s)$, $\delta_{\textbf{1}, s}$ ascends via different combinations $\mathcal{A}^{[k, j]}$.
Specifically, for the case, where $k = 4$, $\delta_{\mathbf{1}, s}$ ascends via $\frac{1}{2} \left(\mathcal{A}^{[4, 0]} + \mathcal{A}^{[4, 2]} \right)$.
Alternatively, for $k = 5$, it ascends via either $\frac{1}{2} \left(\mathcal{A}^{[5, 0]} + \mathcal{A}^{[5, 2]} \right)$ or $\frac{1}{2} \left(\mathcal{A}^{[5, 1]} + \mathcal{A}^{[5, 3]} \right)$.
Finally, for $k = 6$, the relevant ascension superoperator is $\frac{1}{3} \left( \mathcal{A}^{[6, 0]} + \mathcal{A}^{[6, 2]} + \mathcal{A}^{[6, 4]} \right)$.
Essentially, each of these ascension superoperators are averages over either even $j$ or odd $j$.
The ``scaling dimensions'' extracted using the even and odd combinations of the $k = 5$ MERA are reported in Fig.~\ref{fig:selecavg} and while they broadly have the same low energy features, we remark that they have some unphysical scaling dimensions at higher descendents.
These unphysical (non-CFT) scaling dimensions presumably vanish when utilzing larger bond dimension MERAs.

\subsubsection{Computation of Radial Potential Assuming Average Ascension}

To gain some conceptual insight into the form and origin of the interaction potential, it is useful to make the simplification that $\delta_{\textbf{1}, s}$ ascends via the average ascension superoperator.
Then the $2$-hologron energy can be expressed as
\begin{align}
    \frac{4\pi}{N}\,E_{2\text{h}}(\rho_{1}, \rho_{2}) &= \frac{4\pi}{N}\,E_{1\text{h}}(\rho_{2})  + \sum_{s \in \text{bdy} \cap \text{LC}_{\mathbf{1}}} \left[ 2^{-\Delta_{T} (\hat{\rho}_{1} - 1)} \langle \delta_{\mathbf{1}, s} \rangle_{\psi^{\rho_{1}}_{1\text{h}}(\rho_{2})} \right] \\
    &= \frac{4\pi}{N} (E_{1\text{h}}(\rho_{2}) + E_{1\text{h}}(\rho_{1}))  + \sum_{s \in \text{bdy} \cap \text{LC}_{\mathbf{1}}} \left[ 2^{-\Delta_{T} (\hat{\rho}_{1} - 1)} 2^{-\Delta_{\varepsilon} (\hat{\rho}_{2} - \hat{\rho}_{1})} \langle \widetilde{\mathcal{X}}_{\mathbf{2}} \delta_{\mathbf{1}, s}^{(\hat{\rho}_{2} - \hat{\rho}_{1})} \widetilde{\mathcal{X}}_{\mathbf{2}} - \delta_{\mathbf{1}, s}^{(\hat{\rho}_{2} - \hat{\rho}_{1})}  \rangle_{\psi^{\rho_{2}}} \right]\,.
\end{align}
From here, we can isolate the interaction potential
\begin{align}
V_{\text{TN}}(\rho_{1}, \rho_{2}) &=  \frac{N}{4\pi} \sum_{s \in \text{bdy} \cap \text{LC}_{\mathbf{1}}} \left[ 2^{-\Delta_{T} (\hat{\rho}_{1} - 1)} 2^{-\Delta_{\varepsilon} (\hat{\rho}_{2} - \hat{\rho}_{1})} \langle \widetilde{\mathcal{X}}_{\mathbf{2}} \delta_{\mathbf{1}, s}^{(\hat{\rho}_{2} - \hat{\rho}_{1})} \widetilde{\mathcal{X}}_{\mathbf{2}} - \delta_{\mathbf{1}, s}^{(\hat{\rho}_{2} - \hat{\rho}_{1})}  \rangle_{\psi^{\rho_{2}}} \right] \\
    &\approx \frac{N}{\pi} \times 2^{-\hat{\rho}_2} \times \left\{ \sum_{\alpha \neq \mathds{1}} b_{\alpha} \langle \mathcal{X}_{\mathbf{2}} \varphi_{\alpha}^{R} \mathcal{X}_{\mathbf{2}} - \varphi_{\alpha}^{R} \rangle_{\psi^{\rho_2}} 2^{-(\Delta_{\alpha} - 1)(\hat{\rho}_2 - \hat{\rho}_1)} \right\}\\
 & =  b_{\text{AdS}}(\rho_{1}, \rho_{2}) \times \left[ C_1 + C_2\,e^{-|\rho_{1} - \rho_{2}|/\ell_{\text{AdS}}} \right]  \quad \text{ for }\quad |\rho_1 - \rho_2| \gg \ell_{\text{AdS}} \label{E:comparetomain1}
\end{align}
where in going from the first to second line we used the fact that the size of $\text{bdy} \cap \text{LC}_{\mathbf{1}}$ is $\approx 2^{\hat{\rho}_1}$.
This is precisely the form of the potential quoted in the main text.
Note that in the last line, we defined coefficients:
\begin{equation}
    C_{\Delta} = \sum_{\{\alpha \,:\, \Delta_{\alpha} = \Delta\}} C_{\alpha} \equiv \sum_{\{\alpha\,:\,\Delta_{\alpha} = \Delta\}} \frac{1}{\pi} \,b_{\alpha} \langle \widetilde{\mathcal{X}}_{\mathbf{2}} \varphi_{\alpha}^R \widetilde{\mathcal{X}}_{\mathbf{2}} -  \varphi_{\alpha}^R \rangle_{\psi^{\rho_{2}}}\,.
\end{equation}
In the next two subsections, we will discuss how to extract these coefficients both from a parameter fit and also through a more principled numerical protocol.
For the latter, it is useful to unpack the definition of $C_{\alpha}$ a bit, yielding: 
\begin{equation}
    C_{\alpha} = \frac{1}{\pi} \,c_{T, \bar{T}}\left[\text{tr}\!\left( \varphi_{\alpha}^L \mathcal{X}_{\mathbf{1}} \varphi_{T + \bar{T}}^R  \mathcal{X}_{\mathbf{1}}\right) - \delta_{\alpha T} - \delta_{\alpha \bar{T}} \right] \langle \widetilde{\mathcal{X}}_{\mathbf{2}} \varphi_{\alpha}^R \widetilde{\mathcal{X}}_{\mathbf{2}} -  \varphi_{\alpha}^R \rangle_{\psi^{\rho_{2}}}
\end{equation}
where recall that $c_{T} = \text{tr}\!\left( \varphi_{T, \bar{T}}^{L} \hat{h} \right) = c_{\bar{T}} = c_{T, \bar{T}}$.
In terms of $C_{\alpha}$, the most general form of the potential between two operators inserted within the network is given by:
\begin{equation} \label{eq-VTNanalytic}
    V_{\text{TN}}(\rho_{1}, \rho_{2}) = b_{\text{AdS}}(\rho_{1}, \rho_{2})\sum_{\alpha \neq \mathds{1}}\, C_{\alpha} e^{-(\Delta_{\alpha} - 1)|\rho_{1} - \rho_{2}|/\ell_{\text{AdS}}}\,.
\end{equation}
which is quoted in the main text.

\subsubsection{Numerical Confirmation from Parameter Fitting}

\begin{figure}[t]
    \centering
    \includegraphics[scale = 0.58]{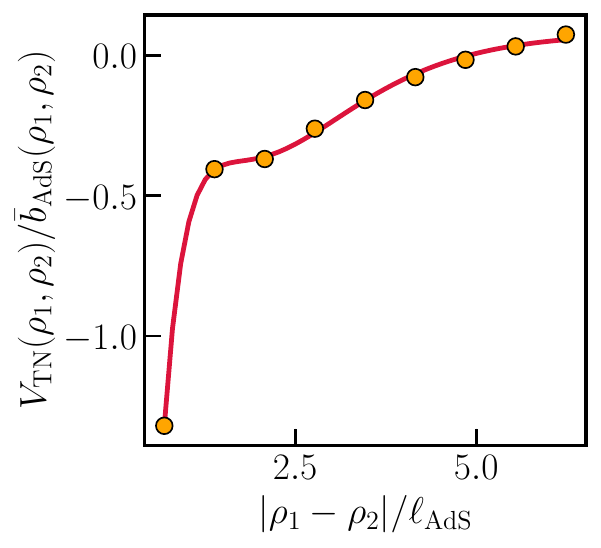}
    \caption{\textbf{Numerically Fitting the $2$-Hologron Potential.} We fit the collapsed $2$-hologron potential with a sum of exponentials as per~\eqref{E:Weq1}, suggested by our ascension superoperator analysis summarized in~\eqref{E:VtildeTN2}.  We find an excellent fit to the entire potential with four linear parameters.}
    \label{fig:enter-label}
\end{figure}

To numerically confirm that Eq.~\eqref{eq-VTNanalytic} is indeed the form of the potential that was numerically computed within the MERA, we first remark that, analytically, we expect: 
\begin{equation}
\label{E:VtildeTN2}
    \widetilde{V}_{\text{TN}}(|\rho_{1} - \rho_{2}|) = \frac{V_{\text{TN}}(\rho_{1}, \rho_{2})}{\bar{b}_{\text{AdS}}(\rho_{1}, \rho_{2})} = \frac{2}{mc^2} \sum_{\alpha \neq \mathds{1}} C_{\alpha} \,e^{-(\Delta_{\alpha} - 1)|\rho_{1} - \rho_{2}|/\ell_{\text{AdS}}}
\end{equation}
where $\bar{b}_{\text{AdS}}(\rho_{1}, \rho_{2}) = \text{min}\{E_{1\text{h}}(\rho_{1}), E_{1\text{h}}(\rho_{2}) \}$.
To see if this is indeed the form found in the numerics, we fit data points obtained from the left hand side of the above (obtained numerically) to:
\begin{align}
\label{E:Weq1}
    W_{A, B, C, D}(|\rho_{1} - \rho_{2}|) &=  A + 4B\,e^{-(\Delta_T - 1)|\rho_{1} - \rho_{2}|/\ell_{\text{AdS}}}+ 2C\,e^{-(2.5 - 1)|\rho_{1} - \rho_{2}|/\ell_{\text{AdS}}} + 6D\,e^{-(3 - 1)|\rho_{1} - \rho_{2}|/\ell_{\text{AdS}}}
\end{align}
where the integer factors in front of $A, B, C, D$ reflect the numerically obtained multiplicity of their associated scaling dimension extracted from diagonalizing $\mathcal{A}$.
Note that the $2.5$ appearing in the exponential next to parameter $C$ is an unphysical scaling dimension that appears when working with selectively averaged ascension superoperators [c.f.~Fig.~\ref{fig:selecavg}]
The resulting parameter fit yields: 
\begin{equation} \label{eq-fit}
    A = 0.08(1) \qquad B = -3.4(1) \qquad C = 25(1) \qquad D = -7.9(4)
\end{equation}
where the error is estimated from the square root of the diagonals of the fit's covariance matrix.

\subsubsection{Protocol for Extracting $C_{\alpha}$ Coefficients}

Having confirmed that the potential we find from our MERA numerics is of the form predicted in Eq.~\eqref{eq-VTNanalytic} via a numerical fit, we now wish to extract the $C_{\alpha}$ coefficients from the expressions predicted in our theory.
Let us recall that these coefficients are defined as
\begin{align}
    C_{\alpha} &= \frac{1}{\pi} \,c_{T, \bar{T}}\left[\text{tr}\!\left( \varphi_{\alpha}^L \mathcal{X}_{\mathbf{1}} \varphi_{T + \bar{T}}^R  \mathcal{X}_{\mathbf{1}}\right) - \delta_{\alpha T} - \delta_{\alpha \bar{T}} \right] \langle \widetilde{\mathcal{X}}_{\mathbf{2}} \varphi_{\alpha}^R \widetilde{\mathcal{X}}_{\mathbf{2}} -  \varphi_{\alpha}^R \rangle_{\psi^{\rho_{2}}} \\ 
    &\approx \frac{1}{\pi} \,c_{T, \bar{T}}\left[\text{tr}\!\left( \varphi_{\alpha}^L \mathcal{X}_{\mathbf{1}} \varphi_{T + \bar{T}}^R  \mathcal{X}_{\mathbf{1}}\right) - \delta_{\alpha T} - \delta_{\alpha \bar{T}} \right] \text{tr} \!\left(\varphi_{\mathds{1}}^{L} \widetilde{\mathcal{X}}_{\mathbf{2}} \varphi_{\alpha}^R \widetilde{\mathcal{X}}_{\mathbf{2}}\right) \quad \text{ for } \quad \rho_{2} \gg \ell_{\text{AdS}} 
\end{align}
where $c_{T} = \text{tr}\!\left( \varphi_{T, \bar{T}}^{L} \hat{h} \right) = c_{\bar{T}} = c_{T, \bar{T}}$.
The above expression reduces the problem of evaluating the coefficients into evaluating traces of small matrices.
The only remaining difficulty is to determine the $\varphi_{\alpha}^{R/L}$ operators.
Such a problem is non-trivial because while different operators $\varphi_{\alpha}^{R/L}$ are distinguished by the ascending superoperator if they have distinct scaling dimensions, if they have the same scaling dimensions (e.g. $T$ vs $\partial \varepsilon$), then they cannot be distinguished by the ascension superoperator and an independent diagnostic is necessary.
To our knowledge, there is no straightforward diagnostic.\footnote{In principle, objects in the $\varepsilon$ conformal family should be odd under the Kramers-Wannier duality which is useful for distinguishing descendants of $\mathds{1}$ from descendants of $\varepsilon$. Unfortunately, this approach does not aid in distinguishing between chiral and anti-chiral descendants.}
Here, we discuss a simple but principled approach that we use to estimate $C_{\alpha}$.

In particular, recall that from the ascension superoperators, we have unique definitions of the primary operators $\varphi_{\mathds{1}}^{R/L},\, \varphi_{\varepsilon}^{R/L},$ and $\varphi_{\sigma}^{R/L}$ since they have unique scaling dimensions and will be obtained uniquely upon any numerical diagonalization of $\mathcal{A}$.
To extract the stress-energy tensor, we first use an ansatz for the chiral and anti-chiral stress tensors motivated from the Koo-Saleur formula~\cite{GuifreKooSaleur}:
\begin{equation}
    \varphi_T^{R, \text{trial}} = \frac{1}{2} \left(h_s + v p_s \right)\,, \qquad \varphi_{\bar{T}}^{R, \text{trial}} = \frac{1}{2} \left(h_s - v p_s\right)\,,
\end{equation}
where $p_s \equiv i[h_s, h_{s -1}]/v$ is the momentum density designed to satisfy the energy-momentum continuity relation $\partial_t h_s = i[H, h_s] = p_{s + 1} - p_s$.
Moreover, in the above $v = 1$ is the ``velocity'' of the CFT, extractable by looking at the spectrum of the Hamiltonian $H$ in the main text and matching it to $E_{\alpha} = v \Delta_{\alpha}$.

While these trial operators are good lattice analogs for the chiral and anti-chiral stress tensors, they are not exactly right eigenvectors of the ascension superoperator.
To remedy this, we define a conformal projection superoperator $\mathcal{P}_{\Delta}$ which projects any input operator into the span of eigenvectors associated with scaling dimension $\Delta$.
Specifically, we define
\begin{equation}
    \varphi_{T}^{R} \equiv \mathcal{P}_{\Delta = 2}[\varphi_T^{R, \text{trial}}]\,, \qquad \varphi_{\bar{T}}^{R} \equiv \mathcal{P}_{\Delta = 2}[\varphi_{\bar{T}}^{R, \text{trial}}]\,.
\end{equation}
Having obtained the stress tensors, it would also be desirable to have a procedure to obtain derivative descendants from our primaries.
To be concrete, we consider the $\varepsilon$ tensor.
To get the chiral and anti-chiral derivative descendent $\partial \varepsilon$ and $\bar{\partial} \varepsilon$, we first must define discrete analogs of the derivative in space and time:
\begin{equation}
\partial_x\varphi_{\varepsilon}^{R} \approx  \varphi_{\varepsilon}^{R}(s + 1) - \varphi_{\varepsilon}^{R}(s)\,,  \qquad \partial_t\varphi_{\varepsilon}^{R} \approx i [H, \varphi_{\varepsilon}^{R}]\,.
\end{equation}
From this we readily obtain ansatzae for the chiral and anti-chiral descendants: $\varphi^{R, \text{trial}}_{ \partial \varepsilon} = \partial_x \varphi_{\varepsilon}^R + \frac{1}{v} \partial_t \varphi_{\varepsilon}^R$ and $\varphi^{R, \text{trial}}_{ \bar{\partial} \varepsilon} = \partial_x \varphi_{\varepsilon}^R - \frac{1}{v} \partial_t \varphi_{\varepsilon}^R$.
With these ansatzes we once again conformally project to define our descendants: 
\begin{equation}
    \varphi_{\partial \varepsilon}^{R} \equiv \mathcal{P}_{\Delta = 2}[\varphi_{\partial \varepsilon}^{R, \text{trial}}]\,, \qquad \varphi_{\bar{T}}^{R} \equiv \mathcal{P}_{\Delta = 2}[\varphi_{\bar{\partial} \varepsilon}^{R, \text{trial}}]\,.
\end{equation}
A similar approach can be used to get higher derivative descendants.
We conclude by remarking that the left eigenvectors can be determined by inverting the matrix whose columns are the appropriate right eigenoperators.
Protocol in hand, we now compute $C_{\alpha}$ for $\alpha \in \{T, \bar{T}, \partial \varepsilon, \bar{\partial} \varepsilon\}$ using the $5$-site average ascension superoperator.
We find that
\begin{equation}
    C_{T} = C_{\bar{T}} \approx -1.151\,, \qquad C_{\partial \varepsilon} = C_{\bar{\partial} \varepsilon} \approx -0.0375\,.
\end{equation}
Evidently, the stress tensor contributions are much larger than the $\partial \varepsilon$ contributions.
We therefore find that the $\Delta = 2$ contribution to the potential is 
\begin{equation}
    C_2 = \sum_{\alpha \in \{T, \bar{T}, \partial \varepsilon, \bar{\partial} \varepsilon\}} C_{\alpha} \approx -2.377\,.
\end{equation}

\subsection{Angular Hologron Potential}

We now turn to deriving the angular potential.
In particular, we consider MERAs with hologrons inserted at positions $(\rho, s_1)$ and $(\rho, s_2)$.
From our earlier discussions, we can already anticipate the form of the two-hologron energies to be
\begin{align}
    \frac{2\pi}{N} (E_{2\text{h}}  + E_{\text{GS}}) = \sum_{s \notin \text{bdy} \cap \text{LC}_{\mathbf{2}}} \varepsilon_s  &+ \sum_{s \in \text{bdy} \cap \text{LC}_\mathbf{1} \cap \overline{\text{LC}_{\mathbf{2}}}} \left[ \varepsilon^{\text{GS}}_s + 2^{-\Delta_T( \hat{\rho} - 1)} \langle \widetilde{\mathcal{X}}_\mathbf{1} \hat{h}_s^{(\hat{\rho} - 1)} \widetilde{\mathcal{X}}_\mathbf{1} \rangle_{\psi^{\rho}} \right] \\ \nonumber
    & + \sum_{s \in \text{bdy} \cap \text{LC}_\mathbf{2} \cap \overline{\text{LC}_\mathbf{1}}} \left[ \varepsilon_s^{\text{GS}} + 2^{-\Delta_T( \hat{\rho} - 1)} \langle\widetilde{\mathcal{X}}_\mathbf{2}\hat{h}_s^{(\hat{\rho} - 1)}\widetilde{\mathcal{X}}_\mathbf{2} \rangle_{\psi^{\rho}} \right] \\
    & + \sum_{s \in \text{bdy} \cap \text{LC}_\mathbf{1} \cap \text{LC}_\mathbf{2}} \left[ \varepsilon_s^{\text{GS}} + 2^{-\Delta_T( \hat{\rho} - 1)} \langle\widetilde{\mathcal{X}}_\mathbf{2}\widetilde{\mathcal{X}}_\mathbf{1}\hat{h}_s^{(\hat{\rho} - 1)}\widetilde{\mathcal{X}}_\mathbf{1}\widetilde{\mathcal{X}}_\mathbf{2} \rangle_{\psi^{\rho}} \right].\nonumber
\end{align}
Using similar tricks to those used to derive the radial potential, we arrive at
\begin{align}
    \frac{4\pi}{N} V_{\text{TN},\,\rho}(s_1, s_2) =  \sum_{s \in \text{bdy} \cap \text{LC}_\mathbf{1} \cap \text{LC}_\mathbf{2}} \left[2^{-\Delta_T( \hat{\rho} - 1)} \langle\widetilde{\mathcal{X}}_\mathbf{2}\widetilde{\mathcal{X}}_\mathbf{1}\hat{h}_s^{(\hat{\rho} - 1)}\widetilde{\mathcal{X}}_\mathbf{1}\widetilde{\mathcal{X}}_\mathbf{2} - \widetilde{\mathcal{X}}_\mathbf{1}\hat{h}_s^{(\hat{\rho} - 1)}\widetilde{\mathcal{X}}_\mathbf{1} - \widetilde{\mathcal{X}}_\mathbf{2}\hat{h}_s^{(\hat{\rho} - 1)}\widetilde{\mathcal{X}}_\mathbf{2}   \rangle_{\psi^{\rho}} \right] .
\end{align}
The key subtlety in the above expression is the size of the lightcone overlap $\text{bdy} \cap \text{LC}_\mathbf{1} \cap \text{LC}_\mathbf{2}$.
From simple geometric considerations, we find
\begin{equation}
    |\text{bdy} \cap \text{LC}_\mathbf{1} \cap \text{LC}_\mathbf{2}| \sim \begin{cases} \mathcal{O}(2^{\hat{\rho}}) & |s_1 - s_2| = 1 \\ \mathcal{O}(1) & |s_1 - s_2| = 2 \\ 0 & |s_1 - s_2| > 2 \end{cases}
\end{equation}
and so we arrive at
\begin{equation}
    V_{\text{TN},\,\rho}(s_1, s_2) \sim \left[\begin{cases}
        2^{\rho} & |s_1 - s_2| = 1 \\ 
        2^{2 \rho}/N & |s_1 - s_2| = 2 \\
        0 & |s_1 - s_2| > 2
    \end{cases}\right] \times \langle\widetilde{\mathcal{X}}_\mathbf{2}\widetilde{\mathcal{X}}_\mathbf{1}\hat{h}_s^{(\hat{\rho} - 1)}\widetilde{\mathcal{X}}_\mathbf{1}\widetilde{\mathcal{X}}_\mathbf{2} - \widetilde{\mathcal{X}}_\mathbf{1}\hat{h}_s^{(\hat{\rho} - 1)}\widetilde{\mathcal{X}}_\mathbf{1} - \widetilde{\mathcal{X}}_\mathbf{2}\hat{h}_s^{(\hat{\rho} - 1)}\widetilde{\mathcal{X}}_\mathbf{2}   \rangle_{\psi^{\rho}}\,.
\end{equation}
For $|s_1 - s_2| = 1$ the above is consistent with the gravitational prediction.

\section{Supplementary Materials E: Additional Remarks Regarding Experimental Implementation}

Below we provide some additional remarks on some details of the experimental implementation.
In particular, we first provide some analytic arguments for the presence of the Ising CFT in the long-range transverse-field Ising model that can be realized in the Rydberg chain and numerically identify its critical point.
Subsequently, we comment on the stability of our experimental proposal to gate imperfections.

\subsection{Effect of Errors on Hologron Energetics}

We now briefly analyze how robust the hologron potentials reported in the main text are to errors.
Specifically, we envision noisy digital state preparation of the two-hologron state and the ground state, where each gate of the MERA is replaced with a quantum channel $\mathcal{E}_u(\cdot)$ and $\mathcal{E}_w(\cdot)$.
Each of these quantum channels can be expressed as
\begin{equation}
    \mathcal{E}_u(\sigma) = \int \left[\prod_{\mathcal{S}} d \theta_{\mathcal{S}} \right] p(\boldsymbol{\theta})\, \widetilde{u}^{\dagger}_{\boldsymbol{\theta}}\, \sigma\, \widetilde{u}_{\boldsymbol{\theta}}\,,\qquad \mathcal{E}_w(\sigma) = \int \left[\prod_{\mathcal{S}} d \theta_{\mathcal{S}} \right] q(\boldsymbol{\varphi})\,  \widetilde{w}^{\dagger}_{\boldsymbol{\varphi}}\, \sigma\, \widetilde{w}_{\boldsymbol{\varphi}}
\end{equation}
where $\mathcal{S}$ labels different two-qubit Pauli matrices, $\theta_{\mathcal{S}}$ and $\varphi_{\mathcal{S}}$ are independent random variables for each $\mathcal{S}$ (and for each location in the network) distributed according to $p$ and $q$ respectively, and $u_{\boldsymbol{\varphi}}$ and $w_{\boldsymbol{\varphi}}$ are deformed versions of the MERA tensors
\begin{equation}
     u_{\boldsymbol{\theta}} =  u \exp\left(i \sum_{\mathcal{S}} \theta_\mathcal{S}\, \mathcal{S} \right)\,, \qquad w_{\boldsymbol{\varphi}} =  w \exp\left(i \sum_{\mathcal{S}} \theta_\mathcal{S}\, \mathcal{S} \right)
\end{equation}
Here, the choice of distribution $p, q$ determines the error model, as they uniquely determine the Kraus operators 
\begin{equation}
    K^u_{\boldsymbol{\theta}} = \sqrt{\prod_{\mathcal{S}} p(\theta_{\mathcal{S}})} \,u_{\boldsymbol{\theta}}^{\dagger}\,,\qquad     K^w_{\boldsymbol{\varphi}} = \sqrt{\prod_{\mathcal{S}} p(\varphi_{\mathcal{S}})} \,w_{\boldsymbol{\varphi}}^{\dagger}  
\end{equation}
Below we consider two types of error models: the first is a channel where errors arise from stochastic control and calibration errors (where each unitary gate is independently rotated away from the ideal gates in a manner that varies independently from trial to trial), and the second is a channel implementing dephasing noise.

Given functions $p$ and $q$, we can compute expectation values in the prepared density matrix by generating MERA states prepared by sampling unitaries $u_{\boldsymbol{\theta}}$ and $w_{\boldsymbol{\varphi}}$ with parameters drawn from $p$ and $q$, respectively.
Expectation values of the hologron potential can then be computed for each of these MERA states and the average will be the hologron potential evaluated within the mixed state. 
Explicitly, let $\sigma$ be the density matrix formed from sending the $\ket{0}$ initialized ancillas and the core state through the noisy MERA quantum channel and $\sigma^{1\text{h}}(\rho)$ and $\sigma^{2\text{h}}(\rho, \rho')$ be the density matrices where the initial state has one and two ancillas initialized in $\ket{1}$.
Then the noisy hologron potential is
\begin{align}
    \overline{V(\rho, \rho')} &= \text{Tr} \left[ \sigma_{2\text{h}}(\rho, \rho') H \right] - \text{Tr} \left[ \sigma_{1\text{h}}(\rho) H \right] - \text{Tr} \left[ \sigma_{1\text{h}}(\rho') H \right] + \overline{E_{GS}} \\
    &\approx \sum_{i = 1}^{N_{\text{samples}}} \bra{\Psi^{2\text{h}}_{\boldsymbol{\theta}_i, \boldsymbol{\phi}_i} (\rho, \rho')} H \ket{\Psi^{2\text{h}}_{\boldsymbol{\theta}_i, \boldsymbol{\phi}_i} (\rho, \rho')} - \bra{\Psi^{1\text{h}}_{\boldsymbol{\theta}_i, \boldsymbol{\phi}_i} (\rho)} H \ket{\Psi^{1\text{h}}_{\boldsymbol{\theta}_i, \boldsymbol{\phi}_i} (\rho)} - \bra{\Psi^{1\text{h}}_{\boldsymbol{\theta}_i, \boldsymbol{\phi}_i} (\rho')} H \ket{\Psi^{1\text{h}}_{\boldsymbol{\theta}_i, \boldsymbol{\phi}_i} (\rho')}  + \overline{E_{GS}}
\end{align}
where $\overline{E}_{\text{GS}}$ is the noisy estimate for the ground state energy and the second line can be computed efficiently from the MERA architecture of the states $\ket{\Psi^{2\text{h}}_{\boldsymbol{\theta}_i, \boldsymbol{\phi}_i}(\rho, \rho')}$ and  $\ket{\Psi^{1\text{h}}_{\boldsymbol{\theta}_i, \boldsymbol{\phi}_i}(\rho)}$ and Monte Carlo sampling $p, q$.

\subsubsection{Depolarizing Errors}

\begin{figure}
    \centering
    \includegraphics[width = 0.8\textwidth]{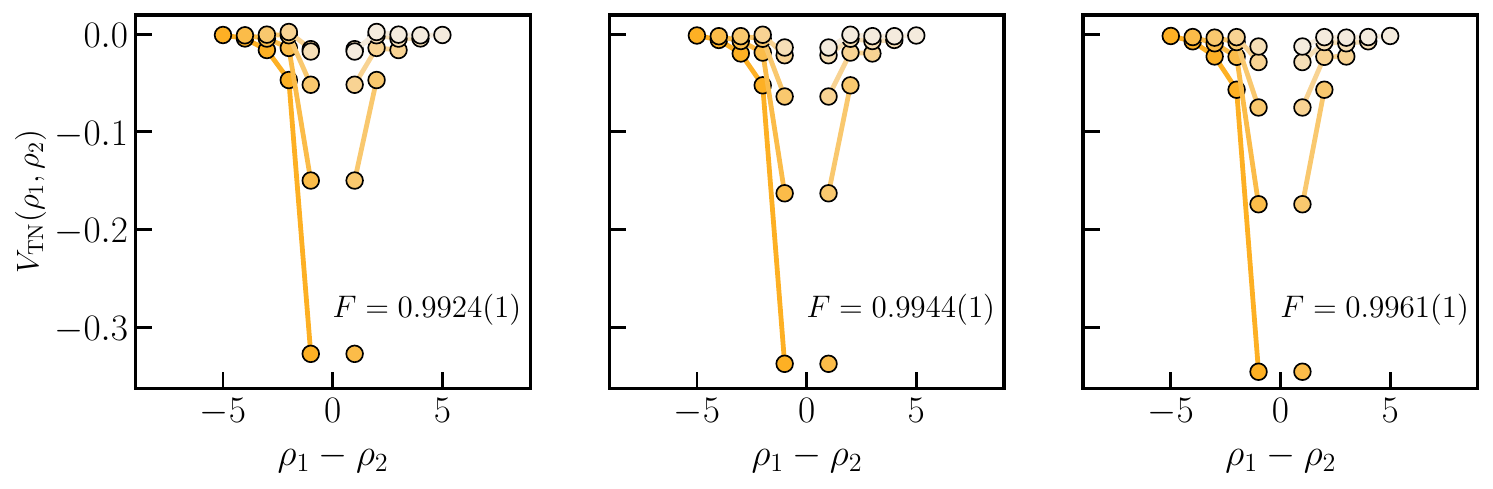}
    \caption{\textbf{Hologron Potential with Depolarizing Errors.} We compute the $2$-hologron potential for a noisy MERA tensor network with depolarization errors.
    Specifically, from left to right, we show the two-particle potential for a MERA with two-qubit gate fidelity given by $F = 0.9924(1), 0.9944(1), \text{ and } 0.9961$ respectively.
    In all cases, we find that the potential is largely attractive.
    }
    \label{fig:gateimperfection}
\end{figure}

\begin{figure}
    \centering
    \includegraphics[width = 0.8\textwidth]{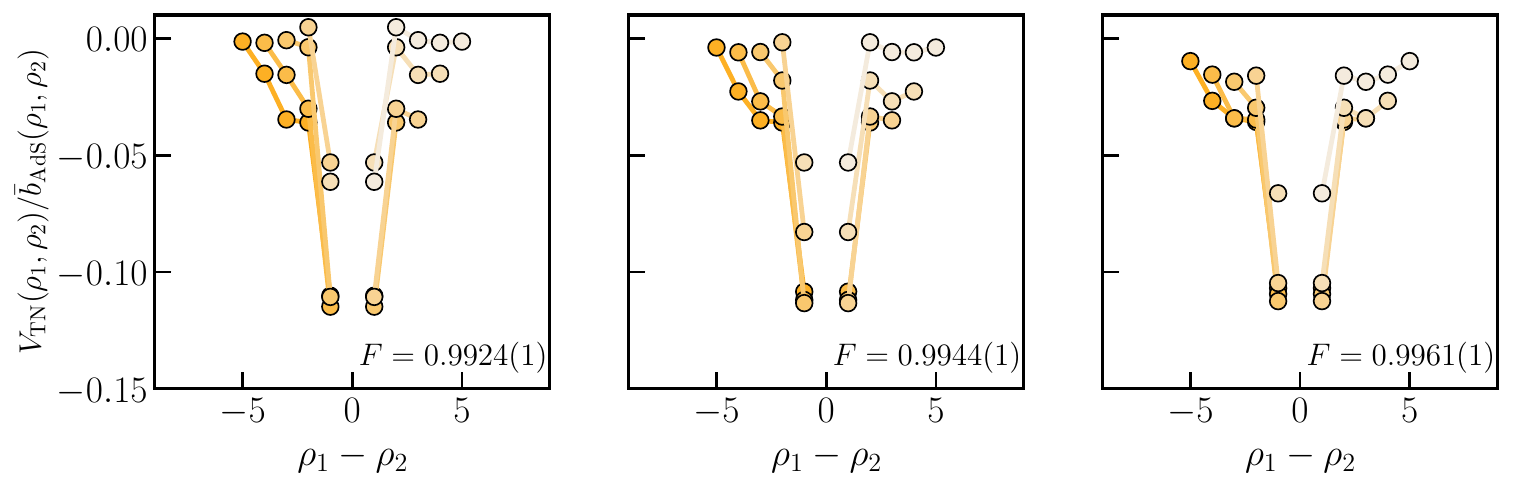}
    \caption{\textbf{Collapsed Hologron Potential with Depolarizing Errors.} We compute the $2$-hologron potential divided by the average boost factor for a noisy MERA tensor network with depolarization errors.
    Specifically, from left to right, we show the two-particle potential for a MERA with two-qubit gate fidelity given by $F = 0.9924(1), 0.9944(1), \text{ and } 0.9961$ respectively.
    As we improve the fidelity, we find that the collapse of the potential improves but remains imperfect.
    }
    \label{fig:gateimperfectioncol}
\end{figure}

We start by considering $2$-qubit errors modeled by independently sampling $\theta_{\mathcal{S}}(\rho, s), \varphi_{\mathcal{S}}(\rho, s)$ from $[0, 2\pi \varepsilon)$ for each Pauli $\mathcal{S}$ and each location $(\rho, s)$ in the network.
Here, $\varepsilon$ is a proxy for the gate fidelity and for each $\varepsilon$, and so we can compute the average two-qubit gate fidelity for $u$ as\footnote{We remark that Eq.~\eqref{eq-Fu} can be easily derived via $
\mathbb{E}_{\psi \sim \text{Haar}}[\rho_{\psi} \otimes \rho_{\psi}] = \frac{ 1 + \mathsf{SWAP}}{ d (d + 1)}$ for $\rho_{\psi} = |\psi\rangle \langle \psi|$.}
\begin{equation} \label{eq-Fu}
    F_u \equiv \mathbb{E}_{\boldsymbol{\theta} \sim p} \mathbb{E}_{|\psi\rangle \sim \text{Haar}}\Big[ |\langle \psi| u^{\dagger} u_{\boldsymbol{\theta}} |\psi\rangle|^2\Big] = \mathbb{E}_{\boldsymbol{\theta} \sim p}\!\!\left[ \frac{ |\text{tr}(u^{\dagger}u_{\boldsymbol{\theta}})|^2 + d}{d (d + 1)}\right],
\end{equation}
where $d = 4$ is the dimension of the matrix $u$.
The average two-qubit gate fidelity $F_u$ for $w$ is defined similarly.
For $\varepsilon = 7 \cdot 10^{-3},\, 6 \cdot 10^{-3},\,  5\cdot 10^{-3}$, we find a two-qubit gate fidelity of $F = 0.9924(1), 0.9944(1), 0.9961(1)$.
The latter two parameters give rise to a two-qubit gate fidelities that are within the range of experimentally achieved fidelity of $99.55\%$  \cite{evered_2023_gates}. 
The numerical results for the two-hologron potential is shown in Fig.~\ref{fig:gateimperfection} for these three fidelities.
We find that qualitatively the potential remains attractive and has roughly the same features as the potentials shown in the main text.
We note that achieving a good collapse (See Fig.~\ref{fig:gateimperfectioncol}) by dividing via the averaged boost factor is challenging because $\ell_{\text{AdS}}$ can easily vary from the theoretically expected value since the Hamiltonian no longer strictly has supported on the $\mathds{1}$-conformal family in the noisy ascension superoperator.
As such, in the noisy system, it may be necessary to find the appropriate noisy value of $\ell_{\text{AdS}}$ that is capable of collapsing the observed potential to gain an appropriate gravitational interpretation of the bulk.

\subsubsection{Dephasing Errors}

\begin{figure}
    \centering
    \includegraphics[width = 0.8\textwidth]{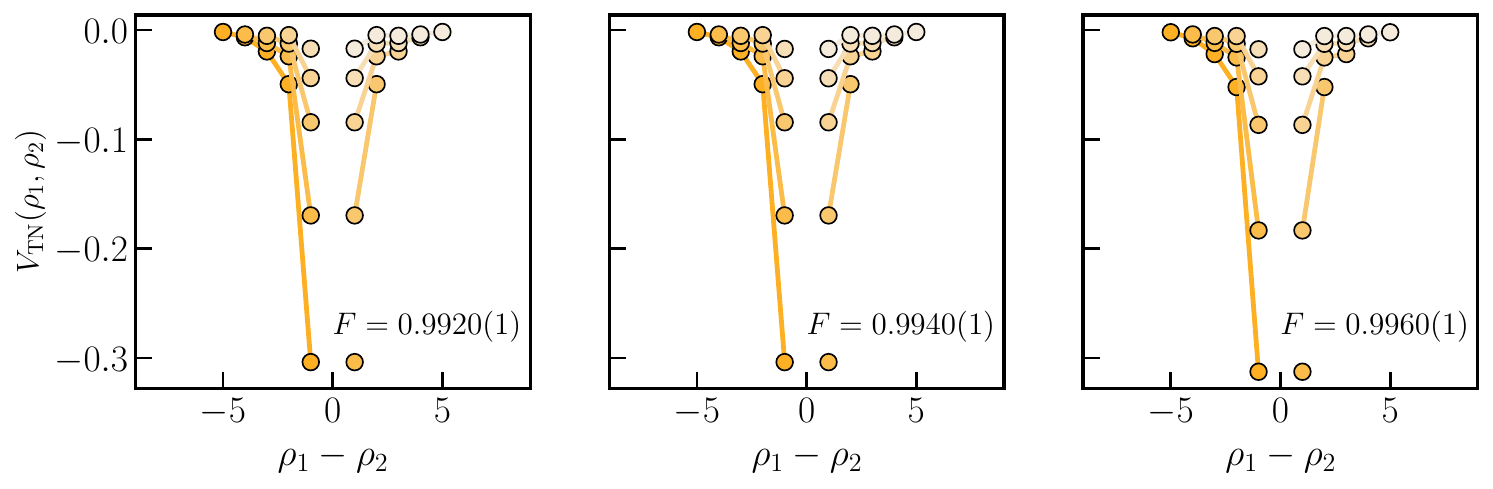}
    \caption{\textbf{Hologron Potential with Dephasing Errors.} Here we compute the $2$-hologron potential for a noisy MERA tensor network with dephasing errors.
    Specifically, from left to right, we show the two-particle potential for a MERA with two-qubit gate fidelity given by $F = 0.9920(1), 0.9940(1), \text{ and } 0.9960(1)$ respectively.
    In all cases, we find that the potential is largely attractive.
    }
    \label{fig:dgateimperfection}
\end{figure}

\begin{figure}
    \centering
    \includegraphics[width = 0.8\textwidth]{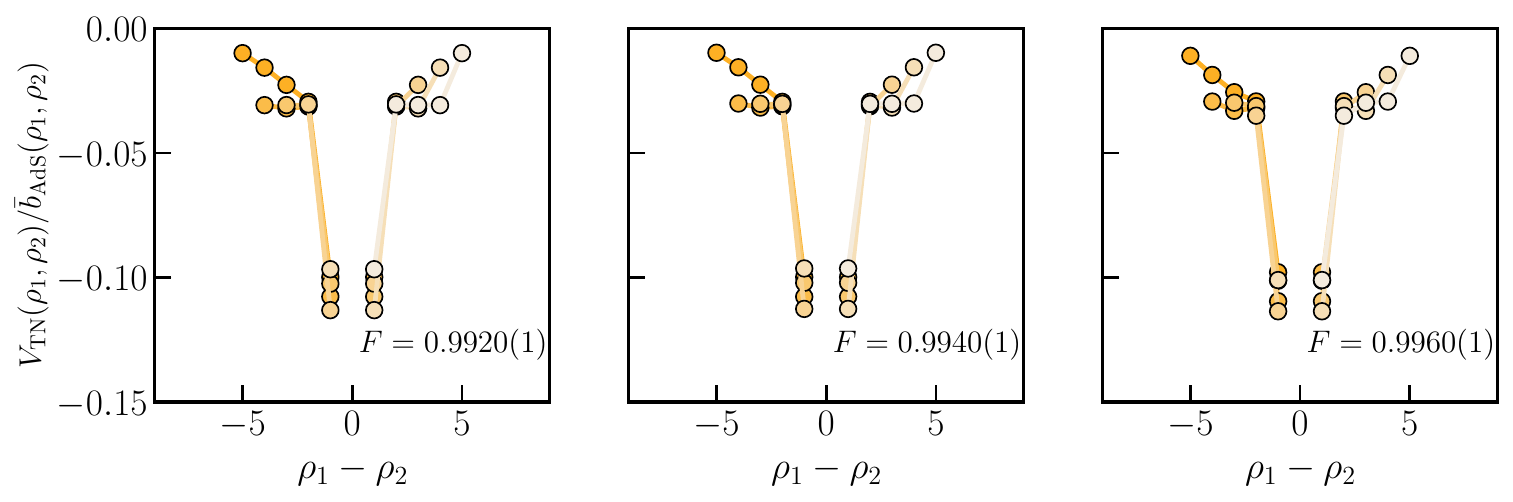}
    \caption{\textbf{Collapsed Hologron Potential with Dephasing Errors.} We compute the collapsed $2$-hologron potential for a noisy MERA tensor network with dephasing errors.
    Specifically, from left to right, we show the two-particle potential for a MERA with two-qubit gate fidelity given by $F = 0.9924(1), 0.9944(1), \text{ and } 0.9961$ respectively.
    As we improve the fidelity, we find that the collapse of the potential improves but remains imperfect.
    }
    \label{fig:dgateimperfectioncol}
\end{figure}

Next we consider dephasing errors.
The Kraus operators are then
\begin{equation} 
    K_{11}^{u} = (1 - \varepsilon)\,  u^{\dagger}(\mathds{1} \otimes \mathds{1})\,, \quad K_{1Z}^{u} = \sqrt{\varepsilon ( 1 - \varepsilon)}\, u^{\dagger}\, (\mathds{1} \otimes Z)\,, \quad K_{Z1}^{u} = \sqrt{\varepsilon ( 1 - \varepsilon)}\, u^{\dagger}\, (Z\otimes \mathds{1})\,, \quad K_{ZZ}^{u} = \varepsilon\, u^{\dagger}\, (Z \otimes Z)\,,
\end{equation}
and similarly for $w$.
Given the Kraus operators, we can compute the many-body fidelity by
\begin{align}
    F_u &= \sum_{a, b \in \{1, Z\}} \mathbb{E}_{\psi \sim \text{Haar}} \left[|\bra{\psi} u^{\dagger} K_{ab}^{u, \dagger} \ket{\psi}|^2\right] \\
    &= \sum_{a, b \in \{1, Z\}} \mathbb{E}_{\psi \sim \text{Haar}} \left\{ \text{tr} \left[ (u^{\dagger} K_{ab}^{u, \dagger}) \otimes (K_{ab}^u u)\,  \rho_{\psi} \otimes \rho_{\psi}  \right] \right\}
\end{align}
where $\rho_{\psi} = |\psi\rangle \langle \psi|$. 
We can simplify the above using
\begin{equation}
    \mathbb{E}_{\psi \sim \text{Haar}}[\rho_{\psi} \otimes \rho_{\psi}] = \frac{ \mathds{1}\otimes \mathds{1} + \mathsf{SWAP}}{ d (d + 1)}
\end{equation}
where $\mathsf{SWAP}$ acts on two copies of the Hilbert space.
With the above formula, we have
\begin{align}
    F_u &= \sum_{a, b \in \{1, Z\}} \frac{p_{a} p_{b}}{d (d + 1)}  \left[|\text{tr}\!\left( \mathcal{S}_{a} \otimes \mathcal{S}_b \right)|^2 + \text{tr}\!\left( \mathcal{S}_{a}\mathcal{S}_b \right) \right] \\
    &= \sum_{a, b \in \{1, Z\}} \frac{p_{a} p_{b}}{d (d + 1)}  \left[ d^2 \delta_{a1} \delta_{b1} + d \right] =  \frac{d( 1- \varepsilon)^2 + 1}{(d + 1)} 
\end{align}
where $\mathcal{S}_1 = \mathds{1}$, $\mathcal{S}_{Z} = Z$, $p_1 = (1 - \varepsilon)$, and $p_Z = \varepsilon$.
An identical expression is obtained for $F_w$.
We choose $\varepsilon = 0.005, 0.0037, \text{ and } 0.0025$ to reproduce two-qubit gate fidelities of $F \approx 0.9920(1), 0.9940(1), 0.9960(1)$ respectively, the former two are within the experimentally achieved gate fidelity of $99.5\%$.
The numerical results for the two-hologron potential is shown in Fig.~\ref{fig:dgateimperfection} for these three fidelities.
We find that qualitatively the potential remains attractive and has roughly the same features as the potentials shown in the main text.
Similar to the noisy calibration case, we do not observe a good collapse under the AdS boost factor as shown in Fig.~\ref{fig:dgateimperfectioncol}.

\end{document}